\documentstyle[12pt,epsf]{article}
 \font\blackboard=msbm10 
 \font\blackboards=msbm7 \font\blackboardss=msbm5
 \newfam\black \textfont\black=\blackboard
 \scriptfont\black=\blackboards \scriptscriptfont\black=\blackboardss
 \def\Bbb#1{{\fam\black\relax#1}}

\font\ninerm=cmr9
\def\uniset{\rlap{\ninerm 1}\kern.15em 1}

\def\mi{{\rm i}}
\def\e{\mathop{\rm e}\nolimits}
\def\sq2{\sqrt{2}}
\def\sqn{\sqrt{N}}
\def\defi{\stackrel{\rm def}{=}}
\newcommand\grsim{\mathrel{\hbox{\lower1ex\hbox{\rlap{$\sim$}\raise1ex\hbox{$>$}}}}}
\newcommand\losim{\mathrel{\hbox{\lower1ex\hbox{\rlap{$\sim$}\raise1ex\hbox{$<$}}}}}

\newtheorem{theo}{Theorem}
\newtheorem{lem}{Lemma}
\newtheorem{conj}{Conjecture}

\title{Chaotic eigenfunctions in phase space}
\author{{\bf S. Nonnenmacher and A. Voros} \\
\\
CEA--Saclay, Service de Physique Th\'eorique\\
F-91191 Gif-sur-Yvette CEDEX (France)\\
{ E-mail : nonnen@spht.saclay.cea.fr, voros@spht.saclay.cea.fr}}

\begin{document}
\maketitle
{\abstract 
We study individual eigenstates of quantized area-preserving maps on the 2-torus
which are classically chaotic. In order to analyze their semiclassical behavior, we use the 
Bargmann--Husimi representations for quantum states, as well as their
stellar parametrization, which encodes states through a minimal
set of points in phase space (the constellation of zeros of the Husimi density). We rigorously
prove that a semiclassical uniform distribution of Husimi densities
on the torus entails a similar equidistribution for the corresponding constellations. We deduce from this property a universal behavior 
for the phase patterns of chaotic Bargmann eigenfunctions, 
which is reminiscent of the
WKB approximation for eigenstates of integrable systems (though in a weaker sense). In order to obtain more precise information on ``chaotic eigenconstellations", we then model their properties by ensembles of random
states, generalizing former results on the 2-sphere to the torus geometry. This approach yields statistical predictions for the constellations, which fit quite well the chaotic data. We finally observe that specific dynamical information, e.g. the presence of high peaks (like scars) in Husimi densities, can be recovered from the knowledge of a few long-wavelength Fourier coefficients, which therefore appear as valuable order parameters at the level of individual chaotic eigenfunctions. 
}

\section{Introduction}

We present a detailed exploration of the eigenfunctions of certain quantum maps,
corresponding to classical {\sl chaotic maps\/} 
defined over a torus phase space.
Those eigenfunctions will be described in phase space 
by means of the Bargmann--Husimi--stellar family of representations.
This study is intended to contribute towards understanding the general issue
of the {\sl semiclassical behavior of individual eigenfunctions\/}, which is 
a basic and largely unsolved problem when the classical dynamics is chaotic.

When this dynamics is integrable, we may consider that the semiclassical
behavior of the eigenfunctions is known: it follows a WKB Ansatz
\begin{equation}
\label{WKB}
\psi \sim \sum_j  A_j \e^{\mi S_j/\hbar}, \qquad \hbar \to 0,
\end{equation}
where $S_j$ are the (finitely many) branches of the classical action at
energy $E$ and $A_j$ are certain invariant 1/2-densities. Thus in principle
(disregarding fine points and technicalities), the eigenfunctions
and all derived quantities are computable in the semiclassical regime
(e.g., the discrete eigenvalues are yielded by EBK quantization formulae). 

By contrast, no definite behavior is known when the classical
dynamics is far from integrable, namely chaotic (especially, ergodic).
The classical ergodic property creates a definite semiclassical constraint,
expressed by the Schnirelman (family of) theorem(s): 
basically, suitable phase-space measures constructed from the eigenfunctions
(e.g., their Husimi densities) must tend towards the classical phase-space
ergodic measure as $\hbar \to 0$. 
However, as far as the effective eigenfunction shapes themselves 
are concerned, this property brings very weak and indirect information, 
except on the negative side: it is strong enough to forbid all
familiar semiclassical patterns like WKB, Gaussian wave packets, etc.;
but at the same time, it yields no affirmative shape prediction (as yet).
Moreover, the theorem allows a tiny fraction of eigenstates to evade ergodic
behavior, and any of these will wholly escape description at this stage.

We therefore intend to refine the description of quantum eigenfunctions
in a classically chaotic regime (``chaotic eigenfunctions", for short), 
and will employ for this purpose 
a holomorphic phase-space representation of quantum mechanics,
which can highlight certain semiclassical features in great detail.

We are going to probe a very restricted class of dynamical models
which minimize accessory complications. 
At the classical level we take chaotic area-preserving maps 
acting on a 2-d torus (= one degree of freedom),
because this is the simplest (compact) phase space to admit of such maps
with rigorously proven chaotic behavior; out of these we select linear
automorphisms (``cat" maps) and the ``baker's" map. In either case,
corresponding quantum dynamics are readily available. 
The semiclassical behavior of the resulting eigenfunctions
can then be examined in this framework, and it already looks quite intricate.
These simple, albeit abstract, models may then lend themselves
more readily to basic exploratory investigations than more realistic problems
plagued with technicalities.

Our results are mixed, ranging from rigorous to empirical
through statistical. 
We definitely do not resolve the internal dynamical structure of eigenfunctions
in the chaotic regime, but gain several new ideas about their behavior. 
We now list these results as we outline the contents of the paper.

The framework just described, and accompanying basic notations, are summarized in Sec.2, with the main emphasis placed upon the {\sl stellar parametrization\/}
to be used throughout, 
i.e., the encoding of a wavefunction in the zeros of its Husimi density. Interesting features of the quantum states
are not easily extracted from this (nonlinear) representation, and some
further study of this technique is in fact a parallel topic of this paper.

Sec.3 reviews the {\sl semiclassical ergodic results\/} mentioned above, e.g. like
Schnirelman's theorem for `chaotic quantum maps', and lists unsolved questions
about the precise way the Husimi densities converge towards the ergodic measure.
Various tools ($L^s$-norms and related functionals) can be used 
to globally measure deviations from strict uniformity,
and we present extensive numerical results about the Husimi densities 
of eigenstates for the quantum cat and baker maps.
In particular we discuss the semiclassical relevance of scars in this context.
Finally, existing ergodicity-related results on the {\sl Husimi zeros\/}
(the constellations, to be further studied) are recalled here.

Sec.4 exposes the {\sl new rigorous dynamical results\/} of this paper.
First, semiclassical convergence of the Husimi densities towards the uniform
ergodic measure is shown to {\sl imply the equidistribution 
of the corresponding zeros\/} over the phase space in the classical limit.
We further state as a conjecture
that the densities of zeros should do more than just equidistribute,
i.e., not only do their Fourier coefficients tend to zero, 
but they should decay faster than a certain specified rate.
Last but not least, and counter to the belief that the phase information
of quantum wavefunctions is irretrievable in presence of classical chaos, 
we deduce (from the equidistribution of zeros) 
a {\sl definite universal behavior\/} for a phase-related quantity, 
the {\sl local momentum of the eigenfunction in Bargmann form\/}.
This result can be cast in a very WKB-like expression for the eigenstate,
but to be taken in a much weaker, measure-theoretical, sense 
than for classically integrable systems.

Sec.5 is devoted to {\sl statistical analyses\/} 
of the distributions of Husimi zeros. 
The discussion continues earlier works based on random ensembles of functions,
in the following new directions. 
The results for random polynomials over the Riemann sphere are systematically
transported to the torus, for {\sl ensembles of theta-functions\/}.
At the same time, the effects of parity and time-reversal {\sl symmetries\/} 
are separately taken into account. Finally, emphasis is put on the resulting
statistical properties of the {\sl Fourier-transformed densities of zeros\/},
with the analytical evaluation of a corresponding {\sl form factor\/} 
(alias structure function).

In Sec.6, miscellaneous properties of these {\sl Fourier coefficients\/}
are observed for eigenstates of the quantum cat and baker map,
and compared with the dynamical and statistical expectations.
We find that the previous random form factor imprints {\sl each eigenstate\/},
through a globally strong suppression of most lower-range Fourier coefficients
(up to a distance $\losim \sqn$). 
Moreover, we identify a specific signature, within this dual approach, 
of high peaks in Husimi densities 
(including {\sl scars\/} by fixed points of the classical map).

Sec.7 provides a summary and some overall conclusions.

\section{Quantum models over the torus}

Generally speaking, the most elementary Hamiltonian dynamical systems 
which can exhibit chaos are area-preserving maps over a 2-d compact surface. 
Because such a phase space is not of the standard type 
(i.e., the cotangent bundle of a position space), 
quantum mechanics is not canonically defined; 
the simplest alternative quantization method needs an
extra 1-d complex (K\"ahler) structure chosen on the surface.
These compact surfaces are topologically classified by their genus $g$. 
The most elementary case $g=0$ (the sphere) is the relevant phase space for 
spin dynamics \cite{leb:vor}.
The next case $g=1$ (tori) will focus all of our attention 
since it is the simplest to carry {\sl proven\/} chaotic maps
(the cat and baker maps), while its kinematics (both the classical
and the quantal) remain manageable. 
(The more intricate $g \geq 2$ cases would also deserve investigation, 
even though the physical relevance of such phase spaces is unclear yet; see subsection \ref{alternative} for a sketchy introduction to quantum mechanics on such surfaces).

\medskip

On a 2-d torus, all inequivalent complex structures 
are labelled by a modular parameter,
$\tau \in \{\Im (\tau) >0\}/{\rm SL}(2,\Bbb Z)$.
Here, however, only one $\tau$-dependent feature will be examined 
(see Sec. \ref{comparison} and App. A). 
Otherwise we will only consider the torus $\Bbb T^2$ built from the unit square,
with canonically conjugate real variables $x \equiv(q,p) \in [0,1)^2$
(mod~1), and fix the complex coordinate $z \defi (q-\mi p)/\sq2 \in \Bbb C$ 
(corresponding to the choice $\tau=\mi$, 
the most suited to the discrete rotational symmetry of the square;
we call $T_{\Bbb C}$ the resulting complex torus,
to differentiate it from the underlying real torus $\Bbb T^2$).
We now briefly recall the basic notations and facts
pertaining to the quantum mechanics over this phase space, fixed once for all.

Admissible quantum wave functions $|\psi\rangle$ are to satisfy two quasiperiodicity relations \cite{leb:vor,bouz}:
\begin{eqnarray}
\label{qper}
\langle q+1|\psi\rangle &=& \e^{2\mi\pi\varphi_1}\,\langle q|\psi\rangle\\
\langle p+1|\psi\rangle &=& \e^{-2\mi\pi\varphi_2}\,\langle p|\psi\rangle
\nonumber
\end{eqnarray}
and they form a Hilbert space ${\cal H}_{N,\varphi}$ of finite dimension $N=1/h$,
a condition which restricts Planck's constant itself ($h=2\pi\hbar)$ 
to inverse integer values;
the Floquet angle pair $(\varphi_1,\varphi_2)$ is arbitrary on the dual torus
and this freedom makes for a family of quantizations which can be labelled by
the complex combination $\varphi \defi \varphi_1+\mi \varphi_2$.
However, any $\varphi$-dependent effects will be neglected here,
being of higher order in $\hbar$ than the leading semiclassical features 
which already challenge our understanding.

\subsection{Holomorphic and phase-space quantum structures over $T_{\Bbb C}$}
\label{holom}
We now quickly review (see \cite{leb:vor} for details) quantum representations 
which explicitly depend upon the choice of complex structure on $\Bbb T^2$, 
here fixed as $T_{\Bbb C}$ (the complex torus defined above).

The Hilbert space of eq.(\ref{qper}), 
${\cal H}_{N,\varphi} \approx  {\Bbb C}^N$, can also be spanned 
by doubly periodicized Gaussian coherent states, i.e., 
\begin{equation}
\label{coh}
|z \rangle_{N,\varphi} = N\sum_{n,m\in{\Bbb Z}} (-1)^{Nmn} \e^{2\mi\pi (\varphi_1 n - \varphi_2 m)}\e^{{\mi\over\hbar}(m\hat{Q} -n\hat{P})} |z\rangle
\end{equation}
where $|z\rangle\defi\exp(\bar z a^\dagger/\hbar)\,|0\rangle$ 
is a Weyl coherent state at the point $(q,p)$ of the plane
($a^\dagger=2^{-1/2}(\hat q-\mi\hat p)$ is the usual creation operator and $|0\rangle$ the ground state of the harmonic oscillator);
the state $|z\rangle$ is thus made to
depend holomorphically upon the coordinate $\bar z \equiv (q+\mi p)/\sq2$,
instead of being normalized (and $\langle z\mid z\rangle= \e^{z \bar z/\hbar}$).
Then for all $\psi\in {\cal H}_{N,\varphi}$, 
$\psi(z) = \langle z | \psi \rangle$ is an entire function of $z$, 
subject to the two quasiperiodicity relations \cite{leb:vor}:

\begin{eqnarray}
\langle z+1/\sq2 |\psi\rangle_{N,\varphi} &=& \e^{2\mi\pi\varphi_1}\e^{\pi N(1/2+\sq2 z)}\ \langle z|\psi\rangle_{N,\varphi}\\
\langle z+\mi/\sq2 |\psi\rangle_{N,\varphi} &=& \e^{2\mi\pi\varphi_2}\e^{\pi N(1/2-\mi\sq2 z)}\ \langle z|\psi\rangle_{N,\varphi}.\nonumber
\end{eqnarray}
These functions $\psi(z)$ also span an $N$-dimensional Hilbert space of entire functions,
constituting the {\sl Bargmann representation\/} of ${\cal H}_{N,\varphi}$.
Over the torus $T_{\Bbb C}$ itself, however, they do not define intrinsic functions,
only holomorphic sections of the complex line bundle specified by the two equations above.
It is then worthwhile to seek other representations based exclusively 
on genuine phase-space entities. 

The most popular such construct is the {\sl Husimi density\/},
\begin{equation}
\label{hus}
H_\psi(q,p)=H(x) \defi |\langle z|\psi\rangle|^2 \,\e^{-{z\bar z/\hbar}}
\end{equation}
which is strictly doubly-periodic hence defines a (positive) density 
on the torus;
the normalization condition for the quantum state $|\psi\rangle$ 
in ${\cal H}_{N,\varphi}$ translates as 
\begin{equation}
\label{husnorm}
\int_{\Bbb T^2} H(x)\, d^2x =1,
\end{equation}
and $H(x)$ can be viewed as a phase-space localization probability 
for the quantum state $|\psi\rangle$. 
The Husimi representation thus maps the unit vectors of ${\cal H}_{N,\varphi}$
into probability measures over $\Bbb T^2$; however, the infinite codimension of
this embedding makes it very uneconomical, and impractical to invert.

{\sl A contrario\/} \cite{leb:vor}, 
we emphasize that $\psi(z)$ can be mapped onto a much 
{\sl leaner\/}
phase-space structure, namely the skeleton of its zeros $\{z_k\}$
(counted with their multiplicities and modulo the periods; 
this is also called the divisor of $\psi(z)$).
Any Bargmann function has exactly $N$ zeros on the torus $T_{\Bbb C}$,
linked by a single linear constraint,
\begin{equation}
\label{sum}
 \sq2\ \sum_{k=1}^N z_k =  N\left({1+\mi\over 2}\right) - \mi\varphi \ \bmod  [1,\mi].
\end{equation}
and conversely, all such $N$-uples constitute an explicit and 1--1 parametrization of this Bargmann space, thanks to this special case of
Hadamard factorization \cite[p.22]{boas} for entire functions of order $\le 2$,
\begin{equation}
\label{fact}
\psi(z) = \langle z|\psi\rangle_{N,\varphi} 
= {\rm const.} \times
\e^{2\pi z \sum_{k=1}^N \bar{z}_k}\,\prod_{k=1}^N \chi\left(z-z_k\right) ,
\end{equation}
where 
\begin{equation}
\label{chi}
\chi(z)=2^{1/4}\e^{\pi z^2}\theta_1(\pi\sq2 z \mid \mi) 
\equiv (2\pi)^{-5/4} \Gamma(1/4)^3\sigma (z \mid 2^{-3/2},2^{-3/2}\mi)
\end{equation}
is the Bargmann function of the ``singlet" state $\chi$, the unique state of 
${\cal H}_{N=1,\,\varphi=(1+\mi)/2}$, possessing a single zero at $z=0$
(we use the notations of \cite{WW} for the Jacobi theta functions
and Weierstrass $\sigma$ function).

This immediately implies a nontrivial result, in the form of
an explicit criterion for a Husimi density to be a pure quantum state! 
Namely, by eqs. (\ref{hus}--\ref{fact}), 
a Husimi density $H_\psi$ is constrained
first to vanish (quadratically) precisely at the $N$ zeros $z_k$ of $\psi(z)$,
then to satisfy the factorization identity
\begin{equation}
\label{hfact}
H_\psi(x) = C_\psi \prod_{k=1}^N\,H_\chi(x-x_k) \qquad
(x_k=(q_k,p_k)= {\rm \  real\ form\ of\ } z_k=2^{-1/2}(q_k-\mi p_k));
\end{equation}
$C_\psi>0$ is a constant factor to be adjusted via the normalization 
eq.(\ref{husnorm}). Conversely, any $H(x)$ having the form (\ref{hfact}),
for an arbitrary subset of $N$ points on the torus,
is the Husimi density of a wave vector, specified by eq.(\ref{fact})
within the space ${\cal H}_{N,\varphi}$ where now $\varphi$ is constrained by
eq.(\ref{sum}). 

A parametrization of pure quantum states is thus provided
by this direct 1--1 correspondence of their Husimi densities 
with all $N$-uples or ``constellations" of zeros, or equivalently 
with the (singular, normalized) 2-d Dirac distributions of these zeros,
\begin{equation}
 \rho(x) \defi N^{-1} \sum_{k=1}^N \delta^2(x-x_k).
\end{equation}
This new density $\rho(x)$ can now be studied in place of, or in parallel with, 
the Husimi density $H(x)$ itself.
We hope that this nonlinear but ``naked" phase-space description 
of pure quantum states, which we call ``stellar representation",
will shed new light on the semiclassical eigenfunction problem.
At the same time, however, we must still improve our understanding
and control of this stellar representation
which has not yet been exhaustively studied.

The basic relationship between the densities $\rho$ and $H$ 
is deduced from the factorization property
(\ref{hfact}), as
 \begin{equation}
\label{Poiss}
\Delta \log H(x)=4\pi N (\rho(x)-1).
\end{equation}
This approach thus privileges the {\sl logarithm\/} of the Husimi density, 
as the electric potential generated by a point-like unit charge at every zero
minus a uniform charge density exactly restoring global electric neutrality.

Now, translating any property of one density concretely in terms of the other
becomes a central task, but an arduous one.
First, the basic relationship (\ref{Poiss}) is highly nonlinear.
Then, each zero is a maximally quantal object 
(contributing one quantum of phase to the Bargmann function $\psi(z)$), 
hence any semiclassical effects (which show up as high-density regions) 
have to be coherent manifestations from a large number of zeros, 
as observations will confirm.

Due to all the aforementioned difficulties, 
the stellar representation still requires development;
the present work should then be viewed not only in terms of raw results 
where it is perhaps exploratory, but also in terms of methods
where it suggests several novel approaches and directions for further study.

\subsubsection{Other Riemann surfaces}
\label{alternative}
As we already remarked, a sphere is the natural phase space for the dynamics of angular momentum (or spin). Its quantization involves finite-dimensional representations of ${\rm SU}(2)$. A stereographic projection of the sphere then allows to represent each state by a polynomial (its Bargmann function), leading to both Husimi and stellar representations \cite{leb:vor}. 
 
Any higher genus ($g\geq 2$) surface $X_g$ can also be given a K\"ahler structure, as a quotient of the Poincar\'e unit disk by some Fuchsian group. Although we are not aware of any interesting classical dynamics on such a phase space, we briefly describe the associated quantum mechanics to be able to generalize our rigorous results of section \ref{equidistribution} to such geometries.

The quantum states on $X_g$ can be defined directly in the `Bargmann representation': they consist of holomorphic sections on some canonically defined line bundles over $X_g$, called automorphic forms; for each allowed value of $\hbar$, they form a finite-dimensional Hilbert space \cite{klimek}, \cite[sec. 2.1.2]{these}. From these holomorphic sections, one can also define Husimi densities, which verify factorization properties similar to eq.(\ref{hfact}), so that a quantum state will be defined uniquely (up to a global prefactor) by its constellation of zeros on $X_g$ \cite[sec. 4.6]{these}.

\subsection{Semiclassical eigenfunction problem in reduced setting}
\label{rsef}
The issue stated in the introduction can now be considered in the special
setting of quantum maps on the torus. Let us mention that similar considerations work on the 2-sphere, which is the phase space of a classical spin \cite{leb:vor}.

We first rapidly comment on the problem of quantizing a given area-preserving map on the torus, i.e. how to associate to this classical transformation a sequence of $N \times N$ unitary matrices $U_{N,\varphi}$ acting respectively on Hilbert spaces ${\cal H}_{N,\varphi}$, with correct semi-classical properties \cite{BBTV,tabor}; the whole sequence of matrices $\{U_{N,\varphi}\}_{N\geq 1}$ is called a {\sl quantum map\/}.

Such a quantization is in general not uniquely defined, assuming there exists one. A systematic way of
quantizing symplectic maps was recently devised by Zelditch, using a Toeplitz operator formalism \cite{zeld}; however, this formalism does not in general lead to closed-form expressions for the $U_{N,\varphi}$, and is currently restricted to smooth maps. On the other hand, if the classical map is piecewise affine, a `down-to-earth' construction provides explicitly a sequence of unitary matrices. We will restrict our numerical investigations to maps of the latter type, namely the baker's and the cat maps, which are already quantized and studied in the literature \cite{hannay:b,bv}. 

Whereas the Floquet angles $\varphi$ will be fixed and subsequently implied, dependence upon the integer $N$ will be a crucial feature for us, with the semiclassical regime corresponding to $N=1/h \to +\infty$.

(From the standpoint of physical systems,
which are described by quantum Hamiltonians as opposed to maps,
the above construction crudely models a quantum-dynamical {\sl reduction\/}
of a 2-d time-independent Schr\"odinger equation to a 1-d quantum map, 
analogous to a Poincar\'e surface-of-section map for a classical flow 
\cite{bogo,prosen1}; our problem is thus `of reduced type'.)

The issue is then to compute (or describe) all (or some) of the
eigenfunctions of the matrix family $\{U_N\}$ asymptotically as $N \to +\infty$,
in connection with the classical dynamics.
We will moreover focus on phase-space descriptions for these wave
functions, privileging the stellar representation for its nonredundancy;
we coin the shortened notations ``eigendensity", resp. ``eigenconstellation",
for the Husimi density of an eigenfunction, resp. its constellation of zeros.

The two quantum maps we select
both correspond to fully chaotic area-preserving transformations 
of the torus in the classical setting.
One is the {\sl baker's map\/}, quantized for even $N$ \cite{bv,saraceno},
and under {\sl antiperiodic\/} boundary conditions 
($\varphi=(1+\mi)/2$ in eq.(\ref{qper})) so as to preserve the classical
parity symmetry P: $\{q \mapsto -q,\ p\mapsto -p\}$. 
The other consists of the {\sl cat maps\/} or linear hyperbolic automorphisms, 
given by matrices 
$\Bigl(\begin{array}{cc}a&b\\c&d\end{array}\Bigr) \in {\rm SL}(2,\Bbb Z)$
(with integer entries and unit determinant) such that ${\rm tr}(S)>2$; these
will be quantized under {\sl periodic\/} boundary conditions (i.e., $\varphi=0$) \cite{hannay:b}.
Both of the resulting quantum models commute with spatial parity,
hence their eigendata separate into even and odd subsets.

As concerns `cat' maps, moreover, we will look at their quantizations only for
{\sl certain prime\/} values of $N$ that are `splitting' \cite{crystal} and such that spectral degeneracy (a potential source of extra complications)
affects just one eigenvalue (of multiplicity 2); for these values, 
semiclassical ergodic behavior is fully proven \cite{DEGI}.
We will however have to switch between two distinct classical cat dynamics, 
of respective matrices
\begin{equation}
\label{cats}
S= \left(\begin{array}{cc}2&1\\3&2\end{array}\right), \qquad
S'= \left(\begin{array}{cc}12&7\\41&24\end{array}\right).
\end{equation}
The reason is that $S$ gives a time-reversal-invariant dynamics 
(a consequence of $a = d$ in that matrix),
and this quantum-mechanically implies {\sl real} eigenfunctions.
While this symmetry (T) does not interfere with most computations, 
it greatly complicates the analysis of random ensembles 
to be used as comparison models (Sec.\ref{stat});
at that point it becomes simpler to work with the other map like $S'$,
which has no anti-canonical symmetry 
and is as easy to quantize, using the formulas in \cite{hannay:b}.
To prove that $S'$ has no extra symmetry (besides P, always present),
we had to study conjugacy classes in ${\rm SL}(2,\Bbb Z)$, 
and found no such map with a trace $<36$. 
More generally, in order to avoid anti-canonical symmetries, 
in practice one has to accept much larger Lyapunov exponents, 
which in turn make numerical studies of semiclassical behavior 
(e.g., the search for scars) much more delicate.
Consequently, the map $S$ will be more suitable for asymptotic verifications, 
and $S'$ for statistical ones.

\section{Measures of semiclassical ergodicity}
\subsection{The Schnirelman property}
\label{schn}
Semiclassical ergodicity reads as follows for quantum maps on the torus.

Let $\{\psi^{(N)}_j\}_{j=1,\cdots,N}$ be orthonormal eigenfunctions of
a quantum map $U_N$ defined over the space ${\cal H}_N$. 
Let $\cal L$ denote the (normalized) Liouville density over the phase space
for the corresponding classical map (namely, the invariant area density). 
If this classical map is ergodic with respect to $\cal L$, then almost any
subsequence of Husimi ``eigendensities" $\{H_{\psi^{(N)}_{j(N)}}\}$
converges to $\cal L$ as $N \to +\infty$, in the weak-$*$ topology for measures.

(Weak-$*$ convergence of a sequence of measures $\{\mu_N\}$ on a compact space
$X$ is defined by the convergence of every scalar sequence $\int_X f \,\mu_N$
where $f$ is any {\sl fixed\/} continuous function \cite[p.113]{reed}.)

We will say that a sequence of measures $\{\mu_N\}\stackrel{{\rm w}-*}{\to}{\cal L}$ ``has the Schnirelman property"---by reference to the name ``Schnirelman's theorem" generically used for the many dynamical versions of the whole statement above (the first one was worded for Riemannian Laplacians \cite{schnirelman}, while we directly stated the quantum-map version).

In the theorem, ``almost any" means that the stated property may be violated by 
an exceptional subset of eigenfunctions of asymptotic density zero, 
i.e., if this subset reads
as $\{\psi^{(N)}_j\}_{j \in \Omega(N)}$ then $\#\{\Omega(N)\}/N \to 0$.

For the torus maps we work with, the invariant measure $\cal L$ is just 
the usual flat measure of uniform density 1 on $\Bbb T^2$. 
Then the Schnirelman property for a family of Husimi densities 
is their uniform spread (as measures) over the torus as $N \to +\infty$.

In full mathematical rigor, the above theorem was proved in \cite{zeld} for arbitrary smooth ergodic classical maps (on the torus), quantized through the Toeplitz formalism; this includes in particular quantum cat maps \cite{bouz}. Moreover, direct calculations establish that the particular sequences of quantum cat eigenstates we will numerically study have the Schnirelman property \cite{DEGI}. By contrast, in the case of the baker's map the theorem is only partially proved \cite{DBDE}, but is believed to hold nevertheless.

We now display (fig.1) a sample of Husimi eigendensities of quantum cat and baker's maps.
Already at a glance, the situation shows a wealth of complex behaviors
which cannot be exhausted by just invoking the Schnirelman property. 
The densities do show an overall tendency to
uniformization but only in a very coarse sense and at quite slow rates.
Subdominant structures are quite significant, in fact their intricacy and
profuseness wildly grow with $N$. Certain patterns seem to occur recurrently
but with no definite regularities; in particular, periodic orbits are
often scarred but not in an obviously systematic way.

Our problem of concern will be, in general terms, 
to find various ways to tackle on a finer scale such eigenstates 
obeying the Schnirelman theorem
(which constrains them on the coarsest scale only).
This includes, e.g., the popular topic of {\sl scarred eigenfunctions\/}:
how they should be derived, and how they fit with the Schnirelman picture.
Unresolved issues in this area are:

1) sharpen the Schnirelman property:

\noindent 1a) assess convergence in stronger topologies;

\noindent 1b) seek equivalent conditions upon the eigenfunctions themselves;

2) describe the finite-$N$ situation:

\noindent 2a) estimate the convergence rates;

\noindent 2b) analyze and classify the quantum corrections $H_{\psi^{(N)}_j}-\cal L$
and in particular any exceptional, non-ergodic ones.

\subsection{Norms and other invariant measures of density fluctuations}
\label{norms}
One natural way to estimate the fluctuations of a density $H(x)$ 
away from flatness is by $L^r$-norms and related functionals. 
Keeping in mind that $H(x)$ is a nonnegative density on the torus $\Bbb T^2$
(and that the latter has unit area), one defines, following \cite{prosen-QSS}
\begin{equation}
||H||_r \defi \left[ \int_{\Bbb T^2} d^2x\, H(x)^r \right]^{1/r} \quad (r>0). 
\end{equation}
For $r\geq 1$ these quantities define norms, called $L^r$,
out of which the $L^1$-norm is taken by eq.(\ref{husnorm}) to fix the normalization, as $||H||_1 =1$; then by convexity
(Jensen's inequality, see for instance \cite[p.70]{rudin}), $||H||_s \geq ||H||_r \geq 1$ for $s>r>1$,
with equality precisely characterizing the flat case $H(x) \equiv 1$. Finally, those (and derived) quantities are geometric or shape invariants.

Intuitively, fluctuations of $H(x)$ push $||H||_r$ upwards, 
more so when $r$ is larger. 
Ultimately, $||H||_\infty\ (\equiv H_{\rm max}$, the sup-norm)
records the highest enhancement only. 
A more global assessment of the non-uniformity requires an intermediate norm,
of which the simplest is the $L^2$-norm; since $H$ is quadratic in the wave
vector, $||H||_2^2$ is an integral quartic in $\psi$, 
like the inverse participation ratio commonly 
used as a measure of quantum localization (but often in configuration space).

The weakest norm, hence ``least remote" from the weak-$*$ topology, would be
the $L^1$-norm, had it not already been used up for normalization.
The closest approach then considers $||H||_r$ as $r \to 1^+$, 
or rather the functional (non-negative, by convexity)
$${\rm H}[H] \defi \left[{d \over dr} ||H||_r \right]_{r=1} 
\equiv \int_{\Bbb T^2} d^2x\, H(x) \log H(x) $$
which has another important meaning, as the information content carried 
by the probability density  $H$ (Boltzmann's H-function, or negative entropy). This quantity is related to the `classical-like' entropy of a quantum pure state (more generally, of a quantum density operator) introduced by Wehrl \cite{wehrl}, then studied (on the plane and the 2-sphere) by Lieb and Lee \cite{lieb}. This function can be used to define the `dynamical quantum entropy' of a unitary map \cite{zycz}, which is supposed to generalize the Kolmogorov--Sinai entropy of a classical system. 
   
Last but not least, since the Husimi functions of pure quantum states
(the only ones of interest here) have a multiplicative structure,
the {\sl geometric mean\/} of the density $H$ is a natural functional to use
\cite[p.70]{rudin},
\begin{equation}
 {\rm GM}[H] \defi \exp  \int_{\Bbb T^2} d^2x\,\log H(x) \quad
( =\lim_{r \to 0^+} ||H||_r )
\end{equation}
since it is itself multiplicative 
(${\rm GM}[H_1H_2] \equiv {\rm GM}[H_1]{\rm GM}[H_2]$). 
Thus, for instance, the evaluation of GM on both sides of the Husimi factorization formula eq.(\ref{hfact})
fixes the normalization factor $C_\psi$, resulting in a more precise
expression for this factorization as
\begin{equation}
\label{hfactp}
H_\psi(x)={\rm GM}[H_\psi]\prod_{k=1}^N\,\hat H_\chi(x-x_k); \qquad
\hat H_\chi(x) \defi H_\chi(x)/{\rm GM}[H_\chi]
\end{equation}
(with ${\rm GM}[H_\chi]=(2 \pi)^{-3/2}\Gamma(1/4)^2$: see App.A.1 and table \ref{tab} below). More generally, a Husimi density will be denoted $\hat H$ 
when we normalize it {\sl by its geometric mean\/} (instead of $L^1$-norm).

$ {\rm GM}[H]$, like $||H||_r$ for $r<1$, is not a norm (all convexity-related
inequalities get inverted with respect to $r>1$); 
it satisfies $0 < {\rm GM}[H] \leq 1$, with equality in the flat case
whereas fluctuations of $H$ drive it downwards.

In summary, the $L^2$- and sup- norms, the H-function, and the geometric mean,
all give useful size estimates for the fluctuations displayed by finite-$N$ Husimi densities. Although they yield largely similar density rankings,
the various functionals are inequivalent;
at the same time we see no compelling argument to favor any single one of them.

\subsection{Comparison states}
\label{comparison}

As we intend to probe the finer structure of Husimi eigendensities, 
beyond the Schnirelman property 
$H \stackrel{{\rm w}-*}{\longrightarrow} \cal L$, 
and since those are pure states, we simply cannot dismiss
their {\sl multiplicative\/} structure at the finest quantum scale:
any such Husimi density is the product of $N$ translates of a single
elementary building block, and is thereby determined by its constellation 
of $N$ zeros (up to normalization: cf. eq.(\ref{hfactp})). 
Thus, quantum mechanics imposes
a {\sl granular microstructure\/} upon these densities at finite $N$,
forcing them to vanish at precisely $N$ points. 
If pure-state Husimi densities uniformize as $N \to \infty$, 
it can then happen in a coarse-grained sense only;
in no way can they look smoother and smoother, ``classical-like",
in the semiclassical limit, 
as their inhomogeneities will only experience a shrinkage of their geometric ($x$-) scale and not of their intensity. 
Any fine-scale semiclassical theory must then heavily acknowledge  
such rigid structural constraints. 

A first step in this direction is to gather data about various patterns of zeros
that are at the same time characteristic enough and analytically tractable. 
Such reference constellations will then provide a comparison scale:
we proceed to define a few of them now (up to unspecified translations).

\subsubsection{Lattice states}

Intuitively speaking, the most equidistributed Husimi densities subject to the
constraints (\ref{hfactp}) must be found among lattice states, 
whose zeros form a 2-d lattice $\Lambda$ with a fundamental cell of area $1/N$. 
Such displays of zeros have maximal, ``solid-state" rigidity.
Some of them occur in exceptional eigenstates of cat maps.

Up to scale invariance, the {\sl shape\/} of a lattice $\Lambda$
can be described by a complex number $\tau$ with $\Im(\tau)>0$ (the modulus)
such that the torus ${\Bbb C}/\Lambda$ is conformally equivalent 
to the complex torus $T_\tau$ of generators $\{1,\tau\}$.
A Husimi density specified by the set of zeros $\Lambda$
can be directly written in terms of the (properly rescaled and displaced)
elementary ($N=1$) Bargmann function $\chi(z \mid \tau)$
quasiperiodic w.r.t. $T_\tau$ and its Husimi density $H_{\chi(\tau)}(q,p)$,
which are expressed by the formulae \cite{KP}
\begin{eqnarray}
\label{chitau}
\chi(z \mid \tau)&=&(2\Im(\tau))^{1/4}\,\e^{\pi z^2\over\Im(\tau)}\theta_1(\pi \sq2 z \mid \tau),\\
H_{\chi(\tau)}(q,p)&=&\sqrt{2\Im(\tau)}\;\e^{-2\pi p^2\over\Im(\tau)}\;\Bigl|\theta_1\left(\pi(q-\mi p) \mid \tau\right)\Bigr|^2, \nonumber
\end{eqnarray}
the normalization being 
\begin{equation}
{1\over\Im(\tau)} \int_{T_\tau} H_{\chi(\tau)}(q,p) dq\, dp =1 \qquad
(\Im(\tau) \equiv {\rm area}(T_\tau)).
\end{equation}

Although the present article restricts to quantum mechanics 
on the {\sl square} torus $T_{\Bbb C}\equiv T_\mi$ 
(for which we recover our standard $\chi(z)$ and $H_\chi(x)$), 
this extension of scope to functions quasiperiodic w.r.t.
any torus $T_\tau$ is useful for three reasons.
Firstly, Husimi densities of the form $H_{\chi(\tau)}(Nq,Np)$
for rational values $\tau=(m+\mi n)/N$
have the square torus periodicity too,
and lattice eigenstates of quantum cat maps have precisely this form 
(section \ref{lattice}).
Secondly, in section \ref{counter} we will define Husimi densities 
whose constellations are deformed lattices, and their asymptotic expressions
will make use of the function $H_{\chi(\tau)}(q,p)$ for variable $\tau$.
A third reason (already stated but excluded from the scope of the present work)
is that varying $\tau$ (over the modular domain) also amounts to exploring
the range of complex structures over a fixed real torus,
or all possible Bargmann--Husimi representations for a prescribed problem.

For lattice states, the various norms and functionals
are functions of $\tau$ alone (by scale invariance).
With the phase space itself being of square shape, 
the invariants for $\tau=\mi$ arise the most readily 
(they describe the elementary Husimi factor $H_\chi$ of eq.(\ref{hfact}):
see figs.2a, left and 4, left);
however, we expect the optimal equidistribution to be attained by the closest
packing of the zeros overall, namely the {\sl equilateral triangular\/}
lattice (i.e., $\tau=\e^{\mi\pi/3}$).
While this geometry cannot be realized on a square torus, it can be approached 
in the limit $N \to +\infty$ 
(scale-equivalent to the infinite-plane limit; cf. fig. 2a, middle),
so we conjecture that {\sl its\/} invariants will be the good ultimate bounds. 

(In fact, we seem to encounter natural extensions of the close-packing 
class of problems: to find the constellation(s) of $N$ points on the torus 
that optimize any one of the invariants;
the difficulty may increase by changing the surface, 
as the review \cite{sph} on the spherical case suggests.)

We can already prove (see Appendix A) that two of the above invariants 
(namely, the geometric mean and the $L^2$-norm) are optimized 
by the equilateral lattice within all lattice states. 
Curiously, such invariants are related to spectral determinants of tori,
studied for instance in \cite{osgood}. 

\subsubsection{Localized states}

At the other end, we should consider patterns with highly concentrated zeros.
Examples forming an important class 
are the eigendensities of classically integrable systems,
for which both the Husimi densities and the densities of zeros 
respectively concentrate on {\sl lines\/}:
one prototype is a plane wave (or delta wave, up to $\pi/2$ rotations),
which is in fact just a highly anisotropic lattice state (having $\tau=\mi/N$:
fig.2b, left).
Still more localized is a coherent wave vector defined by eq.(\ref{coh}),
with its Husimi density concentrated around a point in phase space;
its zeros lie on {\sl two\/} perpendicular axes 
(those which meet at the antipodal point: fig.2b, middle). 
The most extreme concentration is however realized by
the completely degenerate pattern in which all $N$ zeros {\sl coincide\/}
(fig.2b, right); even though this state is mainly a curiosity 
and has a Husimi density barely different from the previous one,
we include it for the sake of completeness.

\subsubsection{Statistical states}

In-between the previous two classes, we may consider statistical distributions
of zeros for a random polynomial ensemble, inasmuch as these reproduce
quite well (albeit empirically) the local fluctuations seen for zeros 
in eigenfunctions of classically chaotic maps.
As in the random-matrix modeling of spectra,
the dynamical symmetry of the problem must be taken into account.
The random coefficients have to be taken fully complex for a model without
any antiunitary symmetry (a sample on fig.2a, right), 
and real otherwise (see Sec. \ref{stat} for details).

\subsubsection{Analytical results}

The norms and invariants of the previous comparison states can be computed:
either exactly, or asymptotically for large $N$, or just numerically for a few.
The results are listed in table \ref{tab} 
(upon setting all $L^1$-norms to unity, as usual).

The lattice-state invariants are computed long-hand using classical 
theta-function identities (see also App. A). 
A tricky point for the sup-norm is that the maxima 
of the Husimi density $H_{\chi(\tau)}$ {\sl depart from\/} 
the dual lattice when $\tau \neq \mi$ 
(for $\tau=\e^{\mi\pi/3}$, they lie at the centers
of the equilateral triangles of zeros).

The concentrated state examples (last three lines) are also expressed 
using theta-functions, and furthermore all required integrations 
are performed asymptotically (by stationary phase).
Remark: the study of the Bargmann representation over the whole plane 
(i.e., for the Hilbert space $L^2(\Bbb R)$)
shows \cite[Theorem 3]{lieb} that for each $r\geq 1$, the $L^r$-norm 
of $H_\psi$ is well-defined for all $\psi\in L^2(\Bbb R)$ 
and attains its maximal value when $\psi$ is a Weyl coherent state
(and likewise for the H-function). 
It is reasonable to expect these results to extend to the torus, 
at least semi-classically. 
Then, with our normalization, those invariants would be bounded as follows over ${\cal H}_N$:
\begin{eqnarray}
\forall \ 1 < r \leq +\infty, \quad && 
\sup_{||\psi||_1=1} ||H_\psi||_r \defi M_r(N) \sim {N\over(Nr)^{1/r}}
\quad (N\to\infty) \\
&& \sup_{||\psi||_1=1} {\rm H}[H_\psi]\defi M_1(N) \sim \log N-1 
\quad (N\to\infty)\nonumber
\end{eqnarray}
and these maxima $M_r(N)$ should be reached at (or near) a torus coherent state;
indeed, their asymptotic estimates agree with those of the corresponding
coherent-state invariants in table \ref{tab}. 

\begin{table}
\begin{tabular}  {lllll}
\hline
& & & & \\[-8pt]
state & sup-norm & $L^2$-norm & \hfill H-function \hfill & Geometric Mean\\[2pt]
\hline
& & & &  \\[-6pt]
equilateral \hfill & $\frac{3\Gamma(1/3)^3}{4 \pi^2}$ &
$\frac{\sqrt 3\, \Gamma(1/3)^{3/2}}{2^{7/6} \pi} $ &
& $\frac{\sqrt 3\, \Gamma(1/3)^3}{4 \pi^2} $ \\
\hfill lattice & \hfill $\approx 1.4610$ & \hfill $\approx 1.0768$ & 
\hfill $\approx 0.1016$ & \hfill $\approx 0.8435$ \\[6pt]
square \hfill & $\frac{\Gamma(1/4)^2}{\sqrt 2\,\pi^{3/2}} $ &
$\frac{\Gamma(1/4)}{\sqrt 2\,\pi^{3/4}} $ &
& $\frac{\Gamma(1/4)^2}{(2 \pi)^{3/2}} $ \\
\hfill lattice & \hfill $\approx 1.6693$ & \hfill $\approx 1.0864$ & 
\hfill $\approx 0.1110$ & \hfill $\approx 0.8346$ \\[6pt]
random & $\leq 2\log N$ & $\sqrt{2}\pm {3\over 2\sqn}$ & 
\hfill $(1-\gamma)\pm\sqrt{(2-\gamma)^2+{\pi^2\over 2}+\zeta(3)-6\over N}$ & \hfill $\e^{-\gamma}(1\pm\sqrt{\zeta(3)\over N})$ \\[10pt]
plane wave & $\sim \sqrt{2N}$ & $\sim N^{1/4}$ & $\sim {1\over 2}[\log(2N)-1]$ &
$\sim \sqrt {2 N} \e^{-\pi N/6}$ \\
coherent & $\sim N$ & $\sim \sqrt{N/2}$ & $\sim \log N -1.$ &
$\sim N \e^{-\pi N/6}$ \\
degenerate &  $\sim N$ & $\sim \sqrt{N/2}$ & $\sim \log N -1$ &
$\sim N \, 2^{-N}$ \\[5pt]
\hline
\end{tabular}
\caption{Values of the various fluctuation measures for several comparison states, from the most ergodic (first two rows) to the most localized (last three rows). In the third row, we give typical values for a random state ($\gamma$ is Euler's constant), as well as the standard deviations (numerical values of the averages appear on fig. 3a).}
\label{tab}
\end{table}

The random-model averages are computed from the distribution law 
for the Husimi density values,
$P(H) \sim \e^{-H}$ in the $N \to +\infty$ limit 
(for either the real or the complex Gaussian ensemble). 
We can then compute any integral in average as
\begin{equation}
 \left\langle \int_{\Bbb T^2} f(H(x)) d^2x \right\rangle_N 
\sim \int_0^{+\infty} f(H) \e^{-H}dH 
\end{equation}
and this yields the constant values listed as typical for the ensemble. The standard deviations from the average values are estimated using integrals of the type:
$$
\left\langle \int_{\Bbb T^2} H(x)^s d^2x \int_{\Bbb T^2} H(y)^t d^2y\right\rangle_N
$$
for real positive $(s,t)$ (see Sec. \ref{stat} for details).
The sup-norm expectation value
diverges in this regime, and needs a more delicate finite-$N$ analysis
(App. C).

The ``most naive" ergodic assumption ($H \sim 1$ pointwise) would set all
invariants to unity (except the H-function, to zero). 
By contrast, the tabulated values for the lattice, resp. random, models mark 
the incompressible, resp. the typical, levels of quantum inhomogeneity.
Far from vanishing in the semiclassical limit, these levels are 
independent of $N$: this rules out convergence towards classical ergodicity 
in any of the corresponding stronger topologies. Furthermore, the random invariants are substantially above the optimal lattice values, confirming the
visual impression from the two rightmost plots on fig. 2a that random densities
are quite non-uniform.

\subsection{Invariant data for eigenfunctions}
\label{data}
Concerning now the invariants of eigendensities corresponding to classically
chaotic dynamics, our results consist of numerical data displayed graphically. 
We have computed the full set of invariants for all eigenstates,
at selected values of $N$, of two quantum maps with 
highly chaotic classical limits, i.e., the baker's map and the cat map $S$
specified in Sect. \ref{rsef}.
We plot and compare the distributions of 
the invariants for the resulting eigenstate families.
Fig. 3a plots the values of four functionals for the even-parity eigenstates 
of the quantization of the cat map $S$,
with the geometric mean plotted downwards for full visual consistency
(the odd-parity plots look much the same, and are not shown); 
Fig. 3b plots the values of two functionals for both parity
eigenstates of the quantum baker's map.

\subsubsection{General observations}
The figures first show the values of the invariants to be substantially
depressed in comparison to the concentrated state examples,
but still broadly scattered, with relatively few isolated values or clusters:
this confirms the visual impression that individual eigenfunctions within 
a single matrix (a quantum map at given $N$) retain widely differing global fluctuation patterns. 
For each eigendensity the value of a particular invariant is determined by 
its $N$ zeros, but large values of this number $N$ do not suffice to enforce
a statistical averaging towards a universal behavior in this respect.
Besides, that ordering of eigendensities according to their global contrast 
is generally barely sensitive to the particular choice of functional.
At a more detailed level: the cat fluctuation values are much more squeezed 
than the odd baker values, themselves more squeezed than the even baker ones.
This difference could be related to the singular nature of the baker dynamics,
whose discontinuity will be more felt by even-parity states
(especially those scarred above the irregular fixed point (0,0) which straddles
the discontinuity, whereas the period-2 point is regular).

We next see that, upon further averaging over the entire eigenfunction set
(for a given $N$), each invariant then comes fairly close 
to the typical random value (third line of table 1). 
This convergence is striking for the cat map, whose eigenfunctions
thus appear to fulfill generic expectations (of compliance to a random model)
somewhat better than the baker's map in spite of its arithmetic idiosyncrasies. 
(This is the exact opposite of the situation for the eigenvalues: cat map spectra are highly non-generic, whereas the baker map spectrum is fairly GOE.)
The reason may again be that the baker eigendensities tend to be unusually bumpy
above the map discontinuities. By contrast, the cat map possesses
a few specially regular lattice and crystal states
(always yielding values of the functionals among the lowest-lying),
but these atypical states do not seem to drive the averaged behavior 
away from genericity (at least for our selected values of $N$).

We stress again that those density fluctuations (of the same order as in 
the random ensemble) already represent an appreciable and invariant
($N$-independent) degree of spatial inhomogeneity.
We can raise to a conjecture the empirical observation that 
{\sl in average, the invariants appear to be reliably constrained
to the universal random model values\/}. This entails that 
the average convergence of eigendensities towards uniformity is truly poor.
Eigendensities displaying a reasonably high level of homogeneity or rigidity
must then have invariants well below that average 
(for example, the lattice eigenfunctions in cat maps),
and for the sake of global balance some other states will then have to
display higher inhomogeneity or localization than if they were random
(such as numerous states scarred by periodic orbits).

Concerning $N$-related behaviors: 
most fluctuation sizes grow fairly slowly with $N$, especially for the cat map;
$N$-dependences are erratic as a rule; 
their irregularity is reduced but not wholly suppressed
when a restriction is made to special values of $N$ (primes, powers of 2, etc.) 
and/or to specially selected states (most scarred, or most ergodic, etc.). 

All this confirms that the individual behaviors of eigendensities 
may not be easy to characterize or classify, whereas 
the quantum fluctuations seem to follow the random model 
not only microlocally, but also globally once averaged upon all eigenstates.

\subsubsection{About point scars in eigenfunctions}

The plots allow us to start a quantitative analysis of {\sl scars\/} 
in eigendensities.
The sup-norm, and to a lesser extent the $L^2$-norm, are the best invariants to
select the states with the strongest local density enhancements.
One currently debated issue concerns the semiclassical weight of scars:
in particular, do scarred states still follow the Schnirelman property or not? 
This question is hard to settle partly 
because this convergence property is specifically weak-$*$ (and not metric,
for instance), hence it is difficult to establish
a single and intrinsic general scar strength scale.
 
We can however get a clearer idea about the simplest, {\sl point-like\/} scars.
Scars are features localized near unstable periodic orbits,
and in reduced dynamics these consist of discrete points.
Quite a few Husimi eigendensities of maps do present sharp enhancements that are
very localized near periodic points, and we can focus upon the most
strongly scarred states of this type. 
We always find the widths of such {\sl point scars\/} to scale, very reliably,
like $N^{-1/2}$ in all phase space directions 
(matching the distances of the nearest zeros when these distribute uniformly, 
as the figures confirm). 
The {\sl value of the Husimi density at the scarring point\/}  
(the sup-norm if the highest scar is under scrutiny, as will be the case here)
is then a good strength indicator: the area of such a scar being of order $1/N$,
a safe criterion for it to produce an asymptotically vanishing contribution 
(as a measure) in a sequence of densities $\{H_N\}$ is
$$\alpha = \limsup_{N \to \infty} (\log ||H_N||_\infty / \log N ) < 1 $$
hence $||H_N||_\infty = o(N)$: the {\sl height\/} of the scar should grow 
strictly less than the sup-norm of a coherent state (Table 1). 
Observing each log-log plot of sup-norms, 
we can then compare the growth trend (with $N$) of the uppermost points 
in the cloud of sup-norm values 
against the (unit) slope of the straight line of coherent state sup-norms,
to get a (necessarily rough) idea about the asymptotic persistence of such scars.

For the cat map,
those uppermost points are irregularly distributed with respect to $N$,
nevertheless heuristic upper bounds for the slope $\alpha$ are 0.5 
for odd states (not shown) and 0.34 for even states; 
this is consistent with the proven fact that {\sl all\/} cat map eigenstates
used here satisfy the Schnirelman property. 
For the baker's map, the most strongly scarred states form more regular families
and estimates look more robust; they yield slopes below 0.85 for odd states 
and 0.8 for even ones (but the even-state scars are stronger in the range of 
$N$ under display). The slopes for the baker's map are thus much closer 
to the critical value $\alpha = 1$, while we cannot control the error range
of our crude estimation process.
Hence all we say is that the plots slightly favor the hypothesis
that even the most strongly point-scarred baker eigenstates still satisfy
the Schnirelman property. 

We must however add that 
there also exist eigenstates which are strongly enhanced along whole stretches 
of stable and unstable manifolds (or neighboring arcs of hyperbolae);
the baker's map shows many such instances,
especially about the (singular) fixed point (0,0) (fig.1b, right).
Because of the extended character of these scars, 
their sup-norms need not lie anywhere as high as those of point scars 
for them to have stronger classical imprints; at the same time, the functionals
used above may not suffice to quantify this strength precisely enough.

\subsection{Ergodicity and the zeros}
\label{ergodicity}

A deeper way to incorporate the multiplicative structure (\ref{hfact})
of eigendensities into their analysis is however to handle the eigenfunction
problem entirely within the stellar representation, 
or to describe directly the phase-space densities of zeros 
$\rho(x)$ of individual eigenfunctions.
In particular, the semiclassical eigenfunction problem is then 
to unravel the large-$N$ behaviors of these ``eigenconstellations".

Ideally, we would like to be able to to compute or control $\rho(x)$ (the zeros)
{\sl ab initio\/}, directly from the underlying quantum-dynamical equations, 
and we have begun investigating this approach as well. 
However, the simplest Schr\"odinger operator already yields for the zeros 
a dynamical system of nonlinear equations
involving strong $M$-body interactions \cite{leb}, 
where $M$ ranges from 1 to min(degree of potential,$N$); 
apart from a few trivial cases 
(always yielding completely rigid motions for the constellations),
it is currently unclear how such a system can be handled.
In our reduced setting, moreover, the dynamics is that of a quantum map, 
i.e., it is no longer local, meaning that zeros must fare even worse; and
indeed, we cannot in the least simplify the equations of discrete-time motion
in terms of the zeros (except for certain `lattice states' under cat maps).

That leaves us with indirect approaches where we try to relate
the stellar representation to more familiar ones, especially Husimi's.
Similar questions can be asked about $\rho(x)$ as were listed 
for Husimi densities $H(x)$ in Sec.\ref{schn};
now, moreover, any answers will directly refer to the eigenfunctions 
via the stellar parametrization, implicitly settling question 1b). 
However, each property now requires translation from one density to the other:
in particular, the Schnirelman property, the scar phenomenon etc., should be
redescribed in terms of the zeros, with the hope of a better understanding
later. Concrete problems of this sort are, typically:

1c) describe the densities of zeros and their higher-order correlations,
with their dependence on classical dynamics;

2c) classify the fluctuations of their distributions as dynamical (e.g., scars?)
or statistical (and: universal?).

\subsubsection{Existing observations}

Some features of eigenconstellations have already received 
dynamical interpretations \cite{leb:vor}--\cite{shukla}. 

For integrable systems, eigenfunctions follow a WKB-type Ansatz
like (\ref{WKB}) in the Bargmann representation too, from which it follows that
eigenconstellations have to coalesce as $N \to \infty$ onto fixed curves, 
namely certain anti-Stokes lines of the complex classical action in the $z$
variable, along which the zeros moreover distribute regularly 
with spacings of order $1/N$.

In chaotic systems, by contrast,
numerical computations of Husimi eigendensities 
indicate that eigenconstellations roughly equidistribute in (almost)
all of phase space, with spacings of order $1/\sqrt N$; 
moreover, higher-order correlations between the zeros are found to 
very accurately follow a universal model, namely the statistics
of the zeros of a Gaussian random ensemble of polynomials (which acts nicely
as an analog for eigenfunctions of random-matrix ensembles for eigenvalues).
All those findings about chaotic systems have remained mostly empirical.

More systematic observations of eigenconstellations at increasing $N$ (fig.1)
show a proliferation 
of vaguely lookalike but not uniform patterns in want of ordering.
It is specially challenging to try to correlate 
some spectacular fluctuations of eigendensities, 
like scars above unstable periodic points, with specific patterns of zeros. 
Around every high peak of the Husimi density there is necessarily a region
devoid of zeros, and sometimes the nearest of these vaguely affect 
a hyperbolic distribution, but the connection is tenuous: 
the decrease of the density of zeros under a scar is quite moderate and,
conversely, regions with an equally low density of zeros 
do not build high Husimi densities above them as a rule.

\subsubsection{Some directions of analysis}
A more systematic description of the eigenconstellations on the one hand,
of the high-density fluctuations (e.g., scars) within the Husimi densities
on the other hand, and of explicit relationships between the two,
is still lacking. 
We just know as a matter of principle 
that the zeros have to embody a full description of the states,
which then includes their semiclassical behavior.
The issue is then to locate and decipher the dynamical information 
({\sl order parameters\/}) buried in the zeros. 

At best, each zero in an eigenconstellation might follow some
predictable rules, at least semiclassically;
but even if we restrict to exceptional subsequences of states
or zeros this may be too demanding, as the dynamics of the zeros
is exceedingly complicated. At worst, the information could be so scrambled up
that only quasi-random features could be asserted.
We will seek to identify intermediate types of information within the zeros,
weaker and thereby hopefully more accessible than the individual locations,
but still dynamical (i.e., stronger than statistical). These could lie
within collective coordinates of some sort (e.g., we recall that 
semiclassical effects can only be collectively generated).

One possibility inspired by the integrable situation (regularly spaced zeros on 
a curve produce WKB-like wave functions) would be to build up wave functions
through patching together local 2-d patterns of zeros of a few definite types.
This approach is valuable if forms of short-range order can be identified
in eigenconstellations even when the classical dynamics is chaotic.
Such regularities are actually not infrequent, but are not systematic either:
chaotic eigenconstellations mostly look like gases or liquids of zeros,
as opposed to solids (except for the few lattice eigenstates in cat maps).
Zeros thus appear to substantially use their 2-d freedom of relative motion,
making it difficult to greatly reduce the number of their parameters.

If individual zeros prove difficult to isolate, 
a valuable complementary approach is to analyze their density $\rho(x)$
in the dual {\sl Fourier space\/} which, for our torus phase space,
is the lattice of integer vectors $k=(k_q,k_p) \in \Bbb Z^2$:
the Fourier coefficients of the density are
\begin{equation}
\label{four}
\rho_k= \frac{1}{N}\sum_{j=1}^N \e^{-2\pi\mi k . x_j} \qquad k \in \Bbb Z^2 ,
\end{equation}
under the normalization $\rho_0 \equiv 1$. 
Some valuable features of this approach are, for instance:

1) the basic relationship (\ref{Poiss}), from the logarithm of the Husimi
function to the density of zeros, becomes diagonal in Fourier space: denoting
\begin{equation}
\label{tildeh}
\tilde h(x)\defi N^{-1}\log \hat{H}(x)= \sum_{k \in \Bbb Z^2} h_k \e^{2\pi\mi k . x} \quad ({\rm with\ } h_0 \equiv 0),
\end{equation}
we find that eq.(\ref{Poiss}) is mapped to the simple relation 
\begin{eqnarray}
\label{hk}
\rho_k -\delta_k &=& -\pi(k_q^2+k_p^2)h_k = -\pi |k|^2 h_k \qquad(\forall k \in \Bbb Z^2),\\
{\rm where} \quad \delta_k & \defi & 1 {\rm \ if\ } k=(0,0), 
\mbox{ else 0 (Kronecker symbol)}; \nonumber
\end{eqnarray}

2) if the zeros form a lattice, then the Fourier transform $\rho_k$ is simply 
the Dirac delta distribution on the dual lattice 
(this expresses the Poisson summation formula); 

3) if the zeros form a more disordered distribution of points,
this Fourier transformation can be well controlled statistically 
in terms of structure functions, alias form factors (see Sec. \ref{stat}).

However, this Fourier analysis is deferred to Sec. \ref{Fouri} of this paper, 
because some critically relevant information still has to be gathered before: 
some rigorous dynamical results about the zeros are  presented next, 
whereas a statistical approach is developed in the Section thereafter.

\section{Dynamical results}
\label{equidistribution}

In this section we study the analytical properties of sequences of 
states with the Schnirelman property, i.e. such that their Husimi densities $H_N$ converge weak-$*$ towards the Lebesgue measure $\cal L$ on the torus. We prove that the constellations also equidistribute in the weak-$*$ sense, and we deduce from this some universal properties of the phases of their Bargmann functions. We also give examples showing the limitations of this rigorous approach, as well as stronger results for a class of eigenstates of quantum cat maps (i.e. lattice eigenstates).

\subsection{Equidistribution of the eigenconstellations}
Using the fact that the Husimi density is the square of a holomorphic function
(up to a trivial Gaussian factor), it is possible to extract valuable information about the densities of zeros $\rho_N(x)$ from Schnirelman's
property. 

The link between both densities is provided by the {\sl logarithm} of the Husimi density. Namely, the factorization property (\ref{hfactp}) implies the very simple relation
\begin{equation}
 \log \hat H(x)=  \sum_{j=1}^N h(x-x_j)=N(h*\rho)(x),
\end{equation}
where the function $h(x)$ is the logarithm of the building block $H_{\chi}$, shifted such that its average over ${\Bbb T}^2$ vanishes: $h(x) = \log \hat{H}_\chi(x)$. This function corresponds (up to the sign) to the electric potential generated by a delta-like unit charge at the origin, balanced by a uniform negative charge to
ensure global electric neutrality. Precisely, we have 
\begin{equation}
\Delta h(x)=4\pi (\delta(x)-1),
\end{equation}
from which we easily recover eq.(\ref{Poiss}).

\medskip

Using the above representation of the Husimi functions, we will now prove that Schnirelman's property for a sequence of Husimi densities implies the weak-$*$ convergence of the corresponding densities of zeros to the Lebesgue measure ${\cal L}$ as $N\to\infty$. 

\smallskip
 
The densities of zeros are positive and normalized, so they belong to a compact set of measures in the weak-$*$ topology (this is the Banach--Alaoglu theorem \cite[p.105]{reed}). Therefore, we can extract a subsequence 
converging to a positive normalized measure $\rho_\infty$ (the elements of this subsequence will henceforth be noted $\rho_N$).

\smallskip

Our first task is to compare as precisely as possible the finite-$N$ and limiting electric potentials $h_N=h*\rho_N$ and $h_\infty=h*\rho_\infty$. 
The weak-$*$ property $\rho_N\to\rho_\infty$ obviously implies $h_N\stackrel{{\rm w}-*}{\to}h_\infty$, but we need information in stronger topologies.

$h(x)$ is at the same time upper semi-continuous (u.s.c) \cite[p.37]{rudin}
and in $L^s({\Bbb T}^2)$ for all $1\leq s<\infty$. This, combined with the fact that the measures $\rho_{\#}$ are normalized and positive (${\#}$ stands for $N$ or $\infty$), implies that the potentials $h_{\#}$ are also u.s.c. and in $L^s({\Bbb T}^2)$, with $||h_{\#}||_s\leq ||h||_s$ \cite{hayman}; furthermore, they are bounded above by $M\defi \sup h(x)$ (all these properties can be shown by considering
a decreasing sequence of continuous functions converging pointwise to $h(x)$, as in \cite[theorem 3.6]{hayman}).

These properties, combined with the weak-$*$ convergence of the $\rho_N$, imply  \cite[p.209]{hayman}:
\begin{eqnarray}
\label{hay}
\forall x,\quad \overline{h}_\infty(x)&\defi&\limsup_{N\to\infty} h_N (x)\quad\mbox{exists},\nonumber\\
\mbox{and}\ \overline{h}_\infty(x)&\leq&h_\infty (x).
\end{eqnarray}
 Equivalently, $\forall x$, $h_N(x)$ is smaller than $h_\infty(x)+\epsilon$ for $N$ large enough. This inequality can be proven to be {\sl uniform} w.r.t. $x$ by using {\sl subharmonicity properties} of the potentials. The functions $h_N$ themselves are not subharmonic, because their Laplacians involve a uniform negative charge (independent of $N$). Removing this constant charge amounts to systematically adding the function $\pi x^2$ to the potentials. The resulting (non-periodic) potentials $g_{\#}(x)=h_{\#}(x)+\pi x^2$ are uniformly bounded above over the square $[0,1]^2$ (like $h_{\#}$), but they are also subharmonic. Therefore, we can apply to the sequence $\{g_N\}$ several lemmas pertaining to subharmonic functions (see for instance \cite[\S 1.8]{lelong} or \cite[chapter 3]{doob}), and then transfer them back to the $h$-potentials by subtracting the $\pi x^2$ term. 

To obtain the uniformity in eq. (\ref{hay}), we need to consider a decreasing sequence of continuous functions $\{h_{\infty,m}\}_{m\in{\Bbb N}}$ on ${\Bbb T}^2$ converging pointwise to $h_\infty(x)$ (the existence of such a sequence is due to the upper semi-continuity of $h_\infty$). Then, we have Hartogs' lemma \cite[theorem 1.31]{lelong}:

\begin{lem}
\label{hartogs}
$\forall m\in{\Bbb N},\ \forall\epsilon >0,\ \exists N(m,\epsilon)$ s.t. 

$ N\geq N(m,\epsilon)\Longrightarrow \forall x\in {\Bbb T}^2,\ h_N(x)\leq h_{\infty,m}(x) +\epsilon$.
\end{lem} 

Before using this uniformity result, we state additional properties of the function $\overline{h}_\infty(x)$, also due to the subharmonicity and the uniform upper-boundedness of the $g$-potentials. 

\begin{lem}
The function $x\mapsto\overline{g}_\infty(x)=\overline{h}_\infty(x)+\pi x^2$ is "almost subharmonic".
\end{lem}
 This means that it is equal almost everywhere to a (unique) subharmonic function  $\overline{g}_\infty^*$. This function can be defined as an upper regularization of $\overline{g}_\infty$: $\overline{g}_\infty^*(x)= \lim_{r\to 0} {1\over\pi r^2}\int_{\{|y-x|<r\}} \overline{g}_\infty(y) dy$, which yields the property $\overline{g}_\infty(x)\leq \overline{g}_\infty^*(x)$ on ${\Bbb T}^2$. 

$\overline{g}_\infty^*$ can also be defined as the lowest subharmonic majorant of $\overline{g}_\infty$. Therefore, from the inequality (\ref{hay}) we deduce
\begin{equation}
 \overline{h}_\infty\leq\overline{h}_\infty^*\leq h_\infty.
\end{equation}
This, combined with the weak-$*$ convergence $h_N\to h_\infty$, entails the identity $\overline{h}_\infty^*=h_\infty$. Therefore, (\ref{hay}) is actually an equality for almost all $x$.

\bigskip

In a second step, we use lemma \ref{hartogs} to show that Schnirelman's property implies $h_\infty\equiv 0$ on ${\Bbb T}^2$. The proof proceeds ab absurdo. 

Let us assume that $h_\infty\not\equiv 0$. Using the notation $f^{-}(x)=\min(f(x),0)$, this entails $\int_{\Bbb {\Bbb T}^2}h^-_\infty (x) dx = -A<0$ (we recall that $h_\infty$ is in $L^1({\Bbb T}^2)$, and $\int_{{\Bbb T}^2}h_\infty(x)dx =0$ by construction). Then, the theorem of dominated convergence implies that $\int_{{\Bbb T}^2}h^-_{\infty,m}(x) dx <-A/2$ for $m$ large enough, hence the open set $E_{m,-A/3}=\{x\in {\Bbb T}^2\mid h_{\infty,m}(x)<-A/3\}$ has a non--zero Lebesgue measure. Now, for any $\epsilon>0$, lemma \ref{hartogs} implies that for all $N\geq N(m,\epsilon)$,  $x\in E_{m,-A/3}\Longrightarrow h_N(x)\leq -A/3 +\epsilon$, and therefore 

\begin{equation}
\label{lower}
\int_{E_{m,-A/3}} \e^{N\,h_N(x)} dx \leq {\cal L}(E_{m,-A/3}) \e^{N(-A/3+\epsilon)}.
\end{equation}
(We choose $\epsilon <A/3$ for the sake of the proof).

On the other hand, the compact set $F_+=\{x\in {\Bbb T}^2\mid h_\infty(x)\geq 0\}$ has non-zero Lebesgue measure, since $\int_{F_+} h_\infty(x) dx =A$. We want to estimate the size of its subsets $F_{N,+}=\{x\in F_+\mid h_N(x)\geq 0\}$. The weak-$*$ convergence $h_N\to h_\infty$ entails that for $N$ large enough (say, $N\geq N_o$), $A/2\leq \int_{F_+}h_N(x) dx$, itself less than
or equal to $\int_{F_{N,+}}h_N(x) dx$. Since all the potentials $h_N$ are uniformly bounded by $M$, we obtain the lower bound
${\cal L}({F_{N,+}})\geq A/2M$ for all $N\geq N_o$. Therefore,

\begin{equation}
\label{upper}
\int_{F_+}\e^{N\,h_N(x)} dx\geq \int_{F_{N,+}}\e^{N\,h_N(x)} dx\geq {\cal L}({F_{N,+}})\geq A/2M.
\end{equation}
We now combine the equations (\ref{lower},\ref{upper}) and obtain 
\begin{equation}
\forall N\geq \max(N(m,\epsilon),N_o),\quad \frac{\int_{F_+} H_N(x)dx}{\int_{E_{m,-A/3}}H_N(x)dx}\geq \e^{N(A/3-\epsilon)} {A\over 2M {\cal L}(E_{m,-A/3})}.
\end{equation}
 Such a sequence of Husimi densities $\{H_N\}$ obviously violates Schnirelman's property, since the above ratio diverges in the limit $N\to\infty$, instead of converging towards ${{\cal L}(F_+)\over {\cal L}(E_{m,-A/3})}$. This concludes the proof ab absurdo.

We have proven that for any sequence $\{H_N\}$ weak-$*$ converging to ${\cal L}$, the only possible accumulation point (in the weak-$*$ topology) of the corresponding sequence $\{\rho_{N}\}$ is ${\cal L}$. Since this sequence stays in a compact set, we deduce that it converges weak-$*$ to ${\cal L}$. 
 
\begin{theo}
\label{equi}
For any sequence of Husimi densities $\{H_{\psi^{(N)}_{j(N)}}\}_{N\in {\Bbb N}}$ weak-$*$ converging to the Lebesgue measure ${\cal L}$ on the torus in the semi-classical limit, the corresponding densities of zeros $\{\rho_{\psi^{(N)}_{j(N)}}\}_{N\in {\Bbb N}}$ also equidistribute in the weak-$*$ sense.
\end{theo}
This theorem concerns Husimi densities defined on the torus, but it can be easily generalized to sequences of Husimi densities (and the associated constellations) living on compact Riemann surfaces of any genus, using the formalism sketched in section \ref{alternative} \cite{these}. The theorem has also been recently generalized \cite{schiff} to holomorphic sections $s_N(z)$ on the powers $L^{\otimes N}$ of a positive hermitian line bundle $L$ over a compact K\"ahler manifold $(X,\omega)$ of arbitrary dimension; the analog of the Husimi density is the hermitian metric of the section $H_{s_N}(z,\bar z)\defi\|s_N(z)\|^2$; the zero set of $s_N$ defines a $(1,1)$-current on the manifold, which is shown to converge weak-$*$ to the K\"ahler form $\omega$ in the limit $N\to\infty$ as long as the Husimi densities become uniform. 

In the case of the 2-sphere, the asymptotic equidistribution of chaotic eigenconstellations had already been noticed from numerical calculations \cite{leb,leb:shukla}. The data actually showed a more precise phenomenon: not only do the zeros spread throughout the whole phase space, but they also seem to {\sl repel each other} at distances of order $1/\sqrt{N}$ (see section \ref{stat} for a more quantitative statement of this phenomenon). Is such a repulsion another consequence of Schnirelman's property?  

\subsection{Further considerations on Schnirelman's property}

\subsubsection{Fourier coefficients}

In the toral geometry, the weak-$*$ convergence of a sequence of positive normalized measures $\rho_N$ to $\rho_\infty$ is equivalent to the convergence of each Fourier coefficient (eq. (\ref{four})); we have just proven that Schnirelman's property for a sequence $\{H_N\}$ implies 
\begin{equation}
\label{weakcv}  
\forall k \in{\Bbb Z}^2,\qquad \rho_{N,k }
\stackrel{N\to\infty}{\longrightarrow} \delta_{k}.
\end{equation}
($\delta_k$ is the Kronecker symbol as in eq.(\ref{hk})).
We would like to know better the rate of decrease w.r.t. $N$ of these Fourier coefficients, as well as their dependence on $k$, for $N$ fixed. Apparently, this kind of information is not easy to extract from weak-$*$ estimates. However, by formally looking at the relation 
\begin{equation}
\label{epitom}
\hat H_N(x)= \exp\left(-\sum_{k \neq 0 }{N\rho_{N,k }\over{\pi k }^2}\e^{2\mi\pi k .x}\right),
\end{equation}
(implied by eqs.(\ref{tildeh})--(\ref{hk})), we expect the following:

\begin{conj}
\label{conj}
For any sequence $\{H_{\psi^{(N)}_{j(N)}}\}_{N\in {\Bbb N}}$ with the Schnirelman property, the $k\neq 0$ Fourier coefficients of the constellations 
decay as $\rho_{N,k }=o(N^{-1})$ for $N\to\infty$.
\end{conj}
We are presently unable to prove such an assertion, even with $N^{-1}$ replaced by some other definite $o(1)$ function.

\subsubsection{Upper bound on the potentials}
 Alternatively, we would like to control more precisely the semi-classical properties of the electric potentials $h_N$. Via a Parseval--Plancherel formula, we control their $L^2$ norms:
\begin{eqnarray}
\int_{{\Bbb T}^2} |h_N(x)|^2\,dx&=&\sum_{k\neq 0}|h_{N,k}|^2\\
&=&\sum_{k\neq 0}{|\rho_{N,k}|^2\over \pi^2 k^4}\nonumber\\
&\leq&\sum_{0\neq |k|\leq K_1}{|\rho_{N,k}|^2\over \pi^2 k^4}+\sum_{|k|> K_1}{1\over \pi^2 k^4},\nonumber
\end{eqnarray}
since all Fourier coefficients are normalized. Adjusting the cutoff $K_1$ and using the convexity property $0<r\leq s\Longrightarrow ||.||_r\leq ||.||_s$, we prove the
following:
\begin{theo}
\label{Ls}
Let $\{\psi_N(z)\}_N$ be a sequence of Bargmann functions s.t. their densities of zeros $\{\rho_N\}_N$ weak-$*$ converge to $\cal L$. Then, for any $1\leq s\leq 2$, the corresponding potentials $\{h_N\}$ tend to zero in the $L^s$ norm, as $N\to\infty$.
\end{theo}
In the general case studied by Schiffman and Zelditch \cite{schiff}, the above convergence was
proven for the $L^1$ norm. 

To obtain pointwise information on the potentials, we can use lemma \ref{hartogs} 
with the knowledge that $h_\infty(x)\equiv 0$: this yields
\begin{equation}
\forall \epsilon,\ \exists N(\epsilon)\ \mbox{s.t.}\ \forall N\geq N(\epsilon),\forall x\in{\Bbb T}^2,\ h_N(x)\leq\epsilon.
\end{equation}
We then naturally tried to estimate the rate of increase of $N(\epsilon)$, as $\epsilon\to 0$. In other words, we searched for a universal increasing function $f(N)$, such that the functions $f(N)h_N(x)$ are uniformly bounded above by a constant independent of $N$, as long as $\{H_N\}$ have the Schnirelman property. 
The following example of sequences of Husimi densities shows that we cannot do better than $f(N)\equiv 1$ in general.
At the same time, it shows that Schnirelman's property does not preclude high degeneracies of the zeros.

\subsubsection{Example of singular Husimi function with the Schnirelman property}
\label{singschni}
We present below sequences of Husimi densities which look quite singular as far as their smoothness is concerned, although they weak-$*$ converge to ${\cal L}$.
These Husimi functions are built by rescaling and taking a certain power of the elementary $H_\chi(x)$, so their zero constellations will consist of square lattices, with each zero multiply degenerate. For any couple of integers $(N_1,N_2)$, we consider the Husimi function
in ${\cal H}_N$, ($N=N_1^2N_2$) defined as
\begin{equation}
\hat H_N (x)\defi \e^{N_2 h(N_1 x)}.
\end{equation}
On the one hand, the $1/N_1$ periodicity ensures that the Fourier coefficients $H_{N,k }$ and $\rho_{N,k }$ vanish unless $N_1$ divides both entries of ${k }$. Therefore, if we consider a sequence $\{N=N_1^2 N_2\}$ s.t. $N_1\to\infty$, Schnirelman's property holds for the functions $\{H_N\}$ (and at the same time, $N\rho_{N,k }\to 0$ for all $k $). 

On the other hand, all invariant functionals take the same values for $H_N$ and the completely degenerate state of order $N_2$ (cf. fig.2b, right). If $N_2$ gets large, we can use the asymptotic results from the last line of table \ref{tab}, replacing $N$ by $N_2$. In that case, the Husimi densities $H_N$ asymptotically look like delta peaks concentrated on the lattices dual to their constellations, which explains the large deviations of the invariants from the totally flat case (i.e. $H\equiv 1$).

The electric potentials of the constellations are given by
$h_N(x)={1\over N_1^2}h(N_1 x)$, bounded above by $M/N_1^2$. If $N_2$ increases much faster that $N_1$ (e.g. $N_2\sim \e^{N_1}$), then a function $f(N)$ as defined above has to increase slower than $N_1^2$ (e.g. $\sim \log^2 N$). There is therefore no universal strictly increasing $f(N)$.

At the same time, we notice that such a singular behavior of the invariants, and of the shape of the densities, comes along with a high degeneracy of the zeros. In that sense, the strict {\sl repulsion} between the zeros of chaotic eigenconstellations, conjectured in \cite{leb,leb:shukla}, cannot be deduced from Schnirelman's property. 

\subsubsection{Example of non-ergodic Husimi functions with equidistributed zeros}
\label{counter}

We exhibit a sequence of Husimi functions showing that the converse of theorem \ref{equi} is false. We use constellations in the shape of deformed lattices, defined as follows. We start from a smooth separable density on ${\Bbb T}^2$, $\overline{\rho}(x)=\rho_q(q)\rho_p(p)$, with the normalizations $\int_0^1 \rho_i(v) dv =1$, for $i=q,p$. Separability allows us to integrate this density, i.e. to change coordinates $x\mapsto {\cal N}(x)$, s.t. $\overline{\rho}(x)=\det{D{\cal N}\over Dx}(x)$ (we just take ${\cal N}_i(v)=\int_0^v \rho_i(v')dv'$ for $i=q,p$). 

Now, we build the Husimi function $H_{\overline{\rho},N}(x)$ with zeros at the $N=M^2$ points $x_{i,j}={\cal N}^{-1}(i/M,j/M)$, through
\begin{equation}
\tilde h_{\overline{\rho},N}(x) \defi \log \hat H_{\overline{\rho},N}(q,p) \defi \sum_{i,j=1}^M h\left(q-{\cal N}^{-1}_q(i/M),p-{\cal N}^{-1}_p(j/M)\right).
\end{equation} 
We want an approximate formula for $\hat H_{\overline{\rho},N}(x)$, in the limit $M\to\infty$. For that, we transform the right-hand side by the Poisson summation formula at fixed $x$, which yields two types of terms. The zero Fourier coefficient (Weyl's term) is $M^2 h*\overline{\rho}(x)$, whereas the others (oscillatory terms) read
\begin{equation}
M^2\int_{{\Bbb T}^2} h(y) \overline{\rho}(x+y)\e^{-2\mi\pi Mk .{\cal N}(y+x)}\, d^2 y, \quad 0 \ne k \in {\Bbb Z}^2.
\end{equation}
For $k \neq  0 $ fixed and $M\to +\infty$, the above integral is dominated by the contribution near the singular point $y=0$. Expanding ${\cal N}$ around $x$ and summing over all $k \neq 0 $, we (formally) obtain  
\begin{equation}
\tilde h_{\overline{\rho},N}(x) -M^2 h*\overline{\rho}(x)
\sim \log \hat{H}_{\chi(\tau(x))}(M{\cal N}_q(x)+\tau(x)M{\cal N}_p(x)),\quad\makebox{\rm with }\tau(x)=\mi\,{\rho_p(p)\over\rho_q(q)}
\end{equation}
($H_{\chi(\tau)}$ is defined in equation (\ref{chitau})).  
The convergence of the Poisson series to this function is uniform as long as
one stays away from the $x_{i,j}$ (notice that their positions change with $M$ !). Anyway, since we subsequently exponentiate this function, we need not pay too much attention to problems of convergence near the zeros. In the large-$M$ limit, we thus obtain a ``WKB-like" expression for 
$\hat H_{\overline{\rho},N}(x)$:
\begin{equation}
\hat H_{\overline{\rho},N}(x)\sim \exp(N h*\overline{\rho}(x)) \;\hat{H}_{\chi(\tau(x))}\left(M{\cal N}_q(x)+\tau(x)M{\cal N}_p(x)\right).
\end{equation}
 Locally, this oscillatory function vanishes on a rectangular lattice of spacing $\propto 1/M$, since the parameter $\tau(x)$ varies slowly compared to this spacing: the large-$M$ constellations thus resemble deformed lattices.

\medskip

We now let the smooth density $\overline{\rho}$ itself depend on $N$: we take the sequence ${\cal N}_N(x)=(q+\lambda_q\sin(2\pi q), p+\lambda_p\sin(2\pi p))$, which yields $\overline{\rho}_N(x)=(1+2\pi\lambda_q\cos(2\pi q))(1+2\pi\lambda_p\cos(2\pi p))$, where the coefficients $\lambda_i$ depend on $N$. If $\lambda_q$, $\lambda_p$ vanish in the limit $N\to\infty$, then $\overline{\rho}_N\to 1$ and $\tau(x)\to \mi$ uniformly on ${\Bbb T}^2$ so that the oscillatory factor has asymptotically the local average $1/{\rm GM}[H_{\chi}]$.

Meanwhile, the convolution under the exponential, $h*\overline{\rho}(x)$, yields the trigonometric polynomial $P_N(x) \defi -2[(\pi\lambda_q\cos(2\pi q)+1)(\pi\lambda_p\cos(2\pi p)+1)-1]$. Then, if the decrease of either $\lambda_i$ is slower than $1/N$, the factor $\exp(N P_N(x))$ is singular in the semi-classical limit.
For instance, if both $\lambda_i$ decrease slower than $1/N$, it gives a peak (an artificial point scar) at the point $(1/2,1/2)$; whereas in the marginal case $\lambda_i =\Lambda_i/N$, with $\Lambda_i$ non-vanishing constants, that factor is asymptotically $N$-independent; $\hat H_{\overline{\rho}_N,N}$ then converges weak-$*$ to $\exp(-2\pi\Lambda_q\cos(2\pi q)-2\pi\Lambda_p\cos(2\pi p))/{\rm GM}[H_{\chi}]$, a non-uniform density.

So, Schnirelman's property does not hold for such sequences $\hat H_{\overline{\rho}_N,N}$, although $\rho_N\stackrel{{\rm w}-*}{\to}\cal L$. However, this counterexample is just consistent with the converse of conjecture \ref{conj}.  

In figure 4 we display Husimi densities of the type described above, for various values of $N$ and $\lambda_i$, together with the corresponding invariants.

\subsection{Phase of the Bargmann eigenfunctions}
\label{Phase}

From the semi-classical equidistribution of the eigenconstellations, one can get
some estimates about the {\sl phase} of the Bargmann wave-function, which appears as a complementary information to the Husimi density (see eq.(\ref{hus})). For this purpose, we compute the derivative of $\log H(x)$ w.r.t. the holomorphic variable $z$, and get the following representations:

\begin{eqnarray}
{\psi'(z)\over\psi(z)}-2\pi N\bar{z}&=&{\partial \over\partial z}\log H(z,\bar{z})\\
&=&N[(\partial_z h)*\rho](z,\bar{z})\\
&=&-\mi N\sq2\;\sum_{k \neq 0 }{\rho_{k }\over k_q-\mi k_p}\e^{2\mi\pi k .x}
\label{d3}
\end{eqnarray}
We will use the second and third equations to prove the following theorem.

\begin{theo}
\label{deriv}
Let $\{\psi_N(z)\}_N$ be a sequence of Bargmann functions s.t. their densities of zeros $\{\rho_N\}_N$ weak-$*$ converge to $\cal L$. Then, for any $1\leq s<2$, the functions ${1\over 2\pi N}{\psi'_N\over \psi_N}(z)$ tend to the function $\bar z$ in the $L^s$ norm, as $N\to\infty$.
\end{theo}
To prove the theorem, our strategy is to transform the series (\ref{d3}) into a finite sum, up to a small remainder, and then use the convergence of the individual Fourier coefficients $\rho_{N,k }$ (cf. eq.(\ref{weakcv})), as in the proof of theorem \ref{Ls}. However, due to the stronger singularities of the logarithmic derivatives, we now need to first regularize the functions $\partial_z h*\rho_N$ by Gaussian convolutions $\delta_K({x})=K^2\e^{-\pi K^2{x}^2}$. In the limit $K\to\infty$, $\delta_K$ converges to the identity kernel $\delta$. For finite $K$, the deviation from the identity is given by the series
\begin{equation}
\label{diff}
[\partial_z h-\partial_z h*\delta_K](q,p)=\sq2 \sum_{n,m\in{\Bbb Z}^2} {\e^{-\pi K^2[(q-m)^2+(p-n)^2]}\over(q-m)+\mi(p-n)},\qquad (q,p)\not\in {\Bbb Z}^2.
\end{equation}
Using the normalization of $\rho_N$ and the equation above, we obtain the following estimates, valid for any fixed $1\leq s<2$:
\begin{eqnarray}
\int_{{\Bbb T}^2}\Big|[\rho_N*(\partial_z h-\partial_z h*\delta_K)](x)\Big|^s dx &\leq&
\int_{{\Bbb T}^2}\Big|[\partial_z h-\partial_z h*\delta_K](x)\Big|^s dx\\
\label{gamma}
&\leq&(2\pi s)^{s/2}{\Gamma(1-s/2)\over s\,K^{2-s}}.
\end{eqnarray}
The first inequality is just due to convexity (it is equivalent to the triangular inequality for the $L^s$ norm); the right hand-side can then be calculated exactly, leading to (\ref{gamma}). The divergence as $s\to 2$ is obviously due to the poles of $\psi'/\psi$.
Once we select the power $s$, we can adjust the cutoff parameter $K$ large enough to make $||\rho_N*(\partial_z h-\partial_z h*\delta_K)||_s$ small. On the other hand, the smoothed term can be now estimated:
\begin{eqnarray}
||\rho_N *\partial_z h*\delta_K||_s&\leq&||\rho_N *\partial_z h*\delta_K||_2\\
&\leq&\left(\sum_{k \neq 0 }\;{2\over k ^2} |\rho_{N,k }|^2 \e^{-2\pi{k ^2\over K^2}}\right)^{1/2}\\
&\leq&\left(\sum_{0\neq|k |\leq K_1}{2\over k ^2}|\rho_{N,k }|^2\e^{-2\pi{k ^2\over K^2}} + (K/K_1)^2\e^{-\pi((K_1-\sq2)/K)^2}\right)^{1/2}.
\end{eqnarray}
In the last inequality, we introduced a cutoff $K_1$, and estimated the remainder of the series $\sum_{|k |>K_1}$ using the uniform bound $|\rho_{N,k }|\leq 1$. For $K_1/K$ large enough, this remainder is small; finally, the finite sum $\sum_{|k |\leq K_1}$ vanishes in the limit $N\to\infty$ due to the equidistribution of zeros. 

Notice that the theorem extends to values of $s$ in the interval $(0,1)$, due to the convexity property $0<r<s\leq\infty\Longrightarrow ||.||_r\leq ||.||_s$. 


\medskip

\noindent {\bf Interpretation}

\smallskip

In view of the formal analog $\mi \hbar(\psi'/\psi)(q) \sim p$ 
--- the momentum --- in the 1-d Schr\"odinger representation, we can define
$\pi(z)\defi \hbar {\psi'\over \psi}(z)$ as the {\sl quantum local momentum\/} of the Bargmann function (whereas $\bar z$ is the classical symplectic conjugate variable of $z$). 

The Cauchy--Riemann relations link the estimates $\pi(z)\sim\bar z$ of theorem \ref{deriv} to the variations of both the modulus and the phase of the Bargmann functions:
\begin{eqnarray}
\label{module}
{1\over\sq2}\left({\psi'\over \psi}(z)-2\pi N\bar z\right)&=&(\partial_q\log|\psi(z)|-\pi N q)+\mi (\partial_p\log|\psi(z)|-\pi Np)\\
\label{phase}
&=&-(\partial_p \arg\psi(z)+\pi Nq)+\mi (\partial_q\arg\psi(z) -\pi Np).
\end{eqnarray}
These two equations correspond to 
interpreting the zeros respectively as 

\begin{itemize}
\item either point-like electric charges, generating a potential $-\log|\psi(z)|$, balanced by a uniform charge distribution of potential $\pi Nx^2/2$ (the total electric charge on the torus vanishes). 
\item or vortices of magnetic flux: the corresponding vector potential is the gradient $\vec{\nabla}\arg\psi$, so that each vortex carries a unit of flux $\phi=-2\pi$. The additional vector potential $\pi N\left(\begin{array}{c}-p\\q\end{array}\right)$ corresponds to a uniform magnetic field $B=2\pi N$ (the total magnetic flux on the torus vanishes).
\end{itemize}

It seems difficult to assert the approximation $\pi(z)\sim \bar{z}$ in stronger topologies than the $L^s$ estimates of theorem \ref{deriv}. Generically, the strict equality $\pi(z)=\bar{z}$ can only hold
at {\sl isolated points} on the torus, due to the analyticity of $\psi(z)$, or equivalently due to its {\sl phase}. Indeed, the right-hand sides of equations (\ref{module},\ref{phase}) have quite different behaviors if we try to set them equal to zero. The equation on the modulus has the obvious smooth solution $\log|\psi(z)|=\pi N z\bar z$, which corresponds to a strictly uniform Husimi density, i.e. the formal Schnirelman limit. By contrast, the equation on the phase reads
\begin{equation}
\label{nabla}
\vec\nabla \arg\psi(z) = \left(\begin{array}{c}\pi Np\\-\pi Nq\end{array}\right)
\end{equation}
which has no solution, since the corresponding magnetic field
$ \vec\nabla\times \left(\begin{array}{c}\pi Np\\-\pi Nq\end{array}\right) =-2\pi N$ does not vanish. The singular behavior of the phase of $\psi(z)$ in the semi-classical limit can be therefore interpreted as a `struggle' to achieve
as well as possible the equality (\ref{nabla}); such a singular behavior implies
a distribution of phase {\sl dislocations} (i.e. of zeros of $\psi$) all over ${\Bbb T}^2$ in the semi-classical limit.  

At first sight, the situation looks quite different in the integrable case,
say for eigenstates of a time-independent Hamiltonian which we express 
in the complex torus coordinates as ${\cal H}(z,\bar z)$. 
Assuming that ${\cal H}$ is an analytic function and that the eigenvalue $E$
is a regular energy value, the WKB form (\ref{WKB}) holds for semiclassical
eigenfunctions in the Bargmann representation \cite{voros89}.
Then, in the $\hbar \to 0$ limit, $\hbar (\log \psi)'(z) \sim S'(z)$ solves 
 ${\cal H} (z,S'(z))=E$, the Hamilton--Jacobi equation in the $z$ variable: hence $S'(z)= y_E(z)$,
where $\{y=y_E(z)\}$ is the classical complex energy curve 
$\Sigma_E^\Bbb C=\{{\cal H}(z,y)=E\}$ solved for $y$, 
the conjugate momentum of $z$. 
These asymptotic forms hold outside the {\sl anti-Stokes lines\/}, 
i.e., the curves where
any two real parts of the (multisheeted) function $y_E(z)$ match.
All in all, the dynamical result for an integrable eigenfunction is then
\begin{equation}
\label{resi}
\pi(z) \sim y_E(z), \quad
\mbox{outside anti-Stokes lines, for } \hbar \to 0.
\end{equation}
Precisely, the {\sl quantum\/} local momentum of the Bargmann function $\pi(z)$
tends to some branch of the {\sl classical\/} local momentum function, $y_E(z)$ in the $z$-representation.

However, the perspective changes if we seek the asymptotic behaviors
{\sl specifically obeyed in the classically allowed part of phase space\/}
(which is the only semiclassically meaningful region, 
where the Husimi density will not become negligible).
Here this region is the {\sl real\/} energy curve
$\Sigma_E^\Bbb R \equiv \{y_E(z) = \bar z\}$ (locally), 
on which eq.(\ref{resi}) implies
\begin{equation}
\label{res}
\pi(z) \sim \bar z \quad
\mbox{in the classically allowed phase space, for } \hbar \to 0.
\end{equation}
We now compare with theorem \ref{deriv} for an ergodic situation:
then the classically allowed region {\sl is\/} the whole phase space,
so we realize that the {\sl same\/} asymptotic result (\ref{res}) 
has become generalized from integrable to chaotic situations, 
in which case it moreover applies almost everywhere in phase space 
--- albeit in a weaker sense ($L^s$) than before (pointwise).
The contrast between implementations of eq.(\ref{res}) for integrable 
vs chaotic dynamics is illustrated by color plots for $\pi(z)$ (fig.5 top); 
we add that the same overall appearance is universally
shown by all such plots made for equidistributed zeros 
(disregarding the precise distribution of the singular points themselves),
whereas it is different and case-dependent for zeros concentrating on curves.

Upon a purely formal integration, eq.(\ref{res}) yields
\begin{equation}
\mbox{`` }\log \psi (z) \sim \frac{1}{\hbar} \int^z \bar z' dz'\mbox{ "},
\end{equation}
but this is as inconsistent as eq.(\ref{nabla}) for specifying $\Im \log \psi (z)$, since the 1-form $\Im(\bar z' dz')$ is not closed. 
Hence the phase of $\psi(z)$ will stay undetermined pointwise;
however, eq.(\ref{res}) specifies as universal its {\sl variation pattern\/}
when zeros are equidistributed,
and this is displayed by color plots for $\psi(z)$ itself in fig.5 bottom.

\subsection{Stronger estimates for lattice eigenstates}
\label{lattice}
We can strengthen the above estimates for a particular class of 
eigenstates of quantum cat maps \cite{hannay:b}, namely eigenstates for which
the constellations form {\sl lattices} on the torus (such a lattice must be invariant under the classical cat map). The construction of these
eigenstates is performed in \cite{DEGI,crystal}. For a given classical cat map $S$, such states exist only for some particular values of the inverse Planck's constant $N$, in which case they are quite scarce (they generate a subspace of ${\cal H}_N$ of dimension small compared to $N$). Moreover, their quasi-energies are usually degenerate. On the other hand, the regularity of the constellations allows us to estimate more precisely the invariant functionals, as well as the phase variations. Indeed, an alternative characterization
of such a lattice state is that its Bargmann function can be written as a single Jacobi
theta function \cite{crystal}. For instance, if the eigenconstellation is the lattice on $T_{\Bbb C}$ generated by the two complex numbers $[v_1,v_2]$ ordered s.t. $\Im(\tau\defi v_2/v_1)>0$, then the corresponding Bargmann and Husimi functions read
\begin{eqnarray}
\label{rescale}
\psi(z)&=&\chi({z\over \sq2 v_1}|\tau),\\
H_\psi(x)&=&H_{\chi(\tau)}(x/\sq2 v_1),\nonumber
\end{eqnarray}
 where the elementary functions $\chi(z|\tau)$, $H_{\chi(\tau)}$ are defined in equations (\ref{chitau}). $\psi(z)$ is then quasiperiodic (and $H_\psi$ is periodic) w.r.t. the lattice $[v_1,v_2]$, which is a stronger property than eq.(\ref{qper}) (the Husimi density of such an eigenstate is displayed in fig. 1a, top right). Precisely, for any point $v=nv_1+mv_2$ of the lattice, we have
\begin{eqnarray}
\psi(z+v)&=&(-1)^{(n+m+nm)}\e^{\pi N|v|^2+2\pi N z \bar{v}}\psi(z)\\
\Longrightarrow {\psi'\over \psi}(z+v)&=&{\psi'\over \psi}(z)+2\pi N\bar v.
\label{fine3}
\end{eqnarray}
A crucial property of these lattice eigenstates is that for a given cat map $S$, the values of the modulus $\tau$ stay inside a compact domain of the upper half-plane, namely a rectangle $|\Re(\tau)|\leq 1/2$, $1/2\leq\Im(\tau)\leq C_S$, with $C_S$ a constant independent of $N$ (see Appendix B). 

The norms and related functionals introduced in section \ref{norms} are invariant
through a rescaling of the variable $z$, so they take the same values for $H_\psi(x)$ of equation (\ref{rescale}) and $H_{\chi(\tau)}(x)$. Since these functionals are smooth functions of $\tau$ (cf. App. A), they are {\sl bounded uniformly w.r.t.} $N$ for all lattice eigenstates $\psi$ of the quantum operators $U_S$ associated to a given cat map $S$ (the bounds depend on $S$). 

\medskip
Similarly, for such lattice states we get fine estimates of the logarithmic derivative of $\psi(z)$, using the quasiperiodicity (\ref{fine3}). We obtain from eq.(\ref{rescale}),
\begin{equation}
\int_{T_{\Bbb C}}\Big|{\psi'\over\psi}(z)-2\pi N\bar{z}\Big|^s d^2 z={1\over 2|v_1|^s}
{1\over\Im(\tau)}\int_{T_\tau}\Big| {1\over 2}(\partial_q+\mi\partial_p)\log H_{\chi(\tau)}(q,p)\Big|^s dq\,dp.
\end{equation}   
As $\tau$ stays inside a compact domain, the integral of $|\partial_z \log H_{\chi(\tau)}|^s$ on $T_\tau$ is bounded. On the other hand, we show in Appendix B that $|v_1|^{-1}=O(\sqrt{N})$, so we may strengthen theorem \ref{deriv}:
\begin{equation}
|| {1\over 2\pi N}{\psi'_N\over \psi_N}(z) -\bar{z}||_s\leq {C_{S,s}\over\sqrt{N}},\qquad\mbox{with }C_{S,s}\mbox{ indep. of }N.
\end{equation}

\medskip
On the Fourier side, to a given cat map $S$ is associated a constant $K_S$ s.t. for any lattice eigenstate $\psi_N$, the Fourier coefficients of the constellation (which are supported by the dual lattice) have the following property \cite{crystal}:
\begin{equation}
\forall k \neq 0 ,\quad |k |\leq K_S\sqrt{N}\Longrightarrow \rho_{N,k }=0.
\end{equation}
Notice that the above equation trivially implies that conjecture 1 holds for a sequence of lattice eigenstates, since $\rho_{N,k }=0$ for $N$ large enough.

\section{Statistical model}
\label{stat}

In section \ref{comparison}, we observed that the values of the different functionals for eigenstates of
quantum chaotic maps were surprisingly close to their average values over a certain ensemble of 
random vectors of ${\cal H}_N$, which indicates that a chaotic eigenstate `looks like' a random state, at least in a certain sense. This remark is linked to the various random matrix conjectures in quantum chaos \cite{bohigas}, since the eigenstate of a random matrix is a random state. We will not try to justify these conjectures in the following, but rather describe further the relevant random states and their properties, especially in the Bargmann--Husimi--stellar framework, in order to compare to them the corresponding quantities computed for eigenstates of quantum chaotic maps. 

\medskip

Such statistical models were already studied in alternative phase spaces, namely the 2-sphere and the plane \cite{hannay,edelman,leb:shukla,BBL,prosen}, where the Bargmann functions are respectively polynomials or entire functions in $z$ of controlled growth. The main observations concerning the statistical constellations were the following: 
\begin{itemize}
\item first, the zeros are on average equidistributed over the phase space, or the classically allowed part thereof. This property matches the equidistribution we proved for the eigenstates with the Schnirelman property (theorem \ref{equi}). 
\item second, zeros at short distance (approximately $\sqrt{\hbar}$) tend to repel each other \cite{leb:shukla,BBL,prosen}. Note that this typical distance coupled with the finite phase space volume imply a certain {\sl rigidity} of the constellation. This phenomenon seems present also for eigenstates of quantum chaotic maps; nonetheless, we already argued that such a repulsion cannot be explained by Schnirelman's property (cf. section \ref{singschni}). To characterize
precisely the rigidity of the random constellations, one could study the statistics of
their Voronoi tessellations (such tessellations are shown in figures 1a and 1b, bottom rows, for chaotic eigenstates). This approach was used for instance to characterize eigenvalues of large complex random matrices \cite{lecaer}. We did not investigate in this direction in the present article.  
\end{itemize}

\medskip
{\bf Symmetries}

\smallskip

The several models we will introduce correspond to vectors with different symmetry properties; symmetries play a fundamental role in the {\sl spectral} properties of random matrices \cite{bohigas}, and their relevance in the description of random states has already been noticed \cite{BBL,prosen}. 

The classical maps we study are all invariant under parity $(q,p)\mapsto(-q,-p)$, and the quantizations we consider then yield either even or odd eigenstates; by linearity, the Bargmann functions of these eigenstates will also be even or odd functions of $z$. As a consequence, we will consider models of even random states (the treatment of odd states is very similar, and we skip it).

 Besides, the classical maps can also have anti-canonical symmetries. The baker's map is invariant under the reflection
$(q,p)\mapsto(p,q)$, and the simplest `cat' map usually considered (i.e. the matrix $S= \left(\begin{array}{cc}2&1\\3&2\end{array}\right)$) has the time-reversal symmetry $(q,p)\mapsto(q,-p)$. On the quantum side, these symmetries manifest
themselves on the coefficients of eigenvectors. For instance, the time-reversal symmetry corresponds to eigenvectors with {\sl real} Schr\"odinger coefficients, giving real Bargmann functions (i.e. $\psi(\overline{z})=\overline{\psi(z)}$). Such a property has to be incorporated in the statistical model as well \cite{BBL,prosen}.

It is possible to build chaotic maps on the torus with no anti-canonical symmetries (some generalized baker's maps, or for instance the cat map $S'$ of eq. (\ref{cats})). However, these systems are in practice much more hyperbolic than the simpler ones cited above, so that their semiclassical properties (e.g. scars) may appear only for very large values of $N$, which is inconvenient numerically. We nevertheless start our study by considering an ensemble of random vectors with no symmetry whatsoever; this model yields tractable analytical results, not all of which can be generalized to more complicated ensembles.

\subsection{Model with no symmetry }
\label{nosym} 

We define a random state in ${\cal H}_{N,0}$ ($\varphi=0$ without loss of generality), as:
\begin{equation}
\label{psia}
\psi_{\{a\}}=\sum_{j=0}^{N-1} a_j |q_j\rangle_{N,0},
\end{equation}
where the Schr\"{o}dinger coefficients $a_j$ are random independent Gaussian complex variables with $\Big\langle a_j\Big\rangle_N=0$, $\Big\langle \bar {a}_j a_k\Big\rangle_N=\delta_{jk}/N$. The states $|q_j\rangle_{N,0}$, $j=0,\ldots ,N-1$, form the orthonormal basis of position eigenstates of ${\cal H}_{N,0}$ \cite{crystal} (but the statistical ensemble is invariant through any unitary change of basis).

The vectors $\psi_{\{a\}}$ are not normalized a priori, but their square norm $n_2=||\psi_{\{a\}}||^2$ has the distribution law 
\begin{equation}
\label{canonical}
{N^N\over (N-1)!}\;n_2^{N-1}\,\e^{-N n_2}\,dn_2\sim \sqrt{N\over 2\pi}\exp\left\{-{N(n_2-1)^2\over 2}\right\}dn_2,
\end{equation}
increasingly peaked around $n_2=1$ in the limit $N\to\infty$.
With this fact in mind, we can still 
calculate the typical values of the different functionals introduced in section \ref{norms}.  

The distribution law of the 
Bargmann function at a given point $z_o$ is the Gaussian ${1\over\pi\sigma_{z_o}}\e^{-{|\psi|^2\over\sigma_{z_o}}}d^2\psi$, with width
\begin{equation}
\sigma_{z_o}={1\over N}\sum_{j=0}^{N-1}|\langle z_o|q_j\rangle_{N,0}|^2.
\end{equation}
This width is actually a particular case of the correlation function
$\Big\langle \psi(z_1)\overline{\psi(z_2)}\Big\rangle_N$ which happens to be the crucial quantity of the statistical model. We give its
value for $N$ an even integer (the formula for $N$ odd is slightly more complicated, but has the same large-$N$ behavior):
\begin{eqnarray}
\label{corr}
\Big\langle \psi(z_1)\overline{\psi({z}_2)}\Big\rangle_N&=&N^{-1}\sum_{j=0}^{N-1} \langle z_1|q_j\rangle_{N,0}\;\overline{\langle z_2|q_j\rangle}_{N,0}\\
&=&N^{-1}\,_{N,0}\langle z_1|z_2\rangle_{N,0}\nonumber\\
&=&\e^{2\pi Nz_1\bar{z}_2}\theta_3(\pi N{z_1+\bar{z}_2\over \sq2}\mid\mi N/2)
\;\theta_3(\mi\pi N{z_1-\bar{z}_2\over \sq2}\mid\mi N/2).\nonumber
\end{eqnarray}
The above formula for overlaps of coherent states on the torus is compatible with the corresponding estimates derived in \cite[\S 8.2]{faure} and \cite[\S 1.4.2]{thesekader}. 
We deduce from this that the distribution law of the Husimi function at a given point $x=(q,p)$ is the exponential 
\begin{equation}
\label{width}
{1\over \tilde\sigma_x} \e^{-H/\tilde\sigma_x}dH,\qquad\makebox{with    }
\tilde\sigma_x=\theta_3(\pi Nq\mid\mi N/2)\;\theta_3(\pi Np\mid\mi N/2),
\end{equation}
and the limit $\tilde\sigma_x\sim 1$ holds exponentially uniformly on ${\Bbb T}^2$ as $N\to\infty$.
Notice that $\tilde\sigma_x$ is also the square-norm of the torus coherent state $|z\rangle_{N,0}$, up to the factor $N\e^{2\pi N z\bar z}$.
From this pointwise distribution, we can derive the following averages, leading to the typical values of several invariants as given in table \ref{tab},
\begin{eqnarray}
\Big\langle \int_{{\Bbb T}^2} H(x)^2\,dx\Big\rangle_N&=&2\,\theta_3(0\mid\mi N)^2\sim  2+8\e^{-\pi N}\\
\Big\langle \int_{{\Bbb T}^2} H(x)^s\,dx\int_{{\Bbb T}^2} H(y)^t\,dy\Big\rangle_N&\sim&\Gamma(1+s)\Gamma(1+t)\left\{ 1 + {1\over N} \int_0^\infty {1\over u} \left[_2F_1(-s,-t;1;u)-1\right] \ du\right\}\\
\label{sec}
\Big\langle \int_{{\Bbb T}^2} \log H(x)\,dx\Big\rangle_N&=&-\gamma+2\log \eta(\mi N/2)+\pi N/12\sim -\gamma-2\e^{-\pi N}\\
\Big\langle \int_{{\Bbb T}^2} H(x)\log H(x)\,dx\Big\rangle_N&=&1-\gamma +2\int_0^1 \theta_3(\pi q\mid\mi N/2)\log \theta_3(\pi q\mid\mi N/2)\,dq\sim  1-\gamma +2\e^{-\pi N}.\nonumber
\end{eqnarray}
($_2F_1(a,b;c;z)$ is the hypergeometric function \cite[vol. 1, chap. 2]{bateman}; the function $\eta(\tau)$ is Dedekind's modular form defined in appendix A; $\gamma$ is Euler's constant). The second identity allows to derive the typical deviations from average values for the $L^2$-norm, the geometric mean and the entropy.

The average sup-norm of a random state cannot be calculated by the same techniques as the other invariants. We can nevertheless obtain an upper bound for it (see Appendix C), as given in table \ref{tab}:
$$\Big\langle||H||_\infty\Big\rangle_N\leq 2\log N.$$
This result (which is the phase-space counterpart of similar estimates for sup-norms in the Schr\"odinger representation \cite{iwaniec}) actually concerns a constrained ensemble of random vectors, namely the set of complex vectors on the unit sphere ${\cal S}^N$ of ${\cal H}_{N,0}$, equipped with its standard measure: 
$$\rho(a_0,\ldots,a_{N-1})={1\over{\rm Vol}({\cal S}^N)} \delta(1-\sum_{j=0}^{N-1}|a_j|^2).$$
According to equation (\ref{canonical}), in the semiclassical regime this (microcanonical) ensemble must yield the same average values as the (canonical) Gaussian ensemble studied so far, so we expect the above upper bound to hold for the Gaussian case as well.

\subsubsection{Statistics of the constellations}

Alternatively, it is possible to extract statistical properties for the constellations of the random Bargmann (or Husimi) functions, as was done in \cite{BBL,hannay} for a spherical phase space. The methods used therein can be transposed to the torus. Indeed, Hannay \cite{hannay} shows that any joint probability function $\rho_k(z_1,z_2,\ldots,z_k)d^2z_1\ldots d^2z_k$ of the zeros of random Bargmann functions can be computed from a unique $2k\times 2k$ correlation matrix, whose entries are the functions $\Big\langle \psi(z_i)\overline{\psi(z_j)}\Big\rangle_N$ and their derivatives w.r.to $z_i$ or $\bar z_j$. Notice that both the Gaussian and microcanonical ensembles of random vectors {\sl exactly} yield the same statistics for the zeros, which are independent of the normalization. 

\medskip
{\bf 1-point function}

The first relevant quantity is the average density of zeros on the torus, defined as
$\rho_1(Z)=\Big\langle N^{-1}\sum_{j=0}^{N-1}\delta(z-Z_j)\Big\rangle_N$, where the $Z_j$ are the zeros (ordered arbitrarily) of a sample $\psi_{\{a\}}(z)$. The $\delta$ functions are two-dimensional, and periodicized w.r. to $T_{\Bbb C}$. These notations will apply as well to the pair correlation function defined below. Hannay's formula reads 
\begin{eqnarray}
\label{rho1}
 \rho_1(z)d^2z&=&{1\over N\pi}{\partial^2 \over\partial z\partial\bar z}
\log\Big\langle \psi(z)\overline{\psi(z)}\Big\rangle_N\; d^2z\\
\Longrightarrow \rho_1(q,p)&=&1+\pi N\sum_{m\geq 1}{(-1)^m m\over \sinh(\pi Nm/2)}[\cos(2\pi Nmq)+\cos(2\pi Nmp)]\\
 &\sim& 1-{\pi N\over \sinh(\pi N/2)}[\cos(2\pi N q)+\cos(2\pi N p)]\quad\mbox{ as }N\to\infty.\nonumber
\end{eqnarray}
Therefore, in the semiclassical limit, the zeros of random Bargmann functions are equidistributed over $T_{\Bbb C}$. The deviations from perfect equidistribution ($\rho_1(x)\equiv 1$) are uniformly exponentially small, and periodic w.r. to the square lattice of side $1/N$: this seems related to the sum rule (\ref{sum}), which prevents the distribution of the constellations from being completely translation-invariant.    

\medskip
{\bf Derivative of $\psi$ at the zeros}

The Husimi density is very oscillatory in the
semiclassical limit, due to the dense distribution of its zeros. In the vicinity of a zero $z_i$, it can be approximated by $H_{\psi}(z,\bar z)\approx |(z-z_i)\psi'(z_i)|^2\e^{-2\pi Nz_i\bar z_i}$. Therefore, the quantity
$f(z_i)\defi |\psi'(z_i)|^2\e^{-2\pi Nz_i\bar z_i}$ seems a reliable scalar to measure the local strength of $H_\psi$ around $z_i$ as built by all the other zeros, and it fluctuates much less than $H_{\psi}(z,\bar z)$. 

In our statistical framework, using the joint probability ${\cal D}(\psi_1,\psi'_1)$ of the Bargmann function and its derivative at a point $z_1$, we can derive the distribution law of $N^{-1}\sum_{i=1}^N f(z_i)$ for $\{z_i\}$ the zeros of $\psi$: in the semiclassical limit, this law happens to be the so-called ``1/2-Poisson law"
\begin{equation}
{f\over (2\pi N)^2}\ \e^{-{f/2\pi N}}df.
\end{equation}

\medskip
{\bf 2-point function}

The exact pair correlation function $\rho_2(z_1,z_2)=\Big\langle N^{-2}\sum_{i\neq j}\delta(z_1-Z_j)\delta(z_2-Z_j)\Big\rangle_N$ would already be too lengthy to write down, even in this model without symmetry. However, the formulae get much simpler in the semiclassical limit, i.e. by neglecting exponentially small corrections: we then recover an expression very similar to the one applying to the spherical phase space. Actually, we use the fact that the fundamental correlation function $\Big\langle \psi(z_1)\overline{\psi({z}_2)}\Big\rangle_N$ is given semiclassically by the kernel $\e^{2\pi Nz_1\bar{z}_2}$, uniformly when 
the separation $\delta z=z_1-z_2$ stays inside a compact square $|\delta q|\leq 1/2 -\epsilon$, $|\delta p|\leq 1/2 -\epsilon$ (in such a square, the theta functions in eq.(\ref{corr}) converge uniformly to 1). In this regime, the simplicity of this kernel yields for $\rho_2$ a formula both translation-invariant and isotropic, identical to the one found for the sphere \cite{hannay,BBL}:
\begin{eqnarray}
\label{rho2}
\rho_2(x_1,x_2)&\sim &  g(\sqrt{\pi N/2}\;|\delta x|)\qquad{\rm as }\quad N\to\infty,\\
{\rm with}\qquad g(r)&\defi &{(\sinh ^2 r^2+r^4)\cosh r^2 -2r^2\sinh r^2\over \sinh ^3 r^2}.
\end{eqnarray}
The function $g(r)$ is displayed on figure 6 (left). It starts quadratically near the origin, and converges to $1\sim\rho_1(x_1)\rho_1(x_2)$ as soon as $|x_1-x_2|\gg 1/\sqrt N$. The zeros (interpreted as interacting particles), repel each other at short distance, and become uncorrelated at distances larger than the mean spacing $1/\sqrt N$. 

In the case where $|\delta q|$ or $|\delta p|$ is near $1/2$, we cannot use the kernel $\e^{2\pi Nz_1\bar{z}_2}$ any more, but a direct estimation of the different terms contributing to $\rho_2$ leads to the large-$N$ uniform value $\rho_2(x_1,x_2)\sim 1$, as was expected from the shape of $g(r)$. The periodicity of $\rho_2$ w.r. to ${\Bbb T}^2$ then yields a uniform semiclassical approximation for it valid everywhere.

\medskip
{\bf Fourier coefficients}

In a series of articles \cite{leb:shukla,shukla,BBL}, formula (\ref{rho2}) was compared with numerical computations using eigenstates of quantum chaotic maps on the sphere or on the torus, and an excellent agreement was found. However, averaging over many eigenstates was necessary to recover the shape of the statistical result. Moreover, such a direct comparison yields information about the short-distance correlations between the zeros, whereas we are also interested in the global properties of the constellations. In fact, the semiclassical properties of eigenfunctions are certainly not given by the 
precise position of a given zero, but rather by interferences between a large
number of them. Consequently, as we already explained in section \ref{ergodicity}, we think that some relevant semiclassical information may lie in the Fourier coefficients of the density of zeros rather than in their individual positions (see section \ref{Fouri} below). 
We can actually estimate these Fourier coefficients in our statistical framework. 

The formula for the average density $\rho_1(x)$ shows that $\Big\langle \rho_{k }\Big\rangle_N=\delta_{k}$ up to exponentially small corrections.  

More information is contained in the second moments of the coefficients. We are thus led to the {\sl form factor} of the random constellations, defined for any $k \in {\Bbb Z}^2$ by
\begin{equation}
\Big\langle |\rho_{k }|^2\Big\rangle_N={1\over N}+\int_{{\Bbb T}^2}d^2x_1\,d^2x_2\,\e^{-2\mi\pi k .( x_1-x_2)}\rho_2(x_1,x_2),
\end{equation}
where $\rho_2$ is the pair correlation function corresponding to random states in ${\cal H}_{N,0}$. Using the asymptotic formula (\ref{rho2}) for $\rho_2$ and integrating over the angular variable, then integrating by parts, we obtain the following integral and series representations, valid as $N\to\infty$ up to exponentially small corrections. To this order, the isotropy of $\rho_2$ in (\ref{rho2}) implies the isotropy of the form factor:
\begin{eqnarray}
\Big\langle |\rho_{k }|^2\Big\rangle_N&\sim& \delta_{k} +N^{-1} F_2(|k|/\sqrt{N})\\
\label{F_2}
{\rm with}\qquad F_2(\kappa)&=&- 2\pi \kappa^2\int_0^\infty y\; J_2(\kappa\sqrt{8\pi y})\,(1-\coth y)\,dy\\
&=& {\pi^2 \kappa^4}\sum_{n\geq 0}{(-\pi)^n\zeta(n+3)\over n!}\kappa^{2n}
\label{series}
\end{eqnarray}
where $J_2$ is the Bessel function and $\zeta(n)=\sum_{j>0} j^{-n}$ the Riemann zeta function.
Rigorously, $\Big\langle |\rho_{k }|^2\Big\rangle_N$ is only defined for integer $k $, but the limit $N\to \infty$ involves the values of $F_2(\kappa)$ for $\kappa\in {\Bbb R}_+$. We were not able to find a simpler expression for the rescaled function $F_2$, which is shown on figure 6 (right). The curve $F_2(\kappa)$ looks vaguely similar to the curve $g(r)$ of figure 6 (left), but the two functions have different behaviors near the origin. Whereas $g(r)\sim r^2$ for small $r$, the series (\ref{series}) shows that for fixed $k \neq 0 $, 
\begin{equation}
\label{asymprho}
\sqrt{\Big\langle |\rho_{k }|^2\Big\rangle_N}\sim \pi\sqrt{\zeta(3)}\;|k|^2\;N^{-3/2} \mbox{   when   } N\to\infty.
\end{equation}
Therefore, the decay property stated in conjecture \ref{conj} is amply fulfilled on average for sequences of random states. 

Notice that we also obtained here the typical dependence of $|\rho_{k}|$ as a function of $k $ for fixed $N$, of which we had no idea from our deterministic point of view of section \ref{equidistribution}. 

The linear relation (\ref{hk}) in Fourier space between the density of zeros and  the logarithm of the Husimi density provides us with some statistical information about the latter:
\begin{equation}
\label{hkk}
\Big\langle |h_{k }|^2\Big\rangle_N\sim N^{-3}{F_2(\kappa)\over \pi^2\kappa^4}\quad\mbox{as}\quad N\to\infty.
\end{equation}
This scaling function is also plotted on figure 6 right (dot-dashed curve). In particular, if one
fixes $0\neq k\in{\Bbb Z}^2$, then 
\begin{equation}
\label{asymph}
N^3\;\Big\langle |h_{k }|^2\Big\rangle_N\to \zeta(3)\quad\mbox{as}\quad N\to\infty.
\end{equation}

\subsection{\bf Even-parity model}

The statistical model we have just studied had the advantage of translation invariance and isotropy, up to exponentially small terms in the semiclassical limit. However, the chaotic maps on the torus that we studied numerically have the parity symmetry, and their eigenstates are either even or odd, in both the Schr\"odinger and
Bargmann representations. This property then has to be incorporated in
the statistical model. We will only write down the statistical properties of random {\sl even} states, and to simplify the formulas we restrict ourselves to the case $N$ even. 

The correct ensemble is built in the 
following way: we take a random Bargmann function $\psi_{\{a\}}(z)$ from
the ensemble with no symmetry (\ref{psia}), and consider the symmetrized function 
\begin{eqnarray}
\psi^{\rm even}(z)&\defi & {\psi_{\{a\}}(z)+\psi_{\{a\}}(-z)\over \sq2}\\
&=&\sq2 a_ 0  \langle z|q_0\rangle_{N,0} +\sum_{j=1}^{N/2-1} {a_j+a_{N-j}\over\sq2}( \langle z|q_j\rangle_{N,0}+\langle z|q_{N-j}\rangle_{N,0})+\sq2 a_{N/2}\langle z|q_{N/2}\rangle_{N,0}\nonumber\\
&=&b_ 0 \langle z|q_0\rangle_{N,0}+b_{N/2}\langle z|q_{N/2}\rangle_{N,0}+\sum_{j=1}^{N/2-1} b_j \Big( \langle z|q_j\rangle_{N,0}+\langle z|q_{N-j}\rangle_{N,0}\Big),
\label{beven}
\end{eqnarray}
so the new random independent Gaussian complex variables $b_j$ have the variances
$\Big\langle |b_0|^2\Big\rangle_N=\Big\langle |b_{N/2}|^2\Big\rangle_N=2/N$, $\Big\langle |b_j|^2\Big\rangle_N=1/N$ for $j=1,\ldots,N/2-1$. For this ensemble too, all interesting quantities can be evaluated by use of the 
correlation function 
\begin{equation}
\label{cruce}
\Big\langle \psi(z_1)\overline{\psi({z}_2)}\Big\rangle_N^{\rm even}=\Big\langle \psi(z_1)\overline{\psi({z}_2)}\Big\rangle_N+\Big\langle \psi(z_1)\overline{\psi(-{z}_2)}\Big\rangle_N.
\end{equation}
Therefore, we have the uniform semiclassical approximation $\Big\langle \psi(z_1)\overline{\psi({z}_2)}\Big\rangle_N^{\rm even}\sim 2\cosh(2\pi N z_1 \bar z_2)$  as long as both $(x_1-x_2)$ and $(x_1+x_2)$ stay inside a square of side $1-\epsilon$ centered on the origin (in this square, this formula holds up to exponentially small corrections). 

The distribution laws of the Bargmann and Husimi functions are slightly changed
from the case with no symmetry. The modification concerns the widths of the Gaussian (resp. exponential) distributions:
\begin{eqnarray}
\sigma_{z}^{\rm even}&=&\Big\langle \psi(z)\overline{\psi({z})}\Big\rangle_N^{\rm even}\\
\Longrightarrow\tilde\sigma_x^{\rm even}&=&\tilde\sigma_x+\e^{-2\pi N(q^2+p^2)}\,\theta_3(\mi\pi Nq\mid\mi N/2)\theta_3(\mi\pi Np\mid\mi N/2).
\label{sigmae}	
\end{eqnarray}
The second term implies that the random even Husimi function has a `bump' of height $2$ and width $\sim 1/\sqrt{N}$ at the four P-invariant points $(0,0)$, $(1/2,0)$, $(0,1/2)$, $(1/2,1/2)$ (i.e. the points where $x=-x$ modulo ${\Bbb T}^2$); at a distance $\gg 1/\sqrt{N}$ from these special points, this term is small, and $\sigma_x^{\rm even}\sim 1$. A consequence of these `bumps' is that the average square-norm of $\psi$ is now $N$-dependent: $\Big\langle ||\psi||^2\Big\rangle_N^{\rm even}=1+2/N$. The average values of the other functionals similarly differ by terms of order $1/N$ from their values without parity.

\medskip
{\bf 1-point function}

The average density of zeros is modified by parity in a manner dual to the `bumps' of the Husimi density, i.e. it decreases in the vicinity of the symmetry points.  Indeed, when $|q|\leq 1/4-\epsilon, |p|\leq 1/4-\epsilon$ and $N$ is large, the approximation $\Big\langle |\psi(z)|^2\Big\rangle_N\sim 2\cosh(2\pi N |z|^2)$ in equation (\ref{rho1}) yields:
\begin{equation}
\label{rho1e}
\rho_1^{\rm even}(x)\sim \tanh(\pi N x^2) + {\pi N x^2\over\cosh^2(\pi N x^2)}
\end{equation}
(we show $\rho_1^{\rm even}(x)$ for small $x$ in figure 7 left, solid curve).
The form of $\rho_1^{\rm even}(x)$ in the remaining part of ${\Bbb T}^2$ can be obtained through its periodicity: $\rho_1^{\rm even}(q,p)=\rho_1^{\rm even}(q+1/2,p) =\rho_1^{\rm even}(q,p+1/2)$. The decrease of the density near the symmetry points can be explained as the repulsion between the zeros $z$ and $-z$ which come close to each other when they approach the symmetry points. 

\medskip
{\bf 2-point function}

The pair correlation function, now defined as $\rho_2^{\rm even}(z_1,z_2)=\Big\langle N^{-2}\sum_{Z_i\neq \pm Z_j}\delta(z_1-Z_i)\delta(z_2-Z_j)\Big\rangle_N$, can be calculated by the same techniques as in \cite{hannay}, using the asymptotic correlation function $\Big\langle \psi(z_1)\overline{\psi(z_2)}\Big\rangle_N^{\rm even}\sim 2\cosh(2\pi N z_1\bar z_2)$. The exact asymptotic result cannot be written down in a concise way (it depends explicitly on both $z_1$ and $z_2$). However, if $x_1$, $x_2$ are not in the vicinity of the {\sl same} symmetry point $x_{\rm sym}$ (i.e. we do not have simultaneously $|x_1-x_{\rm sym}|=O(1/\sqn)$ and $|x_2-x_{\rm sym}|=O(1/\sqn)$), the pair correlation function is approximatively
\begin{equation}
\label{approx}
\rho_2^{\rm even}(x_1,x_2)\sim \rho_1^{\rm even}(x_1)\rho_1^{\rm even}(x_2)\rho_2(x_1,x_2)\rho_2(x_1,-x_2)
\end{equation}
where $\rho_2$ is the 2-point function for random states with no parity.
  
\medskip
{\bf Fourier coefficients}

Due to the inhomogeneity of $\rho_1$, the averages of the Fourier coefficients are no longer exponentially small in $N$, as was the case with no symmetry. The periodicity of $\rho_1^{\rm even}$ entails that $\Big\langle \rho_{k }\rangle_N^{\rm even}=0$ unless $k_q$ and $k_p$ are both even. In that case, we have
\begin{eqnarray}
\label{F1}
\Big\langle\rho_{k }\Big\rangle_N^{\rm even}&=& \delta_{k}+{1\over N} F_1(|k|/\sqrt{N})\\
\mbox{ with}\quad F_1(\kappa)&=&4\kappa\int_0^\infty\sqrt{\pi y}J_1(2\kappa\sqrt{\pi y})\,(\tanh(y)-1)\,dy\nonumber
\end{eqnarray}
This provides us with the following asymptotic behaviors of $\Big\langle\rho_{k }\Big\rangle_N^{\rm even}$ (see fig. 8, dashed curve):
\begin{eqnarray}
\Big\langle\rho_{k }\Big\rangle_N^{\rm even}&=& \delta_{k}-{\pi^3\over 6}{k^2\over N^2}+O(k^4/N^3)\quad\mbox{for small}\quad k/\sqn,\\
\Big\langle\rho_{k }\Big\rangle_N^{\rm even}&=&O(N^{-1})\quad\mbox{as}\quad{k\over\sqn}\to\infty.
\end{eqnarray}
The form factor is now given by
\begin{equation}
\Big\langle |\rho_{k }|^2\Big\rangle_N^{\rm even}={1\over N}+{1\over N}\Big\langle\rho_{2k}\Big\rangle_N^{\rm even}+\int_{{\Bbb T}^2}d^2x_1\,d^2x_2\,\e^{-2\mi\pi k .( x_1-x_2)}\rho_2^{\rm even}(x_1,x_2),
\end{equation}
A simple Riemann--Lebesgue argument for the second and last terms in this formula implies that that $\Big\langle |\rho_{k }|^2\Big\rangle_N  ^{\rm even}$ converges to $1/N$ for large $(k/\sqn)$, as was the case in the former section. On the other hand, using the approximation (\ref{approx}) away from the symmetry points, we obtain the following expression in terms of $k/\sqn$:
\begin{equation}
\label{formeven}
\Big\langle |\rho_{k }|^2\Big\rangle_N^{\rm even}=\delta_{k}+{1\over N}F_2(|k|/\sqn)+{1\over N^2}F_{\rm corr}(k/\sqn)+O(1/N^3),
\end{equation}
where $F_2(\kappa)$ is the function given in equation (\ref{series}). $F_{\rm corr}$ is not known analytically; it is bounded and its Taylor series starts by a term of the form  $\alpha k^2/N$ with $\alpha\geq 0$, but we do not know if higher terms are isotropic. 

Nevertheless, equation (\ref{formeven}) shows that the dominant shape of the form factor as a function of $k/\sqn$ is unchanged from the model with no parity: we can therefore compare this analytical expression to the data for the cat map $S'$ (see section \ref{Fouri} and fig. 9a). Besides, for this model we see that $\Big(\Big\langle\rho_{k }\Big\rangle_N^{\rm even}\Big)^2$ is of higher order in $N^{-1}$ than $\Big\langle |\rho_{k }|^2\Big\rangle_N^{\rm even}$, as functions of $\kappa$, so that $\Big\langle\rho_{k }\Big\rangle_N^{\rm even}$ might be difficult to detect (\ref{F1}) in the numerical data for large $N$.  

On the other hand, if one fixes $k$ and studies the $N$-dependence of the coefficient $\rho_k$ (resp. $h_k$), the large-$N$ asymptotics are slightly different from the asymmetric model (eq. (\ref{asymprho}) and (\ref{asymph})):
\begin{eqnarray}
\Big\langle |\rho_{k }|^2\Big\rangle_N^{\rm even}&\sim& {1\over N^3}( \pi^2 \zeta(3)k^4 + \alpha k^2),\\
\Big\langle |h_{k }|^2\Big\rangle_N^{\rm even}&\sim &{1\over N^3}(\zeta(3)+{\alpha\over\pi^2 k^2}).\nonumber
\end{eqnarray}
We therefore recover the $N^{-3/2}$ behavior, but with different prefactors (unless $\alpha$ vanishes, of course). 

\subsection{Real even-parity model}
\label{real}
In order to fit the numerical data for the simple cat map $S=\left(\begin{array}{cc}2&1\\3&2\end{array}\right)$ or the baker's map, we study a statistical ensemble of
real even states (once more, we restrict ourselves to even values of $N$). This ensemble is still defined by the formula (\ref{beven}), but the $b_j$ are now random Gaussian independent {\sl real} variables, with the same variances as in the complex case. Models of random real {\sl polynomials} (without parity symmetry) were studied in detail in the literature \cite{edelman, BBL, prosen, mezin}.

Due to reality, the distribution laws of the Bargmann function or its derivative do not only depend on the modulus $|\psi|$, but also on its phase. However, these
distributions are still Gaussian in the variables $\Re\psi$, $\Im\psi$, $\Re\psi'$, $\Im\psi'$ \cite{prosen}. On the other hand, the distributions of the Husimi function and the averages of its invariant functionals are unchanged from the ensemble of complex even random states (former section).

\smallskip

By contrast, the reality of the $b_j$ modifies drastically the statistics of the zeros. Since $\psi(z)$ is real and even, its zeros come either in quadruplets of complex numbers $\{z,-z,\bar z, -\bar z\}$, or in couples $\{z_r,-z_r\}$ situated on one of the four symmetry axes $\{\Im z=0\}$, $\{\Im z=1/2\sq2\}$, $\{\Re z=0\}$, $\{\Re z=1/2\sq2\}$. The statistics of these two types of
zeros are quite different, but both are still obtained from the correlation functions $\Big\langle \psi(z_1)\overline{\psi(z_2)}\Big\rangle_N$ and $\Big\langle \psi(z_1)\psi(z_2)\Big\rangle_N=\Big\langle \psi(z_1)\overline{\psi(\bar z_2)}\Big\rangle_N$. Since the $b_j$ have the same variances as in the complex case, these functions are still given by equations (\ref{cruce},\ref{corr}). 

We use the formalism of \cite{prosen} to derive the average density $\rho_1^{\rm cmplx}$ of {\sl complex} zeros: for $|q|\leq 1/4-\epsilon$, $|p|\leq 1/4-\epsilon$, we obtain semiclassically
\begin{eqnarray}
\rho_1^{\rm cmplx}(x)&\sim&G_1(\sqrt{\pi N}q,\sqrt{\pi N}p), \quad\mbox{where   }\\
G_1(Q,P)&=&{1\over d_1(Q,P)^{3/2}}\left\{d_1(Q,P)\sinh(Q^2+P^2) +2\sinh(Q^2-P^2)[Q^2\sinh 2P^2 -P^2\sinh 2Q^2]\right.\nonumber\\
&&\left. -2QP\sinh(Q^2+P^2)\sin(4QP)\right\}\nonumber\\
\mbox{with   }d_1(Q,P)&=&\sinh 2Q^2 \sinh 2P^2 +\sin^2 (2QP).\nonumber
\end{eqnarray}
Figure 7 (right) shows a contour plot of $G_1(Q,P)$ near the origin, in the first quadrant. 
Away from the symmetry points, this formula yields back the expression $G_{\rm RP}$ corresponding to complex zeros of real random polynomials: for instance, near the real axis but for $q$ far from ${1\over 2}\Bbb Z$, we have
\begin{equation}
\label{RP}
G_1(Q,P)\sim G_{\rm RP}(P)={1-(1+4P^2)\e^{-4P^2}\over (1-\e^{-4P^2})^{3/2}}
\end{equation}
(the function $G_{\rm RP}$ is shown in fig. 7 left, dashed curve). Away from the symmetry axes (i.e. at a distance $\gg 1/\sqn$), we recover a uniform density $\rho_1^{\rm cmplx}(x)\sim1$. We get the density on the whole torus using the same periodicity properties as in the complex even case. 

A separate treatment has to be made for zeros on the symmetry axes \cite{BBL,edelman, mezin}. It yields a singular density on these axes: \begin{eqnarray}
\label{rho1sing}
\rho_1^{\rm sing}(x)&=&{1\over\sqrt{\pi N}}\left\{\delta(p)+\delta(p-1/2)\right\}G_1^{\rm sing}(\sqrt{\pi N}q)+(q\leftrightarrow p),\\
\mbox{with}\quad G_1^{\rm sing}(Q)&\defi&\left(\tanh(Q^2) + {Q^2\over\cosh^2(Q^2)}\right)^{1/2}
\end{eqnarray}
(see fig. 7 left, dotted curve).
Among the $N$ zeros of $\psi(z)$, the proportion situated on the symmetry axes is asymptotically $4/\sqrt{\pi N}$, which corresponds exactly to the difference $1-\int_{{\Bbb T}^2}\rho_1^{\rm cmplx}(x)dx$. 

The description of the pair correlation function is too involved to be presented
here. Far from the symmetry points, it reproduces the results of \cite{prosen}, where it was shown that the function $\rho_2^{\rm cmplx}(x_1,x_2)$ takes the isotropic form (\ref{rho2}) when $x_1$ and $x_2$ are far from the symmetry axes, whereas the repulsion takes a different shape near the axes. As for the zeros on the symmetry axes, they repel each other as in the case of real random polynomials \cite[eq. (5.35)]{bleher}, as long as they stay away from $x_{\rm sym}$. 

We did not compute explicitly the form factor for this ensemble, but we noticed
qualitative modifications for the {\sl averages} of the Fourier coefficients, due to
the strong anisotropy of $\rho_1$, and its singular part. Writing $\Big\langle \rho_{k}\Big\rangle_N$ as a function of $\kappa=k/\sqn$, the dominant contribution is not isotropic; 
in case both $k_q$ and $k_p$ are even, we get the formula:
\begin{equation}
\label{rhokreal}
\Big\langle \rho_{k}\Big\rangle_N^{\rm real}\sim \delta_{k} +{2\over\sqrt{\pi N}}\{\delta_{k_q}F_{\rm RP}(\kappa_p)+\delta_{k_p}F_{\rm RP}(\kappa_q)\}+{1\over\pi N} F_{\rm point}(\kappa).
\end{equation}
where $F_{\rm RP}$ is the Fourier transform of $G_{\rm RP}(Q)+\delta(Q)$; $F_{\rm point}$ is unknown analytically, it gives the corrections due to the symmetry points.
The term in $1/\sqn$ is clearly anisotropic. We could not obtain a closed formula for $F_{\rm RP}$, but only the following limits (see fig. 8, solid curve):
\begin{eqnarray}
F_{\rm RP}(\kappa)=O(\kappa^2) &\mbox{as}&\kappa\to 0\\
F_{\rm RP}(\kappa)\sim 1-{1\over 2\pi\kappa^2}&\mbox{as}&\kappa\to\infty.
\end{eqnarray}
Notice that the dominant term of $\Big(\Big\langle \rho_{k}\Big\rangle_N^{\rm real}\Big)^2$ is of the same order $1/N$ as
the form factor derived for the complex model
(\ref{formeven}): although we did not compute the form factor in the real model, we nonetheless expect the anisotropy to be visible in numerical data, especially for $\kappa\grsim 1$. More precisely, it seems reasonable to conjecture that
the {\sl variance} $\Delta\rho_k^2=\Big\langle |\rho_{k}|^2\Big\rangle_N^{\rm real}-\Big(\Big\langle \rho_{k}\Big\rangle_N^{\rm real}\Big)^2$ is not very different from its value in the complex model, in particular it should be almost isotropic. This assumption provides the
following tentative formula for the form factor:
\begin{eqnarray}
\label{conjff}
\Big\langle |\rho_{k}|^2\Big\rangle_N^{\rm real}&\approx& \Big(\Big\langle \rho_{k}\Big\rangle_N^{\rm real}\Big)^2+\Big\langle |\rho_{k}|^2\Big\rangle_N^{\rm even}\\
&\approx& {1\over N}\left({4\over\pi}\left\{\delta_{k_q}F_{\rm RP}^2(\kappa_p)+\delta_{k_p}F_{\rm RP}^2(\kappa_q)\right\}+F_2(|\kappa|)\right)
\end{eqnarray}
In the next section, we will use this approximate formula (see fig. 6 right, dashed curve) to probe the anisotropy of the averaged Fourier coefficients of eigenstates for the cat map $S$ and the baker's map. The fixed-$k$ asymptotics for $|\rho_k|$ given by (\ref{conjff}) is still $\propto N^{-3/2}$:
\begin{equation}
\Big\langle |\rho_{k}|^2\Big\rangle_N^{\rm real}\approx{1\over N^3}\left\{
\pi^2\zeta(3)\,k^4+ {4\beta\over\pi}(\delta_{k_q}k_p^2 +\delta_{k_p}k_q^2)\right\},
\end{equation}
using the asymptotics $F_{\rm RP}(\kappa)\sim \beta\kappa^2$ as $\kappa\to 0$. (The corresponding equation for $h_k$ is easy to derive).

\section{Results and questions about the Fourier coefficients}
\label{Fouri}

We now return in greater detail to the Fourier coefficients of
chaotic eigenconstellations.

The Fourier transform of a density of $N$ points on the torus was defined
in eq. (\ref{four}) as
\begin{equation}
\label{four2}
\rho_k= \frac{1}{N}\sum_{j=1}^N \e^{-2\pi\mi k . x_j}, \qquad k \in \Bbb Z^2 ,
\end{equation}
$\rho_0 \equiv 1$ entirely corresponds to the normalization, but in all other
respects it must be understood that the restriction $k \ne 0$ expressly applies. 
The Fourier coefficients $\rho_k$ have already appeared several times above:

1) the equidistribution of zeros in the phase space was most simply
expressed as the property that each $\rho_k$ {\sl individually\/} tends to zero
as $N \to +\infty$; we further argued that the Schnirelman property could 
amount to a stronger statement about their rate of decrease, 
as being $o(N^{-1})$ (our conjecture \ref{conj}).

2) The simple toy constellation of Sec.\ref{counter} carried a modulation
essentially produced by the lowest Fourier coefficients ($|k|=1$); 
this largest possible wavelength was nevertheless able to excite 
an extremely localized `artificial scar';
this illustrates the idea that semiclassical features of Husimi densities 
are likely to be strongly coupled with collective (rather than individual)
degrees of freedom of the zeros, of which Fourier coefficients are an example.

3) Statistical ensembles of zeros make predictions 
about the moments of the Fourier coefficients; especially, the model without symmetry predicts the quadratic average 
$\Bigl\langle | \rho_k |^2 \Bigr\rangle_N$
to be an isotropic function, universal up to the axis scales,
explicitly describable for small or large $k$; the large-$N$ behavior
$\left[\Bigl\langle |\rho_k|^2 \Bigr\rangle_N\right]^{1/2} \sim
\pi\sqrt{\zeta(3)}\;|k|^2N^{-3/2}$ followed.

4) Eq.(\ref{Poiss}), relating the logarithm of the Husimi function
to the density of zeros $\rho(x)$, was made diagonal 
in the Fourier space by eq.(\ref{hk}), as
\begin{equation}
\label{diag}
h_k =-\frac{\rho_k}{\pi |k|^2}\ ({\rm except:\ } h_0 \equiv 0)\quad{\rm where\ }
\tilde h (x) = \sum_{k \in \Bbb Z^2} h_k \e^{2\pi\mi k . x}.
\end{equation}

We will now try to build a (still partial) Fourier picture of chaotic
eigenconstellations incorporating the previous remarks.
We have just listed several arguments supporting this Fourier approach, 
but difficulties also arise which can however be
worded as interesting problems.

a) lack of physical cogency: any (semi)classical meaning of the Fourier space 
and of the action of classical maps upon it
are presently quite unclear --- save for linear (cat) maps ---, and so are
the quantum and semiclassical dynamics in this Fourier picture; 

b) besides eq. (\ref{diag}), another piece of the link from $\rho(x)$ 
to the Husimi density is the formula $H(x)=c \e^{N \tilde h (x)}$
(with $c \equiv 1/{\rm GM}[H]$), and now this relation is local in the phase space, and also nonlinear; 
its Fourier space translation is then quite intractable, meaning that 
we cannot easily transform the Fourier coefficients of $\log H$ 
(the primary objects of interest, in our approach) 
into those of the Husimi density itself
(which would also relate simply to the Fourier coefficients of the Wigner function);

c) a basic well known difficulty with Fourier transformation is that 
it obliterates the very special nature (positivity and finite parametrization)
of this density $\rho$. The zeros being noteworthy for their 1--1 correspondence
with pure states, it is specially desirable to fight the redundancy now
reinstated by going to a countable set of Fourier coefficients, and this
point is discussed next.

\subsection{Essential Fourier coefficients}
\label{essential}
Problems: can one identify a minimal finite set (of size $O(N)$) 
of Fourier coefficients $\rho_k$ that will optimally encode, 
and robustly restore,
the coordinates of the $N$ zeros in any configuration? 
Such Fourier coefficients (dubbed `essential'), which are in finite number, 
will then determine all the others, in a way which can also be asked.

In one dimension this is a classic problem: the Fourier transformation 
(\ref{four2}) on a 1-d torus, setting $z_j=\e^{-2\pi\mi x_j}$, becomes
$\{z_j\} \mapsto \{\rho_k= \sum_{j=1}^N z_j^k \}$, 
the celebrated mapping from sets of $N$ points to their symmetric functions.
This mapping is known to be algebraically invertible 
using the first $N$ symmetric functions as data; 
the solution relies on certain cumulant expansions,
which also happen to play a role in the semiclassical analysis of determinants 
of quantum eigenvalues.
It is thus interesting to encounter a 2-d analog of this problem in connection
with the corresponding analysis for eigenfunctions.

This problem seems much harder than its 1-d progenitor,
and we do not know if it has been answered.
This difficulty is seen from elementary geometry (area estimates).
The issue is to preserve and recover an amount of information
consisting of $2N$ independent real numbers 
(initially: the coordinates of the zeros, essentially arbitrary).
To optimize isotropy and robustness, we will seek this information only 
in {\sl radially\/} truncated Fourier transforms: 
$\{\rho_k\}_{k\in D_K}$, $D_K=$ the disk $\{|k| \le K\}$;
the symmetry relation $\rho_{-k} \equiv \bar\rho_k \ \forall k$ 
(expressing the reality of $\rho(x)$)
then makes each complex Fourier coefficient in the set count as real, 
while their total number can be estimated by the disk area $\pi K^2$.
All in all, lossless encoding then requires an area of at least $2N$.
It is also hard to conceive how the lowest $2N$ Fourier coefficients 
could generically have `hidden' dependence relations, 
besides any obvious symmetries. 
So we expect the Fourier data to be sufficient starting from the disk radius 
$ K _< = \sqn\, \kappa_<$ with $\kappa_< \defi \sqrt{2/\pi}\approx 0.7979$.

But let us now examine the special family of all square lattices of zeros 
(possibly rotated) for any $N$. 
The Fourier transform of such a lattice is a Dirac delta distribution 
on the dual square $k$-lattice which has generators of length $\sqn$,
so that all its Fourier coefficients vanish identically within any disk of area $<\pi N$,
although this area is $\pi /2$ times the heuristic estimate above.
The same reasoning using the optimizing lattice geometries in the sense of 
Sec.\ref{comparison} 
(lattices becoming equilateral triangular in the limit $N\to\infty$)
further dilates this disk of indetermination up to the area 
$2\pi N/\sqrt 3$.
Within any lesser radius, then, 
some patterns will be totally undetectable and hence undeterminable 
(e.g., almost equilateral-triangular lattices will be undistinguishable 
from any of their translated or rotated images).
Intuitively we also believe this is the worst possible case 
for the present argument.
We thus expect the Fourier data to be unambiguous only from the disk radius 
$ K _> = \sqn\, \kappa_>$ with $\kappa_> \defi \sqrt{2/\sqrt 3}\approx 1.0746$.

So, contrary to the 1-d case where the two radii are reached simultaneously at $N$,
here there is a gap of a factor $\sqrt{\pi/\sqrt 3} \approx 1.3468$ from 
an inner radius $K_<$ from which we naively expect the information about the
zeros to be generically preserved,
to an outer radius $K_>$ needed for the actual recovery of the zeros 
in some cases, and hence for their robust reconstruction in all cases. 
In-between there seems to lie a blurred zone in which one could perhaps invert 
more and more robustly by adding redundancy gradually.
This suggests that there may not exist a unique 'best' Fourier inversion algorithm feeding on finite Fourier data, in contrast to the 1-d case.

It is also interesting to consider this restoration issue
for random data (complex-valued and without any symmetry, for simplicity).
The relevant object is then the form factor $F_2(\kappa)$ with 
$\kappa \defi |k|/\sqn$ shown on fig. 6 right.
For $\kappa \geq \kappa_>$ this form factor is virtually undistinguishable 
from being Poissonian; a {\sl gradual\/} transition sets in around the annulus
$\{\kappa_< < \kappa < \kappa_>\}$,
and below $\kappa_<$ there lies a {\sl basin\/} 
of increasingly damped expectation values as $\kappa \to 0$: 
this feature makes any individual fluctuations much more significant
against the statistical noise in this region.
But precisely, the essential Fourier coefficients all lie within this basin, hence they are the ones statistically enhanced (especially far inside).
By contrast, the outer Fourier coefficients, which look completely like noise
(Poissonian) and also dominate in average size, are devoid of primary meaning
since they ought to be (complicated) functions of the lower ones.
Moreover, the Husimi density itself is controlled, via $\tilde h(x)$,
by the coefficients $h_k$ of eq. (\ref{diag}) which have the form factor
$\pi^{-2} F_2(\kappa)/\kappa^4$ of eq.(\ref{hkk}), also shown on fig. 6 right. 
This form factor now enhances 
the coefficients in the essential region {\sl numerically\/}, meaning that 
a resummation truncated within this region must give a fairly accurate picture
of the function $\tilde h(x)$ 
on a scale where individual zeros cannot be separated
(as will be validated numerically below).

\subsection{Fourier coefficients for fixed $N$}

\subsubsection{Averaged behavior}
We have computed the behavior as a function of $k$ of Fourier coefficients 
$\rho_k$ quadratically averaged over the whole basis of eigenfunctions
at fixed $N$, for a few classically chaotic models. (All of these have 
the parity symmetry, which makes the $\rho_k$ purely real.)
 
We first consider a dynamical model without anti-unitary symmetry: 
the cat map $S'$, because the corresponding statistical form factor is a well
controllable isotropic function in Fourier space.
(This actually holds for the statistical form factor without parity symmetry,
the only one we can fully compute, but parity-induced deviations must be
small for large $N$ and are indeed not visible on our plots.)
Fig.9a shows the mean values of $|\rho_k|^2$ and $|h_k|^2$
along selected directions in Fourier space for this map quantized with $N=149$,
restricting to even-parity eigenfunctions for definiteness. 
We did not compute a theoretical dispersion curve (involving 4-point
correlation functions), but we note that the dispersions stay
practically uniformly within 25\% of the theoretical mean value curve, 
however low the latter becomes. 
We conclude that qualitatively speaking,
the distribution of averages very closely follows a random-like behavior,
in a perfectly isotropic fashion. 

The results of the same calculations are now shown (fig.9b) for the cat map $S$
and the baker's map, which both possess antiunitary symmetries. Random theory
predicts a violation of isotropy already for 
$\Bigl\langle \rho_k \Bigr\rangle_N$
at large $k$ along the directions dual to the symmetry lines, 
i.e., the two axes for the cat map
vs the bisecting diagonals for the baker's map. 
Indeed, the averages $\Bigl\langle |\rho_k|^2 \Bigr\rangle_N$ are
selectively larger in the specified singular directions.
Isotropy improves as $k \to 0$, but the behavior stays systematically above 
the curve without symmetry. As was the case with the invariants, 
dispersions are noticeably larger for the baker than for the cat map.
Still, the gross features remain qualitatively random-like as before.

\subsubsection{Individual behavior}

Firstly, anticipating Sec.\ref{fourn} (fig.13), we note that 
within the set of individual eigenstates of a fixed-$N$ quantum map,
each given Fourier coefficient viewed in isolation will have 
large relative fluctuations, like the invariants we studied before (fig.3).

We now consider the {\sl collective\/} behavior of Fourier coefficients 
within {\sl individual\/} eigenconstellations, by looking at fully 2-d Fourier plots of 
some characteristic states, earlier portrayed by Husimi plots (figs.1, 4).
Although $\rho_k$ and $h_k$ are interchangeable as Fourier representations,
their plots stress quite disjoint features of the solutions.

The striking fact about the $\rho_k$-plots (fig.10 top) is that,
notwithstanding the large individual coefficient fluctuations,
these plots neatly display
{\sl a central region where most Fourier coefficients are heavily depressed\/}, 
very much like the basin of the random model.
This region appears for every single chaotic eigenstate we have examined
in this manner, and its radius remains comparable to $\sqn$.
The presence of this basin reflects, in the Fourier space, 
both the repulsion between the zeros and their equidistribution.
We stress that this is now a {\sl random-like feature generically embodied in individual chaotic eigenconstellations\/}
(still on empirical grounds), and not in integrable ones.

As regards the $h_k$-plots (fig.10 bottom), we argue that they can reveal
localized peaks of the Husimi densities such as scars. 
A measure for the presence of a point scar (taken at the origin,
up to a trivial translation) can be given by an averaged value of
$\tilde h(q,p)$ about that point: 
a relatively large and positive value will signal a scar. 
On the other hand, a convenient average over the correct width $O(N^{-1/2})$ 
is supplied by the partial Fourier sum $\tilde h_K \defi \sum_{|k|\le K} h_k$ 
(truncated in the disk $D_K$) precisely when $K \approx \sqn$: i.e.,
this average is basically controlled by the essential Fourier coefficients.
We can see indeed (fig.11) that as a function of the cutoff radius $K$, 
$\tilde h_K$ gradually drifts towards its limiting value $=\tilde h(0,0)$
while $K$ grows within the basin, and basically stabilizes further out 
(unless a zero happens to accidentally lie very near the origin: then 
$\tilde h(0,0)$ is very large and negative, 
but at the same time it is not a good estimator for the desired average).
Moreover, the correct order of magnitude is often approached 
{\sl well inside the basin\/},
i.e. the lower Fourier coefficients dominate the sum (as predicted by the shape of the
form factor (\ref{hkk}) in the random model). 
In conclusion, a scar or high density peak, taken at (or as) the origin,
is betrayed by
essential Fourier coefficients $h_k$ which, {\sl on a global trend\/}, 
fluctuate away from their averages $ \Bigl\langle h_k \Bigr\rangle_N$
{\sl in the positive direction, especially for low\/} $k$.
(The argument holds for $\Re(h_k)$  when the coefficients are complex.)

Fig. 12 illustrates, now over the whole phase space, three stages of 
the radially truncated Fourier resummation for the scarred $N=128$
baker eigendensity of fig. 1b (center), using the respective
cutoff values $K=2$, $K=9\ (\approx K_<)$, $K=12\ (\approx K_>)$. 
The main scar already begins to emerge for $K=2\ (\approx 0.18\,\sqn)$ which, 
upon desymmetrization, leaves only 4 (real) nonzero Fourier coefficients!
At the other end we may consider the Husimi density as recovered 
(in the sense of measures), reasonably well for $K \approx K_<$ and quite well
for $K=12 \approx K_>$.
By contrast, the individual zeros and the associated factorized structure 
of the Husimi density are washed away by the truncations,
since the Fourier picture is dual (complementary) to the stellar representation.
The next challenge is then to find some effective procedure 
to achieve the {\sl eigenvector\/} reconstruction itself from these data,
i.e., to unravel the zeros which will generate this function $\tilde h(x)$
already recovered on some coarse scale $O(1/\sqn)$.
The shape of the form factor and the example of the lattice states suggest that
Fourier coefficients in the transition annulus $\{K_< \losim |k| \losim K_>\}$ 
may become critical for this purpose, 
all the more so when the zeros' repulsion is stronger and their pattern more rigid,
but this example is special in that its duality is explicitly implementable 
(by the Poisson summation formula); 
for more disordered cases, no similar analytical handle is available 
and the restoration of the individual zeros remains an open problem.

One conclusion of this discussion is that Fourier coefficients of constellations
can provide an imperfect but still effective way of reducing the
number of parameters involved in the description of chaotic eigendensities:
the number of essential Fourier coefficients is roughly comparable to the
number of degrees of freedom of the zeros, but in practice their
contribution to the Husimi density often decreases gradually with $k$ 
to become negligible when or before the basin boundary is reached.

\subsection{Fixed Fourier coefficients for variable $N$}
\label{fourn}

We may now return to the issue of semiclassical ergodicity as expressed
by the decay rates of fixed-$k$ coefficients, say $h_k=-\rho_k/(\pi |k|^2)$,
as $N \to \infty$. We present numerical data concerning the lower such
coefficients $h_{1,0}$ and $h_{1,1}$ in graphic displays 
similar to the earlier plots of invariants (fig. 3).
Fig.13a shows the values of those coefficients for even-parity 
eigenconstellations of the cat map $S$, in both log--linear and log--log plots,
while fig.13b shows $h_{1,0}$ only for the baker's map.
(Other low Fourier coefficients, not shown, behave likewise, whereas
high ($k \gg \sqn$) Fourier coefficients confirm the behavior
$\propto N^{-1/2}$ predicted by eqs. (\ref{conjff}) and fig.8.)

The general comments made about the invariants in Sec. \ref{data} carry over
to these scalar quantities. Here, however, the conjectures linked with
semiclassical ergodicity can be probed more sensitively through the decay rates
of $|h_k|$ for each fixed $k$ and $N \to \infty$: one random model predicted
$\left[\Bigl\langle |h_k|^2 \Bigl\rangle_N \right]^{1/2} \sim \sqrt{\zeta(3)} N^{-3/2}$,
whereas we conjectured the Schnirelman property to hold as long as
$h_k=o(N^{-1})$.
We can now fit various subsets of the above data more or less reliably 
with behaviors of the form $c N^\beta$ for $|h_k|$ 
and estimate the resulting values of $\beta$ (the slopes on the log--log plots).

Cat map $S$: we find that $\beta$ ranges between $-1.4$ and $-1.5$ for the 
(quadratically) averaged coefficients;
as for the uppermost coefficients at each $N$, they decrease with slopes 
reasonably steeper than $-1.1$.

Baker's map: $\beta \approx -1.1$ (resp. $\losim -1.33$)
for the (quadratically) averaged coefficients of states whose parity is even
(resp. odd, not shown);
the uppermost coefficients at each $N$ give overall slopes $\approx -1.$
(resp. $\losim -1.2$) for even- (resp. odd-) parity states:
{\sl the even case seems to be on the borderline of conceivable exceptions
to the Schnirelman property\/}.

The plots also highlight the values corresponding to the eigenstates 
maximizing $H(0,0)$ for each $N$ 
(i.e., those with the strongest scars at the origin).
Then the left-hand-side plots confirm the globally {\sl upward\/} deviation 
of each Fourier coefficient of such a state from the linear average. 
The $N$-dependences are erratic
except, remarkably, for the baker's map restricted to powers of 2 where they
become extremely regular, giving slopes $\approx -1.4$. 
The Fourier coefficients for this, or other suitable, subset(s) of $N$ 
are dynamical quantities which could then perhaps be described asymptotically.

\section{Conclusion}

The results of this article are mainly geometrical and statistical. Indeed, the only role played by the ergodic nature of classical dynamics was to assert, through Schnirelman's theorem,
the weak-$*$ convergence (in the semiclassical limit) of the Husimi eigendensities towards the Liouville measure. Although such a measure-theoretic property seems a priori rather weak, combining it with the analytical properties of the Husimi
densities provided direct information on both the Bargmann eigenfunctions (linearly related to the wavefunctions), and their eigenconstellations (which
parametrize the eigenstates optimally in phase space). These
analytical properties actually yielded results about
the {\sl phase} and the {\sl zeros} of the Bargmann function, which seem
totally immaterial quantities in the framework of Husimi measure theory. 
In the study of eigenstates, the relevant quantity seems to be the (scaled) {\sl logarithm} $\tilde h(x)=N^{-1}\log \hat H(x)$ of the Husimi function, linearly related to the density of zeros. For an integrable system, the strategy of WKB theory is also to build semiclassical approximations for the logarithms of eigenfunctions (see eq. (\ref{WKB})). In the chaotic case, we ended up with a WKB-like `symbolic' description for the Bargmann eigenfunctions (section \ref{Phase}):
\begin{equation}
\label{wkb}
\mbox{\large ``}\ \psi(z)\approx \e^{{1\over\hbar}\int^z \bar z' dz'}.\ \mbox{\large "}
\end{equation}
Unlike its integrable counterpart (\ref{WKB}), this formula is meaningless as a computational tool (the integral in the exponent is ill-defined), but it represents concisely two characteristic properties of chaotic eigenstates.

First, as we pointed out in section \ref{Phase}, the above formal $\bar z$-dependence of the analytic function
$\log \psi(z)$ embodies the necessary asymptotic denseness of its singularities, or Bargmann zeros (the dislocation points of the phase of $\psi(z)$). This phase then forms a fan-like pattern (see fig. 5b) which seems quite universal, in the sense that one cannot easily decipher the particular features of an individual
eigenstate (eg. a scar, or in the opposite a lattice eigenconstellation) by just looking at its phase pattern. Similarly, a peak in the Husimi function cannot be detected through the {\sl local} density of zeros, but rather in
{\sl global} parameters of the whole density (Fourier coefficients, for instance). At the local level, zeros and amplitudes of the Husimi function seem
almost totally uncorrelated.

The second idea carried by the above formula is indeed universality. Unlike the WKB formula (\ref{WKB}), where the classical action in the exponent takes the dynamics explicitly into account, equation (\ref{wkb}) does not depend on
the particular dynamical system we are studying, as long as it is fully ergodic. 

The idea of universality (partly) underlies the models of random vectors introduced in
section \ref{stat}, to which the numerical data of chaotic eigenstates are compared. As we recalled, such models are eigenstate counterparts of the random matrix
models often used to mimic eigenspectra of classically chaotic systems. Until recently, they generated a statistical description of eigenstates in configuration space. We rather 
developed their analysis in the Bargmann--Husimi--stellar representations on the torus phase space, and found a generally good agreement with data from chaotic eigenstates, the comparison being made at different levels. 

We first studied the Husimi functions themselves, using norms and related functionals meant to probe the uniformity of these functions in finer topologies than the original weak-$*$ involved in Schnirelman's theorem. When averaged over all eigenstates, these functionals agree very well with their statistical values. Individual fluctuations away from these average values demonstrate the existence of very ergodic or, on the contrary, more localized (e.g. scarred) eigenstates. 

We then explored the properties of eigenconstellations. As was already noted in \cite{leb:shukla,BBL}, the 1-point and 2-point correlation functions of 
the constellations fit perfectly with the random models, when averaged over
all eigenstates. This corresponds to a universal local interaction
between the Husimi zeros, which (to this extent) resemble particles of a gas. We also obtained statistical predictions for the Fourier coefficients $\rho_k$
of the constellations (and, by linearity, for the coefficients $h_k$ of $\tilde h(x)$), which contain relevant {\sl global}
information. The overall shape of the form factor (i.e. a basin of small Fourier coefficients near the origin, surrounded by a white-noise-like sea for $|k|\geq \sqn$) is clearly reflected at the level of {\sl individual} eigenstates. We get a neater
agreement with the statistical curves when averaging over all eigenstates
(of a given parity). On the other hand, the main features of the Husimi density
of a given eigenstate can be obtained by truncating the Fourier series of its logarithm around 
$|k|\losim \sqn$; we therefore believe that the presence of a scar on an individual state, or more generally, {\sl dynamical\/} information, is linked to the deviations of some low-$k$ (dubbed {\sl essential}) Fourier coefficients from
the statistical averages. In other words, to go beyond the unprecise universal equation (\ref{wkb}), we can hope to `see a scar' directly on a 
finite set of coefficients $h_k$. So far, this observation is not supported by any dynamical explanation for the occurrence of a scar above a periodic orbit on a particular eigenstate (some heuristic arguments for this topic are given in \cite{fishman}). The trouble here is that the translation of the dynamics onto the Fourier coefficients of $\tilde h(x)$ is totally unclear. Otherwise, this analysis resembles the semiclassical theory for
the spectral form factor $K(\tau)$ \cite{berry}: 
this (now 1-d) form factor also exhibits
high Fourier coefficients displaying dominantly statistical universal behavior,
vs low Fourier coefficients concentrating the specific dynamical information 
(imprints of short periodic orbits through trace formulae).

In spite of the aforementioned lack of a direct link between classical dynamics and the observed quantum phenomena, we believe some techniques and concepts presented here might prove fruitful to the further study of chaotic eigenstates.
As explained above, the main new tool we used was the {\sl multiplicative} properties of the Bargmann function induced by its holomorphy; this
directly leads to the stellar representation as well as to the occurrence of logarithmic quantities ($\tilde h(x)$ and the geometric mean ${\rm GM}[H]$). Further progress was made by Fourier-transforming these logarithmic quantities: for instance, their connections to semiclassical features proved more robust in Fourier space. The formula which epitomizes best these methods may be eq. (\ref{epitom}). Similar tools were already used to study, on the one hand eigenfunctions of integrable systems (cf. the remark above), on the other hand spectral determinants, whose logarithms can be expanded (\`a la Fourier) in terms of classical periodic orbits. In these cases, the important features (the large values of the Husimi function, the zeros of the Bargmann eigenfunctions or spectral determinants) live on 1-dimensional curves. Here, we
used the same tools to describe quantities (functions, constellations) living genuinely in 2 dimensions. The problem of recovery of the full constellation
from a finite number of Fourier coefficients shows that the dimensional jump is
not trivial, except maybe in very non-generic cases (for instance, the separable constellations considered in section \ref{counter}, or lattice constellations).

At all levels, the agreement between statistical predictions and exact eigenfunction data was much closer for the cat maps than for the baker's map,
for which large deviations were repeatedly encountered. In particular, the baker's data
did not exclude sequences of exceptional eigenstates, for which Schnirelman's
property would not hold (by contrast, it is proven to hold for all cat eigenstates we considered). We believe that such discrepancies might be due to the
discontinuity of the classical map, whose role had been noticed (see for instance \cite{heller,saraceno}) when studying spectral properties of the quantum map. On the other hand, cat maps have arithmetical properties, which make their spectra very non-generic. It would be interesting to find out if the pseudo-random nature of their eigenstates persists if we perturb the map continuously to make it `generic'.

\bigskip
{\bf Acknowledgements:}

We have benefited from fruitful discussions with many colleagues, among whom M. Bauer, A. Gervois, P. Leboeuf, J.-M. Luck, A. Mezincescu and J. Peyri\`ere. 

\appendix
\section*{Appendices}

\section{Optimization of invariants for lattice states}
In this appendix we derive explicitly the geometric mean and the $L^2$ norm for the Husimi functions of lattice states, defined in sections \ref{comparison} and \ref{lattice}. The scale invariance of these functionals means that we just need to derive them for the basic functions $H_{\chi(\tau)}$, defined in equation (\ref{chitau}). We will then prove that both these functionals take absolute extremal values for the triangular lattice $\tau=\e^{\mi\pi/3}$ (we are indebted to M. Bauer for this proof). 

By using eq.(\ref{chitau}) and standard formulas on theta functions \cite{WW},\cite[vol. 2, chap. 13]{bateman}, one obtains the following formulae:

\begin{eqnarray}
\label{inv1}
{\rm GM}[H_{\chi(\tau)}]&=& \exp\left({1\over\Im(\tau)} \int_{T_\tau}\log H_{\chi(\tau)}(q,p)\, dq\,dp \right)=\sqrt{2\Im(\tau)}|\eta(\tau)|^2\\
\label{inv2}
||H_{\chi(\tau)}||_2^2&=&{1\over\Im(\tau)} \int_{T_\tau} H_{\chi(\tau)}(q,p)^2 dq \,dp=\sum_{l,j\in{\Bbb Z}} \exp\left(-\pi{|l+j\tau|^2\over\Im(\tau)}\right).
\end{eqnarray}

In the formula above, $\eta(\tau)$ is Dedekind's modular form \cite[p.129]{miyake}
$\eta(\tau)=\e^{\mi\pi\tau/12}\prod_{n\geq 1}(1-\e^{2\mi\pi n\tau})$.
Curiously, these quantities are closely related to the determinants of the Laplacians $\Delta_\tau$ on $T_\tau$. Indeed, we have  \cite{osgood,kierl}

\begin{eqnarray}
\det\Delta_\tau&=&\Im(\tau)^2|\eta(\tau)|^4= {\rm GM}[H_{\chi(\tau)}]^2\;\Im(\tau)/2 \\ 
{\rm tr}(\e^{t\Delta_\tau})&=&\sum_{l,j\in{\Bbb Z}} {\Im(\tau)\over 4\pi t}\exp\left(-{|l+j\tau|^2\over 4t}\right),
\end{eqnarray}
so $||H_{\chi(\tau)}||_2^2={\rm tr}(\e^{t\Delta_\tau})|_{t={\Im(\tau)/4\pi}}$.

\medskip
The quantities (\ref{inv1},\ref{inv2}) are obviously modular invariant, so we only need to study them for $\tau$ in a fundamental domain of ${\rm PSL}(2,{\Bbb Z})$ (see fig. 14). We will show in the following that the only extrema for both invariants on this domain are situated at the symmetry points $\tau=\mi$ (B on the figure) and $\tau=\e^{\mi\pi/3}$ (A on figure 14). 

\subsection{Optimization of the geometric mean}
The equation for an extremal point of ${\rm GM}[H_{\chi(\tau)}]$ reads ${\eta'\over\eta}(\tau)={\mi\over 4\Im(\tau)}$, which implies that $\Re({\eta'\over\eta}(\tau))=0$. If one denotes $\e^{2\mi\pi\tau}=r\e^{\mi\theta}$, this equation reads

\begin{equation}
\label{eta}
\sum_{n\geq 1}{n\,r^n\sin(n\theta)\over|1-r^n\e^{\mi n\theta}|^2}=0,
\end{equation}
so it holds trivially for $\theta\equiv 0$ mod $\pi$. For small $r$, the first term of the sum dominates the remainder, unless $\sin(\theta)$ is very small. We actually show that for $r$ smaller than $r_o\approx 0.134$, the only solutions of (\ref{eta}) are the two points $\{\sin(\theta)=0\}$. Indeed, if we isolate the first term in (\ref{eta}), we obtain the inequality:
\begin{equation}
{r|\sin(\theta)|\over |1-r\e^{\mi\theta}|^2}=\Bigl|\sum_{n\geq 2}{n\,r^n\sin(n\theta)\over|1-r^n\e^{\mi n\theta}|^2}\Bigr|\leq\sum_{n\geq 2}{n^2\,r^n\over|1-r^n\e^{\mi n\theta}|^2} |\sin(\theta)|.
\end{equation}
If $\sin(\theta)\neq 0$, a few simplifications yield the inequality
$r(1-r)^5\leq4r^2-3r^3+r^4$, which does not hold for $r\in[0,r_o]$, or equivalently for $\Im(\tau)\geq t_o\approx 0.32$ (on the figure, this corresponds to the dashed horizontal line). Thus, the only extremal points with $\Im(\tau)\geq t_o$ must be on one of the axes $\{\Re\tau=0\}$,$\{\Re\tau=1/2\}$. 

It remains to study the variations of ${\rm GM}[H_{\chi(\tau)}]$ along these two axes in the fundamental domain. On the imaginary axis, we have
$${d\over dt}\log{\rm GM}[H_{\mi t}] = -{\pi\over 12}+{1\over 4t}+2\pi\sum_{n\geq 1}{n\e^{-2\pi n t}\over 1-\e^{-2\pi n t}},$$
which is a strictly decreasing function for $t\in{\Bbb R}^*_+$. On the other hand, this derivative vanishes for $t=1$, since the modular transformation $J=\left(\begin{array}{cc} 0&-1\\ 1&0\end{array}\right)$ maps $\{t\in ]\infty,1]\}$ onto $\{t\in]0,1]\}$, $t=1$ is an extremal point; it is therefore the only one along this axis.

To deal with the second axis $\{\Im(\tau)=1/2\}$, we use the following relation, also due to the transformation $J$: 
$${d\over dt}\log{\rm GM}[H_{1/2+\mi t}] = {2\over t}\;\Re\left({\eta'\over\eta}\left({-1\over 1/2+\mi t}\right)\right)$$
When $t$ describes ${\Bbb R}^*_+$, $\tau={-1\over 1/2+\mi t}$ is on the semi-circle $(\alpha,B",A',\beta)$. Fortunately, we know the sign of $\Re({\eta'\over\eta}(\tau))$ in the domain $\Im(\tau)\geq t_o$, since this quantity only vanishes along the symmetry axes, and in the limit $\Im(\tau)\to\infty$ we have 
$\Re({\eta'\over\eta}(\tau))\sim 2\pi\e^{-2\pi\Im(\tau)}\sin(2\pi\Re(\tau))$.
Therefore, $\Re({\eta'\over\eta}(\tau))$ takes the sign of $\sin(2\pi\Re(\tau))$ as long as $\Im(\tau)\geq t_o$, as indicated on the figure. This provides the variations of 
${\rm GM}[H_{\chi(\tau)}]$ along the line $\{\tau=1/2+\mi t\}$ between the points $\alpha'=J(\alpha)$ and $\beta'=J(\beta)$: it increases from $\beta'$ to $A$, then decreases down to $B'$, where it takes the same value as at the conjugate point $B$; the variations from $B'$ to $\alpha'$ are symmetric. A direct estimation of the series ${d\over dt}\log{\rm GM}[H_{1/2+\mi t}]$ shows that its sign does not change above $\beta'$.

We thus conclude that the invariant ${\rm GM}[H_{\chi(\tau)}]$ has only two extremal points in the fundamental domain of ${\rm PSL}(2,\Bbb Z)$:

--it has a global maximum at $\tau=\e^{\mi\pi/3}$, corresponding to the equilateral triangular lattice, i.e. the closest packing of points on the plane. There, it takes the value ${\rm GM}[H_{\e^{\mi\pi/3}}]={\sqrt{3}\over 4\pi^2}\Gamma(1/3)^3$.

--it has a saddle-point at $\tau=\mi$, i.e. the square lattice. There, we have
${\rm GM}[H_\mi]={\rm GM}[H_\chi]={\Gamma(1/4)^2\over(2\pi)^{3/2}}$.

--it vanishes in the limit $\Im(\tau)\to\infty$. 

\subsection{Optimization of the $L_2$ norm}
The variations of the second invariant $||H_{\chi(\tau)}||_2^2$ are studied similarly. Indeed, the real part of the equation ${d\over d\tau}||H_{\chi(\tau)}||_2^2=0$ reads:
\begin{equation}
\sum_{j,k>0}jk\e^{-\pi\Im(\tau)(j^2+k^2)}\sin(2\pi jk \Re(\tau))=0,
\end{equation}
which yields $\sin(2\pi\Re(\tau))=0$ as long as $\Im(\tau)\geq t_1\approx 0.3067$, so the extremal points in the fundamental domain must also be situated on the two symmetry axes $\{\Re(\tau)=0\},\;\{\Re(\tau)=1/2\}$. The variations along the line $\Re(\tau)=1/2$ are obtained as above, i.e through a modular transformation to the half-circle $(\alpha,B',A',\beta)$: we obtain variations of opposite signs as in the former case (on this axis, $||H_{\chi(\tau)}||_2^2$ has a minimum at the point $A$). The analysis on the axis $\Re(\tau)=0$ can be performed directly on the series. Precisely, from the formula
\begin{equation}
{d\over dt}\log\left(||H_{\mi t}||_2^2)\right) ={1\over 2t\theta_3(it)}\left\{1+2\sum_{j\geq 1}\e^{-\pi t j^2}(1-4\pi t j^2)\right\}
\end{equation}
and the fact that the function $x\mapsto \e^{-x}(1-4x)$ is strictly increasing for all $x> 5/4$, we deduce that the whole derivative increases for $t\geq 1$. Besides, this derivative vanishes by symmetry at the point $t=1$. Therefore, $||H_{\mi t}||_2$ has a single extremum on the imaginary axis, at $t=1$. 

This second invariant is thus quite similar to the geometric mean:

-- it has an absolute minimum for the equilateral lattice  $||H_{\e^{\mi\pi/3}}||_2^2={3\Gamma(1/3)^3\over 2^{7/3}\pi^2}$.

-- it has a saddle-point for the square lattice $||H_\mi||_2^2=||H_\chi||_2^2={\Gamma(1/4)^2\over 2\pi^{3/2}}$.

-- it diverges as $\Im(\tau)\to\infty$.

\section{Invariant lattices of cat maps}
In this appendix we derive a few properties of the invariant lattices of cat maps; these properties are used for the estimations made in section \ref{lattice}. A classical `cat' map is given by a hyperbolic matrix $S=\left(\begin{array}{cc} a&b\\ c&d \end{array}\right)$ in ${\rm SL}(2,\Bbb Z)$ (i.e. $ad-bc=1$, $|a+d|>2$). 

When studying the quantized map $U_S$ on ${\cal H}_N$ in Bargmann's representation \cite{crystal}, one is led to search for sublattices $\Lambda$ of ${\Bbb Z}_N^2$ invariant under $S$. More precisely, $\Lambda$ must be a free principal submodule of ${\Bbb Z}_N^2$, i.e generated by a unique integer vector $V$: $\Lambda=\{\alpha V \mbox{ mod }N,\; \alpha=0,\ldots,N-1\}$, and these $N$ points must be different. 

$\Lambda$ can also be considered as a lattice in ${\Bbb Z}^2$: it then admits the basis $[V, \left(\begin{array}{c}0\\N\end{array}\right)]$. Our main task here is to estimate the {\sl minimal} basis $[V_1,V_2]$ for $\Lambda$, i.e. the basis composed of the two successive shortest (for the euclidean norm) non collinear vectors: $0<|V_1|\leq |V_2|\leq\ldots$ (one can show that $[V_1,V_2]$ is indeed a basis for $\Lambda$ \cite[p.83]{cohn}). 

A general theorem ensures that the shortest non-vanishing vector $V_1$ has its norm $|V_1|\leq \sqrt{2N}$ \cite[p.137]{cohn}. In the following, we will prove that for invariant lattices, this norm also admits a {\sl lower} bound of the type $C\sqrt{N}$. This property will then imply the same estimates for the second basis vector $V_2$, and a control over the size of the Dirichlet--Voronoi cell (in the present context, this cell can be defined as the set of all points in ${\Bbb R}^2$ closer to the origin than to any other point of $\Lambda$).

We will study such invariant lattices in the case where they admit a generator $V$ of the form $V=\left(\begin{array}{c}1\\k\end{array}\right)$ (this is possible if $b$ and $N$ are coprime). The resulting sublattice $\Lambda$ modulo $N$ is invariant though $S$ iff
\begin{equation}
\label{kk'}
bk^2+(a-d)k-c=0\quad\mbox{mod }N.
\end{equation}
To estimate the minimal vector $V_1$, we search for the shortest vector $V_\alpha$ congruent to $\alpha V$ modulo $N$, for $\alpha=1,\ldots,N$ (we can restrict $\alpha$ to values in $0,\ldots,N/2$ by parity). Its square norm is  $|V_\alpha|^2=\alpha^2+ (\beta={\rm min} |\alpha k| \mbox{ mod }N)^2$. 

\medskip

In all our calculations occurs an important quantity related to $S$, its {\sl 
discriminant} $D=({a+d\over 2})^2-1$, whose square-root is irrational because $|a+d|>2$. Equation (\ref{kk'}) leads to the following constraint between $\alpha$ and $\beta$:
$$ (b\beta -\alpha{d-a\over 2})^2\equiv \alpha^2 D\quad\mbox{   mod   }N.$$ 
Since $D$ is not the square of a rational, this equation cannot hold in $\Bbb Z$. 
For $\alpha <\sqrt{N/D}$, it then yields the inequality
$$\beta\geq {1\over|b|}(\sqrt{D\alpha^2+N} -\alpha |{d-a\over 2}|)\defi f(\alpha).$$
We then study the variations of $f(\alpha)$, which depend upon the sign of the product $bc$. 

If $bc>0$, $f$ admits a minimum at $\alpha_1=|{a-d\over 2}|\sqrt{N\over bcD}$, where it takes a value $f(\alpha_1)=C_1\sqrt{N}$.  Therefore, the square norms $|V_\alpha|^2\geq N{\rm min}(C_1,1/D)$. 

If $bc<0$, $f(\alpha)$ decreases monotonically from $\alpha=0$, and vanishes at the
point $\alpha_2=\sqrt{N/|bc|}$. However, we can still bound the norms as
$|V_\alpha|\geq C\sqrt{N}$, with $C$ a positive constant: if $\alpha<{\rm min}(\alpha_2/2,\sqrt{N/D})$, then $|V_\alpha|^2\geq f(\alpha)^2>f(\alpha_2/2)^2$; otherwise, $|V_\alpha|^2\geq {\rm min}(\alpha_2^2/4,N/D)$.

\medskip

From the bounds of the shortest vector $C\sqrt{N}\leq |V_1|\leq\sqrt{2N}$, let us now study $V_2$. On the one hand, it defines a parallelogram with $V_1$, of area $N$, that we choose to orient as $V_1\wedge V_2=-N$ (this is a choice between $\pm V_2$). On the other hand, since $V_2+n V_1$ is also in $\Lambda$, we must have $|V_1.V_2|\leq 1/2|V_1|^2$. Both constraints yield $|V_1|^2\leq |V_2|^2\leq N/2({|V_1|^2\over 2N}+{2N\over |V_1|^2})$. This inequality actually improves the upper bound on the minimal vector: $|V_1|^2\leq N\sqrt{4/3}$; besides, the lower bound on $|V_1|$ implies $|V_2|^2\leq N(C^2/4+1/C^2)$.

\medskip

We can now estimate the radius of the Dirichlet--Voronoi cell centered at the origin, which is a hexagon except in the case $V_1.V_2=0$ (a rectangle). The radius is then the distance of the farthest vertex from the origin. We find the uniform bound
$$R^2\leq {N\over 4}\;{\rm max}( 5/2,C^2+1/C^2).$$

\medskip

The invariant lattice $\Lambda$ is mapped to the constellation of an eigenstate $\psi$ in ${\cal H}_N$ through a rescaling and a complex conjugation; the constellation is the sublattice of $T_{\Bbb C}$ generated by $[v_1,v_2]$, with
$\bar v_i= V_i/(N\sq2)$. The above estimates then yield $|v_i|\sim O(\sqrt{N})$, and the modulus $\tau=v_2/v_1$ is bounded in a compact set: 
$$\sqrt{3/4}\leq\Im\tau=N/|V_1|^2\leq 1/C^2\qquad\mbox{and}\qquad -1/2\leq\Re\tau={V_1.V_2\over|V_1|^2}\leq 1/2.$$

\section{Sup-norm estimates for the statistical model}
\label{sup-norm}
We present the derivation of the typical sup-norm of the Husimi density of a random complex vector, which yields the estimate listed in table \ref{tab}, 3rd line. Such a quantity cannot be deduced from the same methods as the other invariants, but requires a completely different analysis. 

The proof proceeds geometrically. Instead of considering an ensemble of Gaussian random states (\ref{psia}), we constrain the statistical states to be on the unit (complex) sphere ${\cal S}^N$, with the probability density
given by the canonical volume form. In Euclidean coordinates, the volume of this sphere reads
\begin{eqnarray}
{\rm Vol}({\cal S}^N)&=&\int_{D(0,1)}d^2 a_0(1-|a_0|^2)^{N-2}\,{\rm Vol}({\cal S}^{N-1})\\
&=&2\pi\,\prod_{j=0}^{N-2}\int_{D(0,1)}d^2 a_j(1-|a_j|^2)^{N-j-2}\nonumber
\end{eqnarray}
where $D(0,1)$ is the unit disk, and $a_0$ is the overlap between the normalized state $|\psi_{\{a\}}\rangle$ and the first vector of the basis, $|q_0\rangle_{N,0}$; this overlap is related to the Fubini--Study (or Hermitian)
distance $d_{\rm FS}$ between the representatives of these two vectors in the complex projective plane $\Bbb {CP}^N$, which we note $[\psi_{\{a\}}]$ and $[q_0]$:
\begin{equation}
|a_0|=|\langle \psi_{\{a\}}|q_0\rangle|=\cos d_{\rm FS}([\psi_{\{a\}}],[q_0]).
\end{equation}
From this, we obtain the normalized volume of the ball $B_N([q_0],d)$ of radius $d$ in $\Bbb {CP}^N$:
\begin{equation}
{{\rm Vol} ( B_N([q_0],d))\over {\rm Vol}( \Bbb {CP}^N)}={{\rm Vol}(\{\psi_{\{a\}}\in {\cal S}^N \;s.t\;\;|a_0|\geq \cos d\})\over {\rm Vol}({\cal S}^N)}=(1-\cos^2 d)^{N-1}.
\end{equation}
In our statistical framework, this quantity is the probability for a random normalized state $|\psi_{\{a\}}\rangle$ to have an overlap $|\langle \psi_{\{a\}}|\phi\rangle|\geq\cos d$ with a fixed (normalized) state $|\phi\rangle$. 

Now, equation (\ref{corr}) shows that the square-norms of the torus coherent states $|z \rangle_{N,0}$ (see eq. (\ref{coh})) are given by $N\e^{2\pi N\bar z z}$ in the classical limit. Therefore, in this limit the Husimi density $H_\psi(z,\bar z)$ can be written as 
\begin{equation}
  H_\psi(z,\bar z)\sim N|\langle \psi|z\rangle_{\rm norm}|^2
= N\cos^2 d_{\rm FS}([\psi],[z]),
\end{equation}
where $[z]$ is the representative of the normalized torus coherent state $|z\rangle_{\rm norm}$. Therefore, the maximum of $H_\psi(z,\bar z)$ is given by the distance between $[\psi]$ and the set of coherent states in the projective plane $\Bbb {CP}^N$: $\{[z],\;z\in T_{\Bbb C}\}$. To estimate this distance, we first select $M$ points $z_i$ on $T_{\Bbb C}$, which are well-distributed: for instance, we consider the square sublattice of $T_{\Bbb C}$ of side $1/\sqrt{M}$, with $M\sim N^{2\beta}$, for a certain power $\beta>1/2$ to be selected later. Thus, any $z\in T_{\Bbb C}$ will be close to one of the $z_i$:
\begin{equation}
\label{zi}
\forall z\in T_{\Bbb C},\;\exists z_i\;\mbox{s.t.}\;|z-z_i|\leq N^{-\beta}\Longrightarrow d_{\rm FS}([z],[z_i])\leq \sqrt{2\pi}\;N^{1/2-\beta}.
\end{equation}

 For a given distance $d$, we can estimate the proportion of states $|\psi\rangle$ s.t. $\max_i |\langle\psi|z_i\rangle|\geq\cos d$, i.e. $\exists i$, $d_{\rm FS}([z],[z_i])\leq d$:
\begin{eqnarray}
{{\rm Vol} (\bigcup_{i=1}^M B_N([z_i],d))\over {\rm Vol} (\Bbb {CP}^N)}&\leq& \sum_{i=1}^M{{\rm Vol} ( B_N([z_i],d))\over {\rm Vol}( \Bbb {CP}^N)}\\
&\leq&M(1-\cos^2 d)^{N-1}
\end{eqnarray}
Using this inequality, we give a lower bound on the distance $d=d_N$ s.t. the proportion of such states in $\Bbb {CP}^N$ is equal to a fixed number $p\in (0,1)$ (we use the scaling law $M=N^{2\beta}$):
\begin{eqnarray}
&&{{\rm Vol} (\bigcup_{i=1}^M B_N([z_i],d_N))\over {\rm Vol} (\Bbb {CP}^N)}= p\qquad\\
\Longrightarrow d_N&\geq&\arccos\left(\sqrt{2\beta\log N+|\log p|\over N-1}\right)\defi f(p,N)
\end{eqnarray}
Now, if a state $[\psi]$ is {\sl not} in $\bigcup_{i=1}^M B_N([z_i],d_N)$, i.e. if its distance from any $z_i$ is greater than $d_N$, the triangular inequality implies
\begin{eqnarray}
\forall z\in T_{\Bbb C},\qquad d_{\rm FS}([\psi],[z])&\geq& d_{\rm FS}([\psi],[z_i])-d_{\rm FS}([z],[z_i])\\
&\geq&f(p,N)-\sqrt{2\pi}N^{1/2-\beta}\\
\Longrightarrow H_\psi(z,\bar z)&\leq& N\cos^2(f(p,N)-\sqrt{2\pi}N^{1/2-\beta})
\end{eqnarray}
If we fix $p\in (0,1)$ and take $\beta=1$, this uniform bound on $H_{\psi}$ behaves as ${2\log N}$ in the limit $N\to\infty$. This bound is valid for a proportion $(1-p)$ of states in $\Bbb {CP}^N$. From there we easily deduce the inequality stated in table \ref{tab} for the average sup-norm.

\vfill\eject

\vfill\eject

\centerline{\bf Figure captions}
\bigskip

Fig. 1a.
Husimi density plots and stellar representations 
(plus their Voronoi tessellations) for a sample of even-parity eigenfunctions
of the $S$-cat map (eq.(\ref{cats})) quantized at $N=107$, a splitting prime. 
The Voronoi partition of a constellation gives an idea of its local density 
of zeros (as the inverse area of each cell).

{\sl Top 2 rows, from left to right:\/} the most uniformly spread eigendensity; 
one with typical (random-like) values of invariants; 
a lattice state (the only non-real eigenstate; together with the conjugate
lattice state they span the doubly degenerate eigenspace of this operator).
{\sl Bottom 2 rows\/}: the most strongly scarred eigenstate 
(at the fixed point (0,0));
another state scarred around (0,0) and along the stable and unstable lines;
another localized eigendensity, featuring the highest H-function 
and lowest geometric mean.

{\sl General caption for Husimi plots:\/} 
Husimi densities are shown on a linear gray scale, 
common to all plots of each figure to preserve contrast differences 
(with zero as white); a small white triangle locates a maximum, 
and the white circle around it materializes a Planck area (equaling $h=1/N$);
main norms and invariants (Sec.\ref{norms}) are listed at top; 
zeros are located by black stars.

Fig. 1b.
Same as fig. 1a for a sample of strongly scarred odd-parity eigenstates
of the quantum baker's map. {\sl Leftmost plots:\/} two members ($N=48,\ 128$)
of the sequence of those eigendensities with the highest scars 
above the period-2 classical orbit (through (1/3, 2/3)); 
{\sl right:\/} an eigenstate strongly scarred 
along the stable and unstable lines of the singular fixed point (0,0).

Fig. 2a.
Some comparison Husimi densities (non localized) on the square torus;
their constellations of zeros are now shown superimposed. 
{\sl Left:\/} the singlet ($N=1$) density $H_\chi$ (cf. eq.(\ref{chi})), 
also giving the elementary density cell of any square lattice state 
(cf. fig.4, left); 
{\sl middle:\/} an almost optimally equidistributed lattice density, for $N=56$
(zeros form near-equilateral triangles, of horizontal side 1/7 and height 1/8);
{\sl right:\/} a computer-produced sample of the random ensemble 
(without any symmetry: Sec. \ref{nosym}) for the same value of $N=56$.

Fig. 2b.
More comparison Husimi densities (localized) on the square torus. 
{\sl Left:\/} an $N=16$ plane wave; 
{\sl middle:\/} an $N=16$ coherent state (cf. eq.(\ref{coh}));
{\sl right:\/} a degenerate density having a single zero of order $N=16$ (white star),
corresponding to the 16-th power of the singlet density (fig.2a, left).

Fig. 3a
Some invariant fluctuation measures of the even-parity eigenstates for the 
quantization of the $S$-cat map of eq.(\ref{cats}), 
with the following splitting prime values of $N$:
13, 37, 59, 61, 83, 107, 109, 131, 157, 179, 227, 347, 397.
The following functionals of the Husimi eigendensities are shown:
(a) sup-norm; (b) $L^2$-norm; (c) H-function (negative entropy); 
(d) Geometric Mean (flipped upside-down).
The `+' mark all individual values and the thick curve connects their suitable
{\sl averages at fixed\/} $N$ (up to 131).
The bullets pinpoint the values corresponding to 
{\sl the most scarred state above the fixed point\/} (0,0) for each $N$
(meaning for us the state giving the highest Husimi value $H(0,0)$);
the dashed curve connects these values, only for visual emphasis;
all curves drawn here are devoid of significance 
outside of the computed points they interpolate between.
Data for some comparison states of Table 1 are added: 
the equilateral triangle lattice value (long-dashed line),
the random-ensemble value (dot-dashed lines, including the typical deviations around the averages---except for the sup-norm, where the line is only the upper bound of Table 1),
and the two integrable-state examples (solid curves).

Fig. 3b.
Some invariant fluctuation measures of the eigenstates for the quantum 
baker's map, with selected even values of $N$ up to 512. 
The following functionals of the Husimi eigendensities are shown:
(a) sup-norm for even-parity states; (b) same for odd-parity states;
(c) H-function (negative entropy) for even-parity states; 
(d) same for odd-parity states.
The `+' mark all individual values and the thick curve connects their suitable
{\sl averages at fixed\/} $N$ (up to 128).
The bullets mark the values corresponding to 
{\sl the most scarred state above a specific periodic point\/} for each $N$,
namely: the fixed point (0,0) in the even-parity case, and the period-2
point (1/3, 2/3) in the odd-parity case (cf. the 2 leftmost plots of fig. 1b); curves moreover connect two special $N$-subsequences from the latter set, i.e. $N=2^k$ (dashed), 
and $N=3 \times 2^k$ (dotted --- for odd states only). 
The same comments apply to curves as in the preceding figure.

Fig. 4.
Striking effect of a long-wavelength deformation upon a constellation of zeros.
{\sl Left:\/} $N=64$ Husimi density of square lattice, with the periods 
$\{1/8,\ 1/8\}$ 
(each cell reproduces in miniature the singlet density of fig. 2a, left).
{\sl Middle:\/} Husimi density of a smoothly deformed constellation (the mesh
is slightly squeezed towards the edges and stretched towards the center
to reach a sides' ratio of 0.9, almost unnoticeable on the constellation). 
{\sl Right:\/} same deformation for $N=256$ (meaning now a four times smaller
Planck's constant). Notice the large variation of the sup-norms 
(and of the other functionals too); due to this, the three contrast scales
are here exceptionally selected independently of one another. 

Fig. 5a.
Illustration of the semiclassical behavior of the logarithmic derivatives
$\pi(z) \defi {1\over 2\pi N}{\psi' \over \psi}(z)$ of Bargmann eigenfunctions,
using color density plots to represent complex functions 
(a complex value is encoded by intensity for modulus --- with zero as white ---
and hue for phase).
Central plot: the classical conjugate momentum $\bar z$ as a function of $z$
(from which our color encoding of complex phases can be inferred).
Left plot: $\pi(z)$ for a classically integrable example,
an eigenstate of a quantum Hamiltonian 
(15-th state of Harper's operator 
$-(\cos 2\pi \hat p + \cos 2\pi \hat q)$ for $N=31$).
The lines enclose a vicinity of the classical energy curve, here defined to be
the region where the Husimi density exceeds 1/10-th of its maximum value;
the approximation $\pi(z) \sim \bar z$ is good only nearby.
Right plot: $\pi(z)$ for a classically chaotic example,
an eigenstate of the quantum baker's map 
(the strongly scarred eigenstate of fig.1b right, with $N=128$);
the approximation $\pi(z) \sim \bar z$ now holds throughout, 
save very close to Bargmann zeros, 
or poles of $\pi(z)$ (the `pimples' in both quantum plots).  

Fig. 5b.
Plots of Bargmann eigenfunctions $\psi(z)$, 
using intensity to encode the Husimi density and color to encode the phase.
{\sl Left}: the 15-th eigenstate of the quantum ($N=31$) Harper Hamiltonian
(same as in preceding figure), specifically with a binary coding of the modulus
(threshold = 1/10-th of maximum Husimi density) 
letting the phases show through everywhere in the classically forbidden regions.
Middle: an (odd-parity) eigenstate of the quantum $S$-cat map for the same $N$.
{\sl Right}: the eigenstate of the quantum ($N=128$) baker's map of preceding figure
and fig.1b right. A universal radial fan-like pattern for the quantum phases,
having the {\sl regular\/} wavevector distribution 
$(k_q,k_p) \sim \pi N \times (p,-q)$
prevails almost everywhere in the case of classically chaotic systems (barring
dislocation points at the Bargmann zeros),
and only about the classical orbit in integrable cases.

Fig. 6.
{\sl Left}: the rescaled pair-correlation function  $g(r)$ between zeros of a random Bargmann function with no symmetry (eq. (\ref{rho2})). The shape (short-distance repulsion followed by a small bump) is similar to the corresponding correlation for particles in
a dilute gas \cite{bohigas}. The scaling $r=\sqrt{\pi N}\;|\delta z|$ indicates that zeros repel each other up to a distance $\sim 1/\sqn$, and are uncorrelated beyond.
{\sl Right}: various rescaled form factors. {\sl Solid curve}: rescaled form factor $F_2(\kappa)$ for the density of zeros $\rho(x)$ in the model with no symmetry, eq. (\ref{F_2}). {\sl Dot-dashed curve}: rescaled form factor for the logarithm $\tilde h(x)$ (eq.(\ref{tildeh})). These form factors also hold in the semiclassical limit for the model of even-parity complex random states (cf. eq. (\ref{approx})), and conjecturally for the model of even real random states as well, in all directions except the axes (eq. (\ref{conjff})). {\sl Dashed curve}: conjectured rescaled form factor along the two symmetry axes for the density of zeros of even real random states. These curves are compared to numerical data in section \ref{Fouri}. The two dotted vertical lines represent the rescaled radii $\kappa_<$ and $\kappa_>$, described in section \ref{essential}.

Fig. 7.
{\sl Left}: average densities of zeros, for various statistical models. {\sl Solid curve}: average density of zeros for the model of even-parity complex random states, as a function of the rescaled distance from a symmetry point (eq. (\ref{rho1e})). The drop of the density in the vicinity of the origin can be interpreted as
a repulsion between $z_i$ and $-z_i$, whereas the subsequent bump ensures the
global normalization of $\rho_1$. The dotted and dashed curves concern the model of real even random states. {\sl Dashed curve}: density of complex zeros transversally to an axis of symmetry, far from symmetry points, so that parity corrections are negligible (this curve is the vertical cut $\{Q=3\}$ in fig. 7 right). {\sl Dotted curve}: renormalized density of singular zeros (e.g. real zeros) away from a symmetry point (eq. (\ref{rho1sing})).
{\sl Right}: Contour plot of the average density of complex zeros near a symmetry point for the model of even-parity real random states, as a function of the rescaled variables $(Q,P)=\sqrt{\pi N}\times(q,p)$. The contour level spacing is $0.03$. We only show the first quadrant $\{Q>0, P>0\}$, the others being obtained by reflection w.r.t. the axes. The density vanishes along the axes and has the bulk value $1$ far from the symmetry axes. The maximum of $G_1$ ($\approx 1.084$) lies at the point $Q=P\approx 1.042$. As long as $|Q|$ or $|P|\geq 2$, a good approximation is given by the formula for random polynomials (\ref{RP}). The cut of this plot along the axis $\{Q=3\}$ is given by the dotted curve of fig. 7 left.

Fig. 8.
{\sl Solid curve}: rescaled plot of the dominant ($\propto N^{-1/2}$) term in the expansion of the average Fourier coefficients along the symmetry axes for the model of even-parity real random states (eq. (\ref{rhokreal})). This dominant term is of the same order as the quadratic average (\ref{conjff}), so it should be visible on the numerical data. {\sl Dashed curve}: rescaled average Fourier coefficients away from a symmetry point for the model of even-parity complex random states (\ref{rho1e}). This average is of order $N^{-1}$, whereas the quadratic average (\ref{formeven}) is of order $N^{-1/2}$. 

Fig. 9a.
Square averages of the Fourier coefficients of the density of zeros (left), and of the logarithmic Husimi density (right), related by eq.(\ref{hk}), for the even-parity eigenstates of the $S'$-cat map with $N=149$. The abscissa carries two scales: the modulus $|k|$ of the Fourier index (in boldface), and its rescaled value $|k|/\sqn$. The Fourier coefficients are computed for $k$ along selected directions from the origin, labeled by distinctive symbols: $+$ horizontal $k$-axis, $\times$ vertical $k$-axis, $\bigtriangleup$ first diagonal, $\bigtriangledown$ second diagonal. The data are compared to the statistical 
form factors in the complex random state model (see fig. 6 right, and eq. (\ref{formeven})).   

Fig. 9b.
Same as fig. 9a, but for the even-parity eigenstates of the $S$-cat map (top) and baker map (bottom) and also including the data along the direction $(2,1)$ (marked with $\circ$). Form factors of the {\sl real\/} random state model are given both for a generic direction (dot-dashed curves) and along a symmetry axis (long-dashed line) (eq. \ref{conjff}). Due to the large deviations from the statistical curves in the baker's case, we use varying ordinate scales.

 Fig. 10.
Two-dimensional density plots for the Fourier coefficients 
$\rho_k\ (0 \ne k \in \Bbb Z^2)$ (top row) and $h_k \equiv -\rho_k/ \pi k^2$
(bottom row) of some constellations. 
Positive values are in red, negative values in blue (zero is white).
The two circles mark the inner and outer radii, 
$K_< =\sqrt{2N/\pi}$ and $K_> =\sqrt{2N/\sqrt 3}$ respectively.
Density scales are independently selected, so we indicate
the largest absolute value appearing in each plot (`Max').

{\sl Left\/}: most ergodic $N=107$ $S$-cat eigenstate of fig. 1a, top left;
{\sl middle\/}: most scarred $N=107$ $S$-cat eigenstate of fig. 1a, 
bottom left; 
{\sl right\/}: $N=64$ deformed lattice state of fig. 4, middle.

{\sl Top row\/}: the two leftmost $\rho_k$-plots (Max $\approx 0.357$) 
exemplify two generic features of chaotic eigenstates:
the strong squeeze of the coefficient values in the central basin, 
and their noisy look outside; the rightmost $\rho_k$-plot (Max $\approx 0.956$)
is an example where
the essential coefficients begin to rise only in the transition region.

{\sl Bottom row\/}: the Fourier coefficients $h_k$ of 
$\tilde h (x)= N^{-1} \log \hat H(x)$
show distributions of widths $\approx \sqn$.
The leftmost $h_k$-plot (Max $\approx 0.00134$) corresponds to 
a Husimi density with a very moderate value $H(0) \approx 1.397$) and
shows a good balance between positive (red) and negative (blue) values of $h_k$.
The middle (Max $\approx 0.00161$) and right (Max $\approx 0.00837$) $h_k$-plots
correspond to densities very peaked at $x=0$, and
show a marked {\sl overall\/} dominance of the {\sl positive\/} (red) 
values of $h_k$. 

Fig. 11.
Partial sums $N \times \sum_{|k|\le K} h_k$ as functions of (integer) $K$
for the six $N=107$ $S$-cat map eigenstates of fig. 1a (each curve is labeled
by the value of the corresponding eigenangle in units of $2\pi$ marked at the
limiting value $\log \hat H(0)$ on the right edge, except for one state
having $\log \hat H(0) \approx -9.3$). 
The inner and outer radii $K_<$ and $K_>$ are marked by the dotted vertical lines. 
The limiting behaviors are seen to set in gradually before $K$ reaches $K_<$
(the oscillations are a Gibbs phenomenon, which would diminish under smoother
truncation schemes).

Fig. 12.
Coarse-grained Husimi density of the $N=128$ baker eigenstate of fig. 1b (middle),
obtained through partial Fourier summations of its logarithm, truncated at
radius $K=2$ (left), $9 \approx K_<$ (middle), $12 \approx K_>$ (right).
(This is {\sl not\/} a smoothing of the Husimi density itself by convolution.)
The resulting densities are normalized {\sl a posteriori\/} (by brute force),
then the other approximate invariant functionals are computed.

Fig.13a.
The low Fourier coefficients $h_{1,0}$ (top row) and $h_{1,1}$ (bottom row)
of the even-parity eigenstates for the quantization of the $S$-cat map.
{\sl In left column:\/} log--linear scale plots of $N^{3/2} h_k$, 
in the same general conventions as fig.3a. 
{\sl In right column:\/} log--log plots of $|h_k|$; the solid curve is the
quadratic average at fixed $N$, against the dot-dashed straight line
of the random model (without symmetry) which shows the slope $-3/2$. As in fig. 3a, the values corresponding to 
{\sl the most scarred state above the fixed point\/} (0,0) for each $N$
are marked by bullets connected by the dashed curve.

Fig.13b.
Same as fig.13a, but for the even-parity eigenstates of the quantum baker's map,
and only for the lowest Fourier coefficient $h_{1,0}$. 
As in fig.3b (left column), the bullets mark the values corresponding to 
{\sl the most scarred state above the fixed point\/} (0,0) for each $N$,
and the dashed curve connects these values above the special subsequence $N=2^k$. 

Fig. 14.
Poincar\'e's upper-half-plane $\{\Im(\tau)>0\}$. The fundamental domain of ${\rm PSL}(2,{\Bbb Z})$ is bounded by the 3 pieces of hyperbolic geodesics drawn as thick curves. The horizontal dashed line corresponds to $\Im(\tau)=t_o$: above this line, $\Re(\eta'/\eta)$ has the sign of $\sin(2\pi\Re(\tau))$, as marked. The modular transformation $J$ maps the semi-circle $(\alpha,B",A',\beta)$ onto the vertical axis $(\alpha',B',A,\beta')$. Large arrows indicate directions of increasing values of ${\rm GM}[H_{\chi(\tau)}]$ along the semi-circles and the vertical axes $\Re(\tau)=0,\,\pm1/2$. The maxima of ${\rm GM}[H_{\chi(\tau)}]$ are the points $A,A'$ and their modular images. 

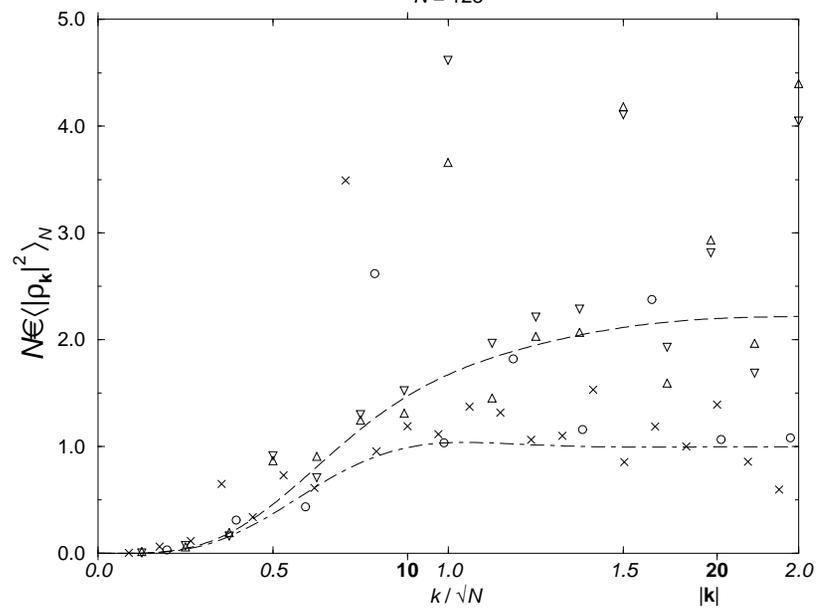
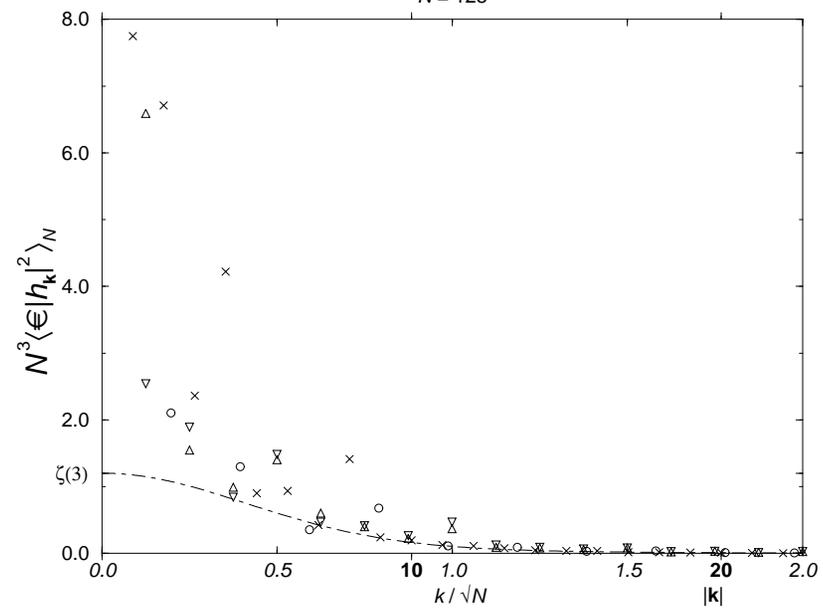
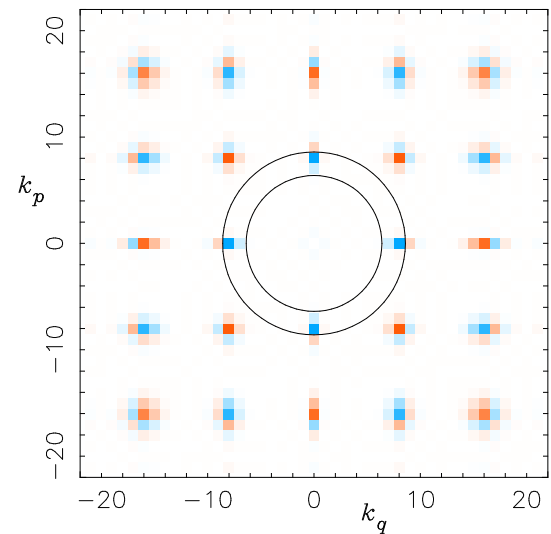
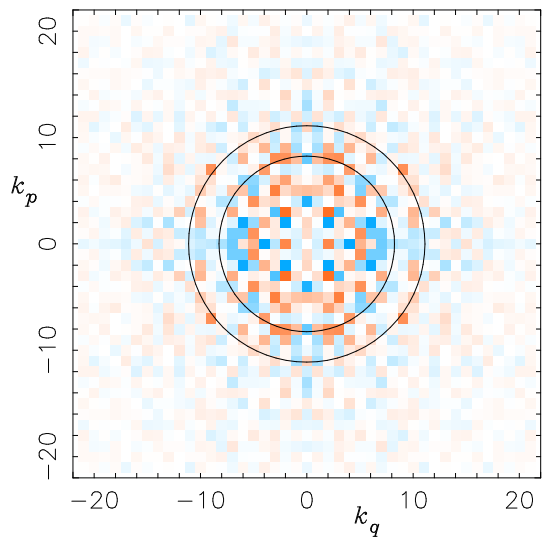
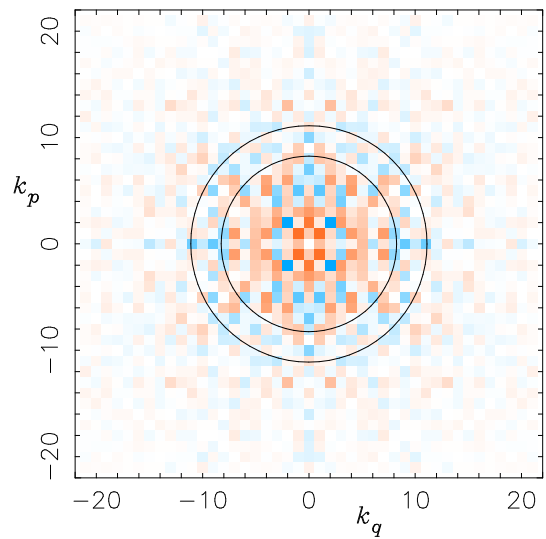
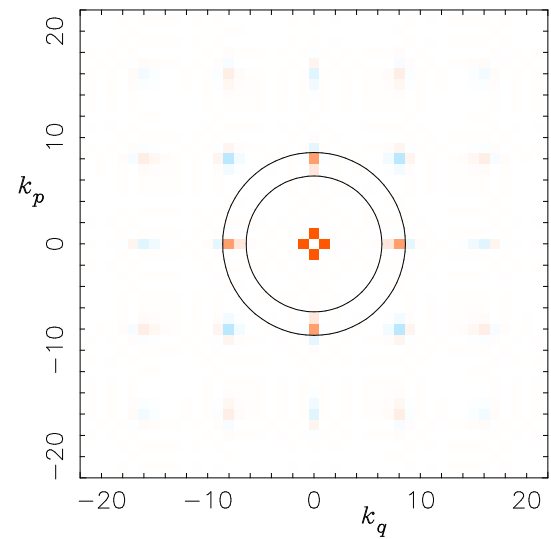
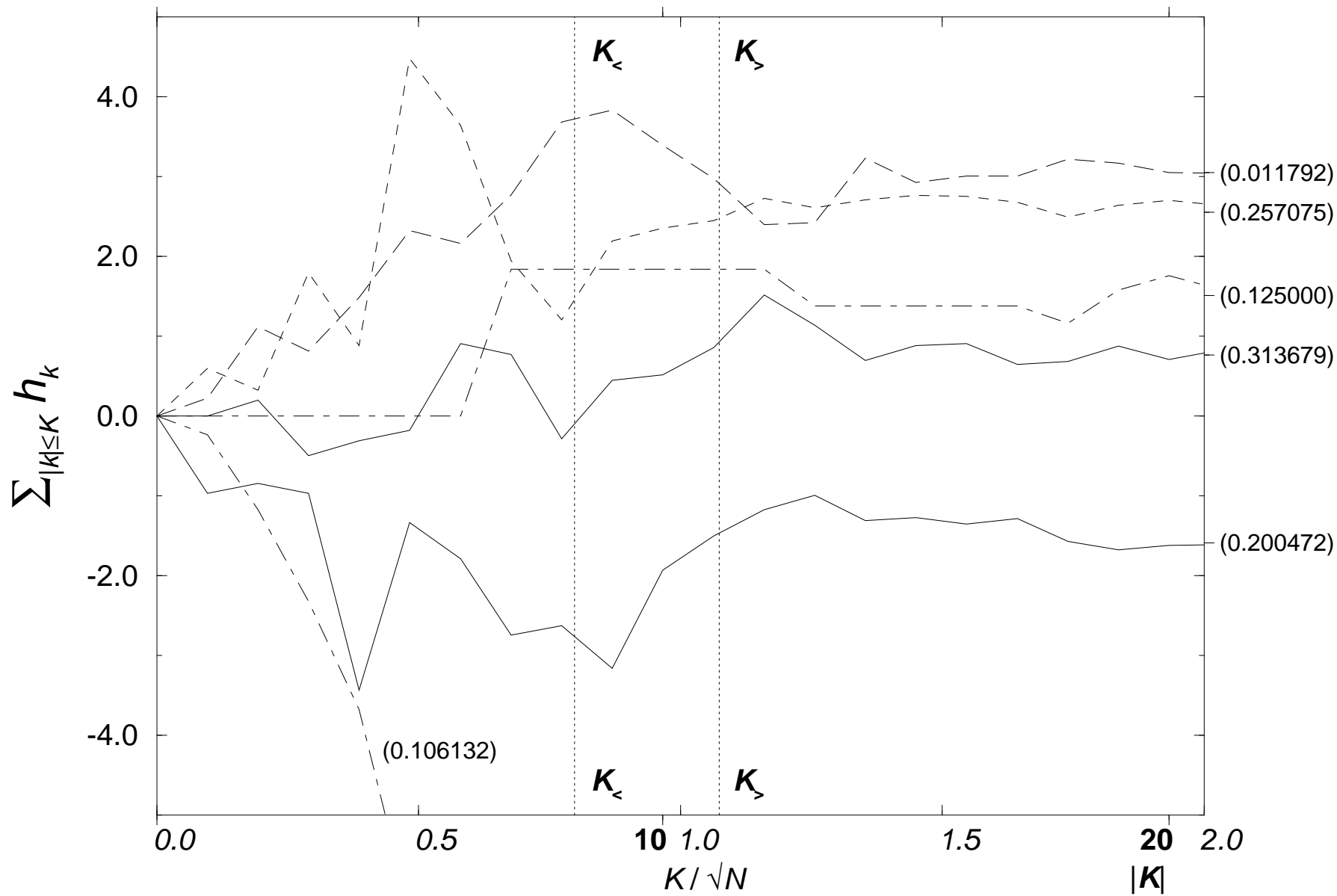
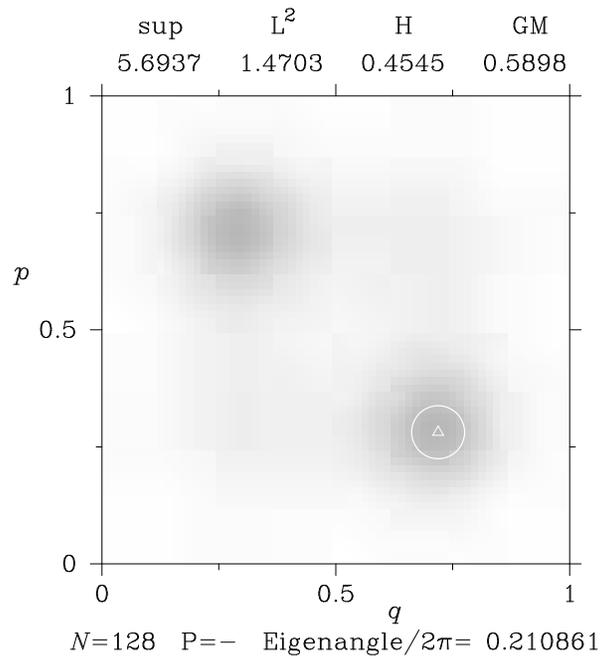
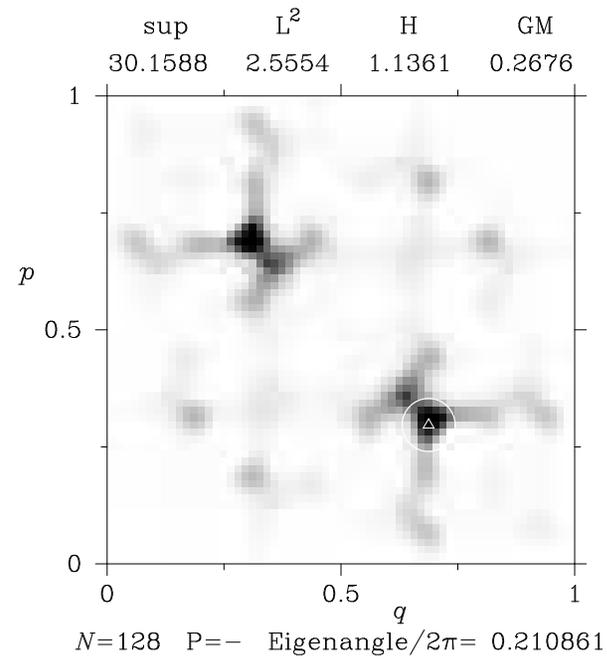
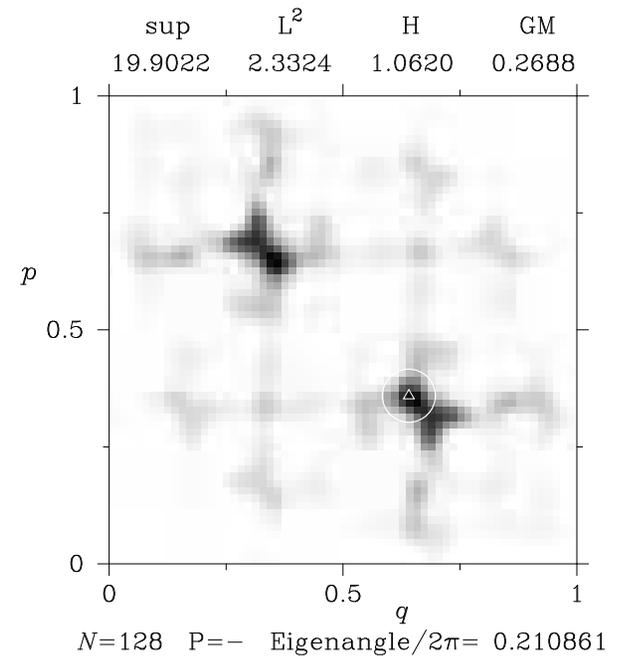
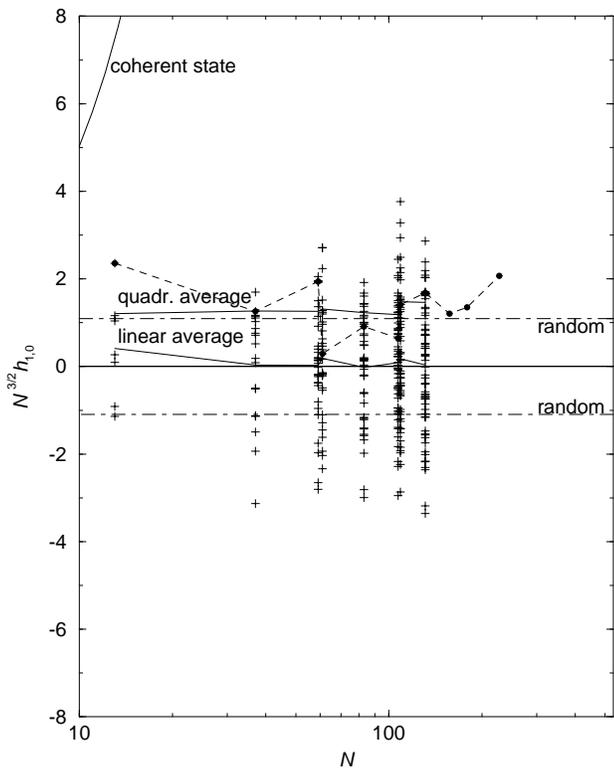
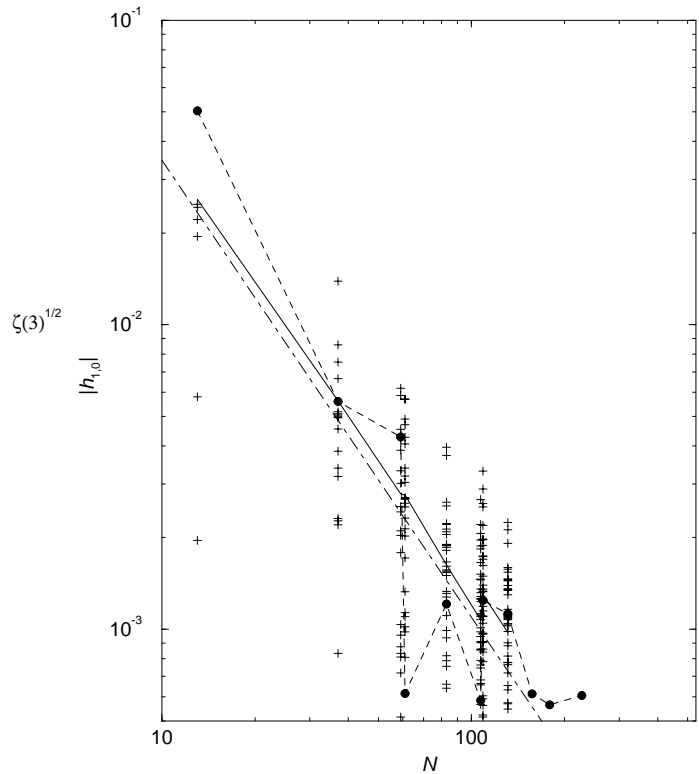
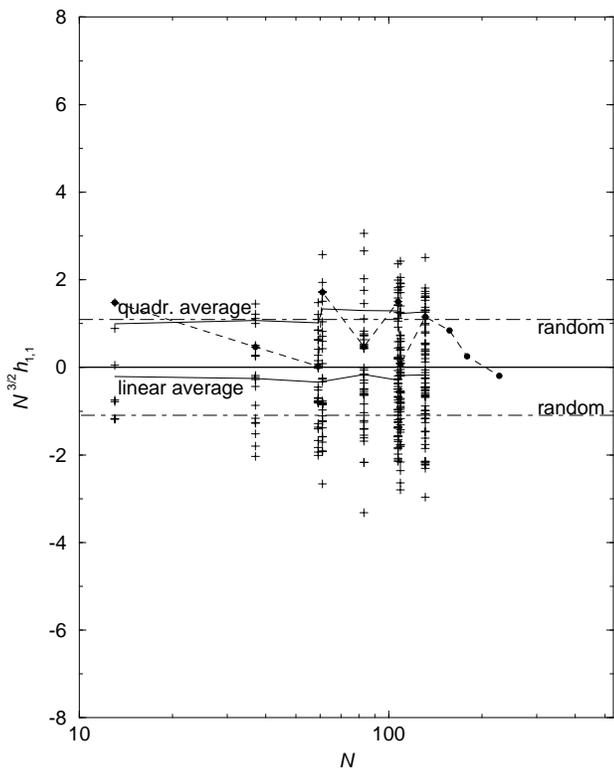
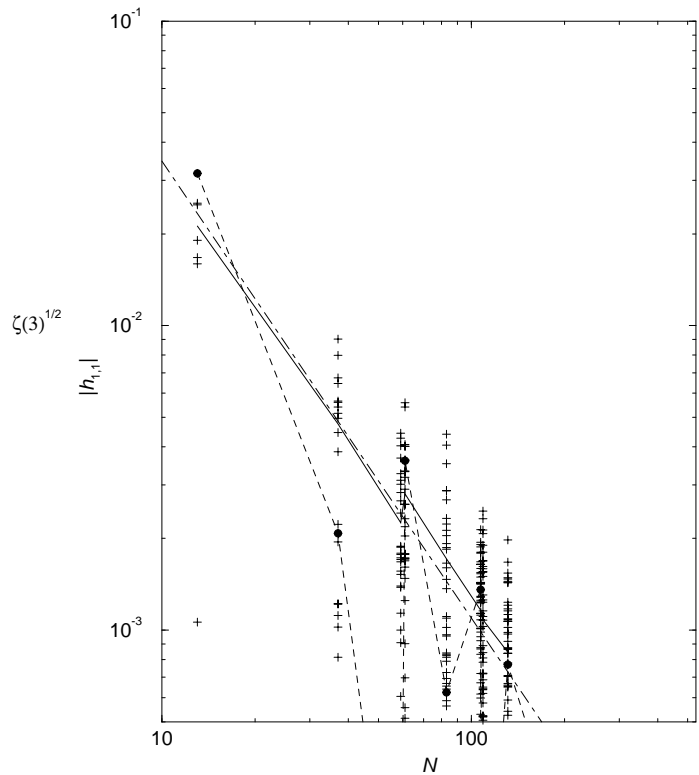
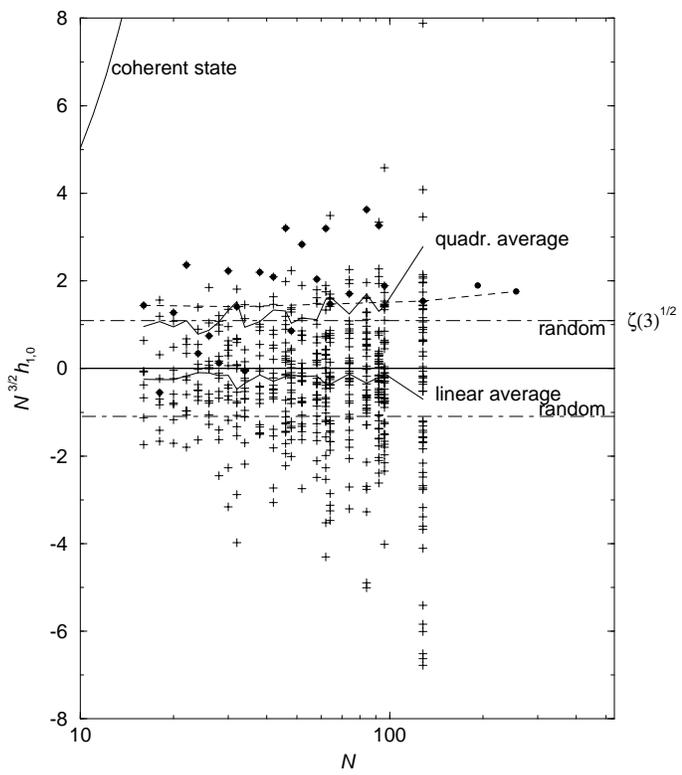
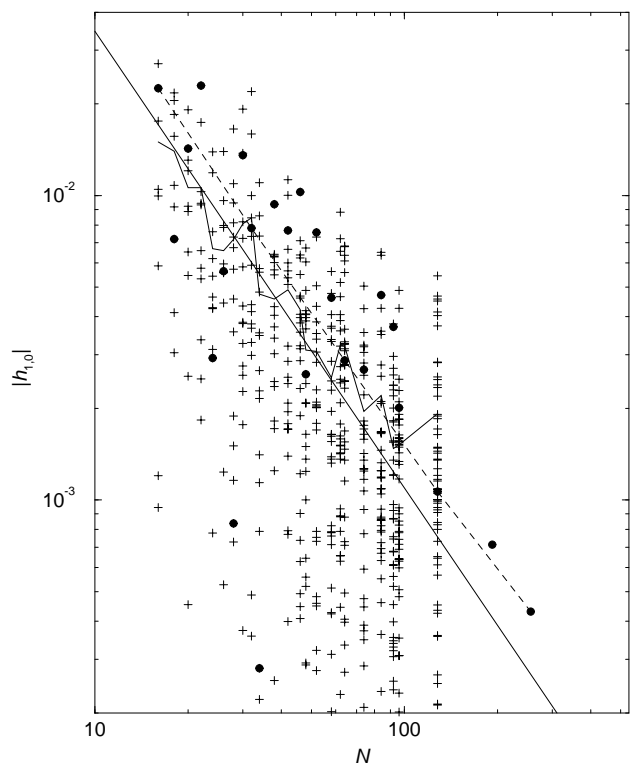
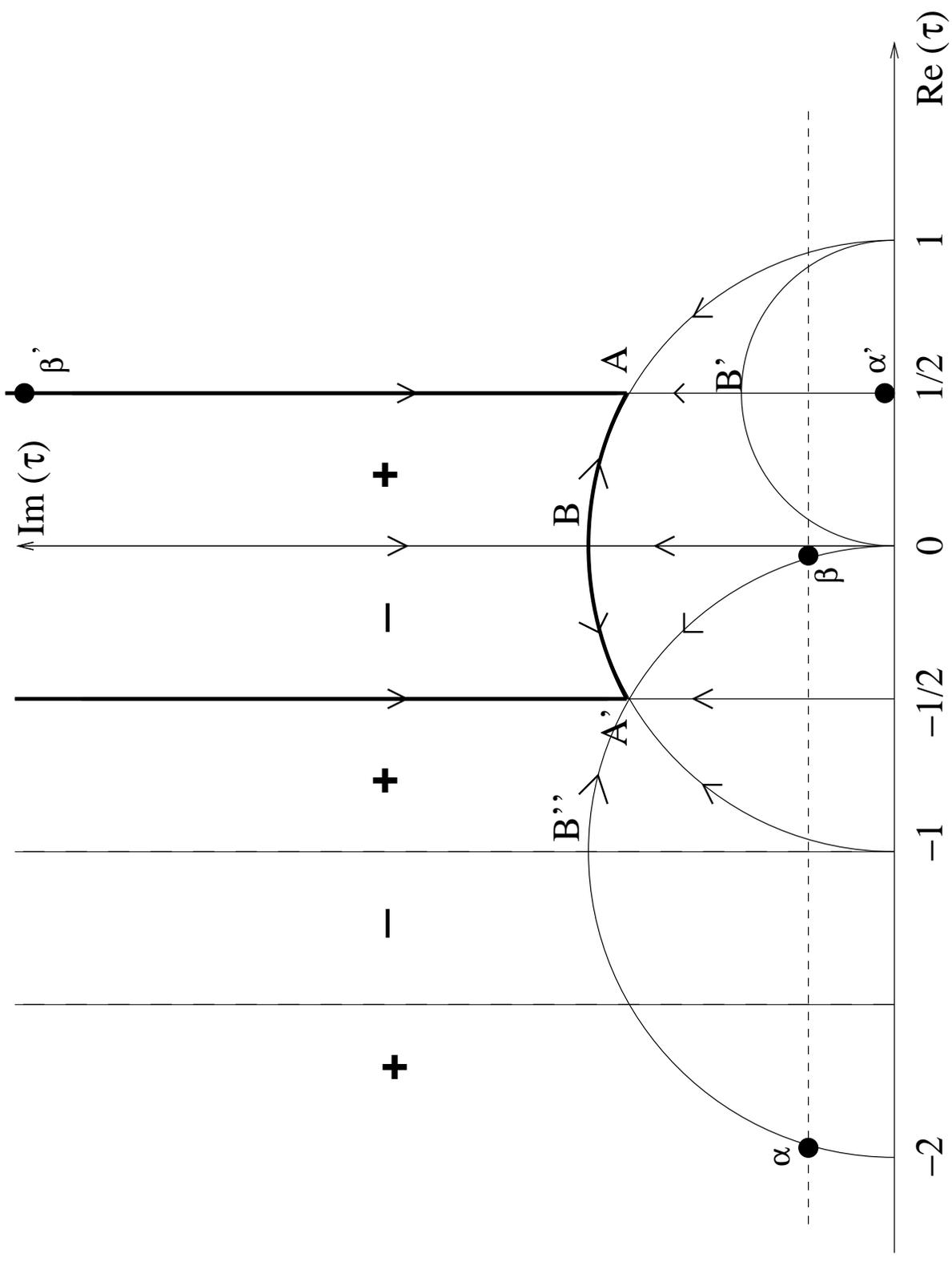
\end{document}